%% file: Driver_Model.tex
\documentclass[11pt,reqno]{amsproc}
\usepackage{longtable}
\usepackage{lettrine}
\usepackage{color}
\usepackage[T1]{fontenc}
\usepackage[semicolon,square,authoryear]{natbib}
\numberwithin{equation}{section}
\usepackage{amscd,amssymb}
\usepackage{epsfig,amsmath}
\usepackage[debug=false, colorlinks=true, pdfstartview=FitV, 
  linkcolor=blue, citecolor=blue, urlcolor=blue]{hyperref}
\usepackage{cite}
\usepackage{epstopdf}
\usepackage{graphicx}
\usepackage{psfrag}
\usepackage{multirow}
\usepackage{fullpage}
\usepackage{subfigure}
\usepackage{enumerate}
\usepackage[small]{caption}
\usepackage{algorithmic}
\usepackage{algorithm}
\newtheorem{theorem}{Theorem}[section]

\newtheorem{remark}[theorem]{Remark}

\usepackage{array}
\newcommand{\PreserveBackslash}[1]{\let\temp=\\#1\let\\=\temp}
\newcolumntype{C}[1]{>{\PreserveBackslash\centering}p{#1}}
\newcolumntype{R}[1]{>{\PreserveBackslash\raggedleft}p{#1}}
\newcolumntype{L}[1]{>{\PreserveBackslash\raggedright}p{#1}}
\usepackage{morefloats}

\newlength{\drop}
\definecolor{amethyst}{rgb}{0.6, 0.4, 0.8}
\definecolor{burgundy}{rgb}{0.5, 0.0, 0.13}

\title{Material degradation due to moisture and temperature \\
  {\small Part 1:~Mathematical model, analysis, and analytical 
    solutions}}
\author{\textbf{C.~Xu}, \textbf{M.~K.~Mudunuru}, and \textbf{K.~B.~Nakshatrala} \\
{\small Department of Civil and Environmental Engineering, 
University of Houston. \\
\textbf{Correspondence to:}~\textsf{knakshatrala@uh.edu}}}

\begin{document}
\date{\today}

%===========================;
%  Title page of the paper  ;
%===========================;
\begin{titlepage}
  \drop=0.1\textheight
  \centering
  \vspace*{\baselineskip}
  \rule{\textwidth}{1.6pt}\vspace*{-\baselineskip}\vspace*{2pt}
  \rule{\textwidth}{0.4pt}\\[\baselineskip]
       {\LARGE \textbf{\color{burgundy}
           Material degradation due to moisture and temperature \\[0.5\baselineskip] 
           {\large
             Part 1:~Mathematical model, analysis, and analytical solutions}}}\\[0.3\baselineskip]
       \rule{\textwidth}{0.4pt}\vspace*{-\baselineskip}\vspace{3.2pt}
       \rule{\textwidth}{1.6pt}\\[\baselineskip]
       \scshape
       %       An e-print of the paper will be made available on arXiv. \par 
       \vspace*{1\baselineskip}
       Authored by \\[\baselineskip]
       
  {\Large C.~Xu\par}
  {\itshape Graduate Student, University of Houston}\\[0.75\baselineskip]
    
  {\Large M.~K.~Mudunuru\par}
  {\itshape Graduate Student, University of Houston}\\[0.75\baselineskip]
    
  {\Large K.~B.~Nakshatrala\par}
  {\itshape Department of Civil \& Environmental Engineering \\
  University of Houston, Houston, Texas 77204--4003 \\ 
  \textbf{phone:} +1-713-743-4418, \textbf{e-mail:} knakshatrala@uh.edu \\
  \textbf{website:} http://www.cive.uh.edu/faculty/nakshatrala\par}
  \vspace{0.1in}
  \begin{figure}[h]
    \includegraphics[scale=0.3,clip]{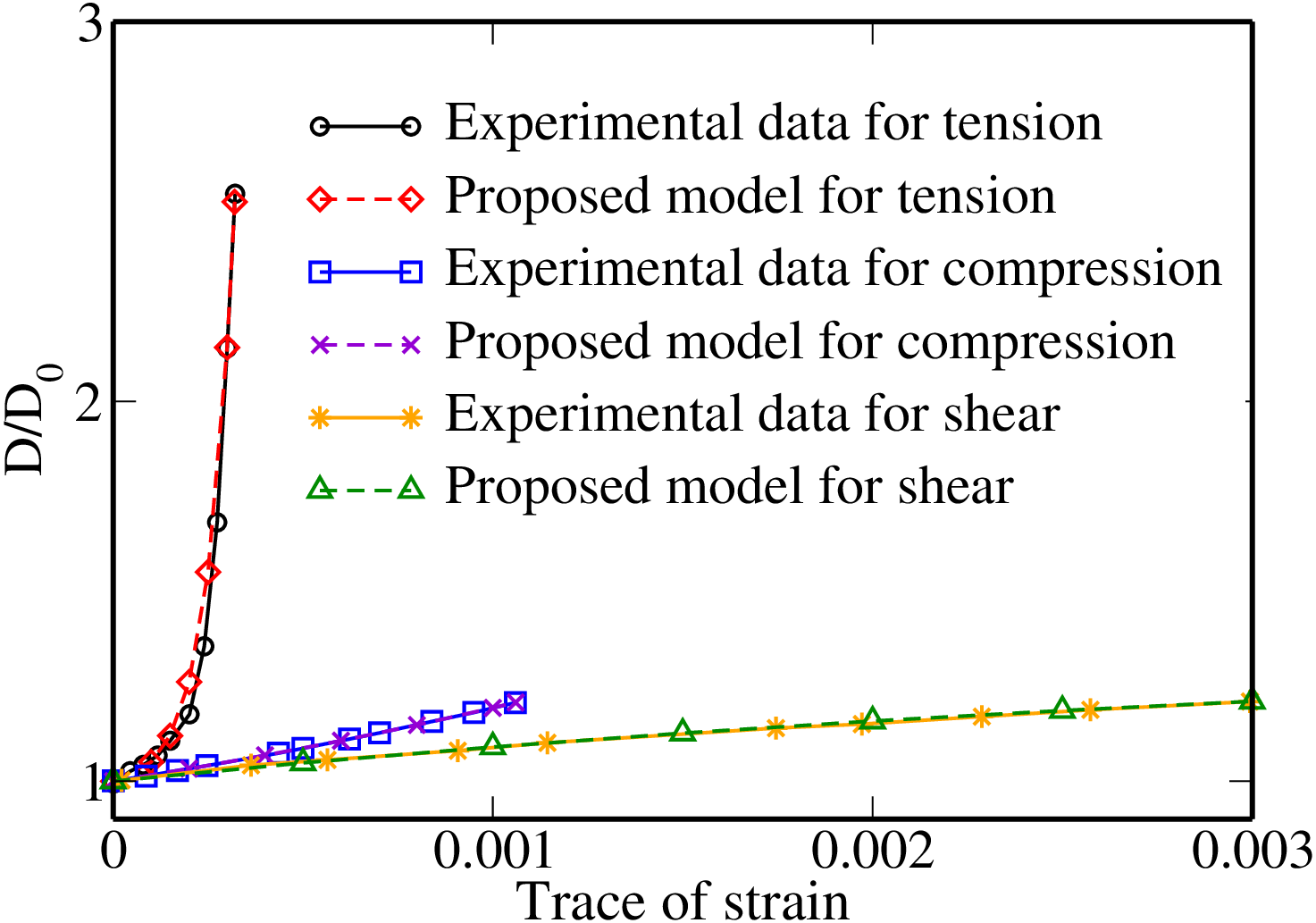}
    \caption*{{\small This figure shows a good agreement 
        between the experimental data and 
        the proposed constitutive model for the diffusivity 
        under tension, compression and shear. Analysis is 
        performed for various extracted sample sizes, and 
        the coefficient of determination is found to be 
        close to 1. \emph{This calibration study provides 
        confidence in the proposed model to apply for brittle 
        materials.}}}
  \end{figure}
  \vfill
  {\scshape 2016} \\
  {\small Computational \& Applied Mechanics Laboratory} \par
\end{titlepage}

%============;
%  Abstract  ;
%============;
\begin{abstract}
  The mechanical response, serviceability, and load bearing 
  capacity of materials and structural components can be 
  adversely affected due to external stimuli, which include 
  exposure to a corrosive chemical species, high temperatures, 
  temperature fluctuations (i.e., freezing-thawing), cyclic 
  mechanical loading, just to name a few. 
  It is, therefore, of paramount importance in several 
  branches of engineering -- ranging from aerospace 
  engineering, civil engineering to biomedical engineering 
  -- to have a fundamental understanding of degradation 
  of materials, as the materials in these applications 
  are often subjected to adverse environments. 
  As a result of recent advancements in material science, new 
  materials like fiber-reinforced polymers and multi-functional 
  materials that exhibit high ductility have been developed and 
  widely used; for example, as infrastructural materials 
  or in medical devices (e.g., stents). The traditional small-strain 
  approaches of modeling these materials will not be adequate. 
  In this paper, we study degradation of materials due to 
  an exposure to chemical species and temperature under 
  large-strain and large-deformations. In the first part 
  of our research work, we present a consistent mathematical 
  model with firm thermodynamic underpinning. We then obtain 
  semi-analytical solutions of several canonical problems 
  to illustrate the nature of the quasi-static and unsteady 
  behaviors of degrading hyperelastic solids. 
\end{abstract}
\maketitle
\setcounter{page}{1}

\input{Sections_Model/S0_Model_Notation}

\input{Sections_Model/S1_Model_Intro}

\input{Sections_Model/S2_Model_GE}

\input{Sections_Model/S3_Model_MODEL}
\input{Sections_Model/S4_Model_Calibration}

\input{Sections_Model/S5_Model_IBVP}

\input{Sections_Model/S6_Model_Canonical}

\input{Sections_Model/S7_Model_CR}

%===================;
%  Acknowledgments  ;
%===================;
\section*{ACKNOWLEDGMENTS}
The authors acknowledge the support from the Department 
of Energy through Nuclear Energy University Programs 
(NEUP). The opinions expressed in this paper are those 
of the authors and do not necessarily reflect that of 
the sponsor(s).

%==============================;
%  Include all the references  ;
%==============================;
\bibliographystyle{plainnat}
\bibliography{Master_References/Master_References,Master_References/Books}
\input{Sections_Model/Model_Figures}
\end{document}

%% file: Sections_Model/S0_Model_Notation.tex
\section*{NOMENCLATURE} 
\label{Sec:S0_Notation}
\begin{longtable*}{|p{.15\textwidth} | p{.50\textwidth}|} \hline
  \small
  $\rho$ & density of solid in deformed configuration 
  $[\mathrm{kg}\,\mathrm{m}^{-3}]$ \\
  $A$ & specific Helmholtz potential 
  $[\mathrm{J}\,\mathrm{kg}^{-1}]$ \\
  $\zeta$ & dissipation functional 
  $[\mathrm{J}\,\mathrm{kg}^{-1}\mathrm{s}^{-1}]$ \\
  $\psi$ & strain energy density functional 
  $[\mathrm{J}\,\mathrm{m}^{-3}]$ \\ 
  $\lambda$, $\mu$ & Lam\'e parameters $[\mathrm{Pa}]$ \\
  $\kappa$ & bulk modulus $[\mathrm{Pa}]$ \\
  $\mathbf{u}$ & displacement $[\mathrm{m}]$ \\    
  $\mathbf{v}$ & velocity $[\mathrm{m}\, \mathrm{s}^{-1}]$ \\    
  $\vartheta$ & temperature $[\mathrm{K}]$ \\
  $c$ & concentration $[1]$ \\
  $R_s$ & specific vapor constant 
  $[\mathrm{J}\,\mathrm{kg}^{-1}\mathrm{K}^{-1}]$ \\ 
  $c_p$ & heat capacity 
  $[\mathrm{J}\,\mathrm{kg}^{-1}\mathrm{K}^{-1}]$ \\ 
  $\mathbf{M}_{\vartheta \mathbf{E}}$ & 
  thermal expansion tensor $[\mathrm{J}\,\mathrm{m}^{-3}\mathrm{K}^{-1}]$ \\
  $\mathbf{M}_{c \mathbf{E}}$ & 
  chemical expansion tensor $[\mathrm{J}\,\mathrm{m}^{-3}]$ \\ 
  $d_{\vartheta c}$ & thermo-chemo 
  coupled parameter $[\mathrm{J}\,\mathrm{kg}^{-1}\mathrm{K}^{-1}]$ \\
  $\varkappa$ & specific chemical potential 
  $[\mathrm{J}\,\mathrm{kg}^{-1}]$ \\
  $\eta$ & specific entropy 
  $[\mathrm{J}\,\mathrm{kg}^{-1}\mathrm{K}^{-1}]$ \\
  $\mathbf{D}_{\vartheta \vartheta}$ & thermal diffusion tensor 
  $[\mathrm{m}^2\mathrm{s}^{-1}]$ \\
  $\mathbf{D}_{\varkappa \varkappa}$ & diffusivity 
  tensor $[\mathrm{m}^2\mathrm{s}^{-1}]$ \\
  $\mathbf{D}_{\vartheta \varkappa}$, $\mathbf{D}_
  {\varkappa \vartheta}$ & Dufour-Soret effect tensors 
  $[\mathrm{m}^2\mathrm{s}^{-1}]$ \\
  $\mathbf{T}$ & Cauchy stress $[\mathrm{Pa}]$ \\
  $\mathbf{h}$ & diffusive flux vector 
  $[\mathrm{kg}\,\mathrm{m}^{-2}\mathrm{s}^{-1}]$ \\
  $\mathbf{q}$ & heat flux vector 
  $[\mathrm{J}\,\mathrm{m}^{-2}\mathrm{s}^{-1}]$ \\
  $h$ & volumetric source 
  $[\mathrm{kg}\,\mathrm{m}^{-3}\mathrm{s}^{-1}]$ \\
  $q$ & volumetric heat source 
  $[\mathrm{J}\,\mathrm{m}^{-3}\mathrm{s}^{-1}]$ \\ \hline
\end{longtable*}

%% file: Sections_Model/S1_Model_Intro.tex
%********************************************;
%                                            ;
%  NAME                                      ;
%    S1_Model_Intro.tex                      ;
%                                            ;
%  WRITTEN BY                                ;
%    Can Xu                                  ;
%    Maruti Kumar Mudunuru                   ;
%    Kalyana Babu Nakshatrala                ;
%                                            ;
%********************************************;
\section{INTRODUCTION AND MOTIVATION} 
\label{Sec:S1_CDD_Intro}
\lettrine[lines=2]{\color{red}M}{aterial} and structural 
degradation is a major problem in infrastructure and various 
other real-life applications. Most of the well-known manifestations, 
such as ``wear out'', ``fracture'', ``spalling'', and ``section loss'', 
are related to the phenomenon of degradation \citep{Batchelor_Lam_Chandrasekaran}. 
Virtually, every material degrades when subjected to hostile 
environment and external stimuli. Importance of this phenomena 
has triggered a surge in research to develop more resistible 
materials. Consequently, understanding the general behavior 
of degrading materials has attracted the interest of researchers. 
A fundamental study of degradation is crucial to several branches 
of engineering:~aerospace, mechanical, civil, and biomedical.
Moreover, some new materials, such as fiber-reinforced polymers 
and multi-functional materials that exhibit high ductility have 
been widely used recently; for example, as infrastructural materials 
or in medical devices (e.g., stents). In order to model these materials, 
the traditional small-strain assumption will not be sufficient anymore. 

In a nutshell, degradation means the loss in either serviceability 
or functionality. To be precise, a material is said to be undergoing 
thermal degradation at a spatial point $\mathbf{x} \in \Omega$ if the 
available isothermal density is lower than the reference available 
isothermal power at that particular point. That is,
%-----------------------------------;
%  Definition: Thermal degradation  ;
%-----------------------------------;
\begin{align}
  \label{Eqn:Thermal_Degradation_Definition}
  \left.\frac{dA}{dt}\right|_{\vartheta>
  \vartheta_{\mathrm{ref}}}\leqslant
  \left.\frac{dA}{dt}\right|_{\vartheta = 
  \vartheta_{\mathrm{ref}}} \ \mathrm{for}
  \ \mathbf{x}\in\Omega
\end{align}
Similarly, the chemical/moisture degradation 
can be defined as follows:
%------------------------------------;
%  Definition: Chemical degradation  ;
%------------------------------------;
\begin{align}
\label{Eqn:Thermal_Degradation_Definition}
  \left.\frac{dA}{dt}\right|_{c>
  c_{\mathrm{ref}}}\leqslant \left.
  \frac{dA}{dt}\right|_{c=c_{\mathrm{ref}}} 
  \ \mathrm{for} \ \mathbf{x}\in\Omega
\end{align}
where $A$ denotes the specific Helmholtz potential of the material. 
$\Omega$ is the degrading body under consideration, $t$ is the time 
of interest, $\vartheta_{\mathrm{ref}}$ and $c_{\mathrm{ref}}$ are 
the specified reference temperature and reference concentration. 
Note that degradation not only reduces the durability of materials 
but also alters material properties. For instance, material damage 
can induce anisotropy in thermal conductivity and diffusivity 
\citep{1975_Peng_Landel_JAPS_v19_p49_p68,
2004_Venerus_Schieber_Balasubramanian_Bush_Smoukov_PRL,Zheng_Tanner_Fan}.

Herein, we develop a coupled continuum mathematical model 
for thermal and chemical-induced degradation of solids, 
which are initially hyperelastic.
We now outline three main reasons for such a need.
%===================================================;
%  Four main reasons and contributions of our work  ;
%===================================================;
\begin{itemize}
  \item There is irrefutable experimental evidence that many 
    modern infrastructural materials used in repair and retrofitting 
    applications exhibit large deformations. For example, the 
    popular high-early-strength Engineered Cementitious Composites 
    (ECC) are capable of delivering a compressive strength of 21 
    MPa within 4 hours after placement. Moreover, the long-term 
    tensile strain capacity of ECC members is more than 2\% 
    \citep{2003_Li_JACT_v1_p215_p230,2006_Wang_Li_ACIMJ_v103_p97_p105}.
  \item In order to understand degradation mechanisms due to 
    moisture, chemical, and temperature, coupling at various 
    levels is needed (which is due to balance laws, material 
    parameters, boundary conditions, and initial conditions). 
    With existing and popular multi-physics packages such 
    \textsf{ABAQUS} \citep{Abaqus_6.14-1}, \textsf{ANSYS} 
    \citep{Ansys_16.0}, and \textsf{COMSOL} \citep{Comsol_5.0-1}, 
    it is possible to couple certain degradation mechanisms 
    to some extent at material parameters, boundary conditions, 
    and initial conditions. However, such packages \emph{do 
    not offer} flexibility to couple important heat and mass 
    transfer terms in balance laws. This is of utmost importance 
    in capturing the effects of chemo-thermo-mechano degradation.
  \item Finally, when a new model or a thermodynamic framework 
    is developed, stability of the solutions for the corresponding 
    initial boundary value problem needs to be shown. However, such 
    an analysis is rarely performed when a new degradation model/framework 
    is developed in literature. Herein, for the proposed degradation 
    framework we shall perform stability analysis in the sense of 
    Lyapunov. Subsequently, this methodology shall be used to construct 
    a robust computational framework in the part-II of the paper.
\end{itemize}
Hence, due to the above reasons small strain assumptions to 
model degradation and healing behavior of these infrastructural 
systems are rarely valid. The proposed framework takes in to
account the underlying degradation mechanisms. Correspondingly,
the respective parameters have a physical meaning and can be
calibrated through experiments.

It should be emphasized that elasticity is an idealization. 
There is no material whose response is perfectly elastic. 
But there are situations in which the response of certain 
materials under normal conditions can be idealized to be 
hyperelastic. For example, large blood arteries and rock. 
Many of these materials function in hostile environments, 
and are constantly subjected to adverse external stimuli. 
One often is interested in the unsteady response of the 
bodies made of hyperelastic materials subjected to degradation/healing.
The application areas in mind are the response of high 
performance cementitious materials (which undergo large 
strains and large deformations) and several important 
coupled deformation-thermal-transport processes in 
biomechanics and biomedicine. In the next couple of 
subsections, we shall discuss various degradation 
mechanisms and the deficiencies in the existing 
frameworks in modeling chemo-thermo-mechano degradation.

%======================================;
%  Subsection: Degradation mechanisms  ;
%======================================;
\subsection{Degradation mechanisms}
\label{SubSec:mechanisms}
There are many mechanisms that can result in the degradation of 
materials and structures. In general, the degradation mechanisms 
can be divided into four categories: mechanical processes, chemical 
reactions, biological degradation \citep{1998_Gu_etal_IBB_v41_p101_p109}, 
and radiation \citep{Kaplan}. 
For mechanical processes, the performance of materials can be affected 
adversely by fatigue \citep{2000_Jung_etal_JDR_v79_p722_p731}, pressure 
loading \citep{2007_Rajagopal_Srinivasa_Wineman_IJP_v23_p1618_p1636},  
and swelling of solid mixtures \citep{2011_Buonsanti_Leonard_Scoppelliti_AMM_v52_p389_p392}. 
Examples of chemical degradation include humid and alkaline effects 
\citep{2003_Bjork_etal_CBM_v17_p213_p221}, exposure to chlorides and 
carbon-dioxide \citep{2008_Glasser_Marchand_Samson_CCR_v38_p226_246}, 
and calcium leaching \citep{2009_Gawin_Pesavento_Schrefler_CMAME_v198_p3051_3083}. 
Biological degradation refers to the dissolution 
of materials by bacteria or other microorganisms. 
Degradation induced by radiation includes radiation 
damage as well as other mechanical and chemical 
processes triggered by radiation. 

%--------------------------------------------------;
%  Table: Degradation mechanisms and consequences  ;
%--------------------------------------------------;
\begin{table}[h]
  \centering
  \caption{Various degradation mechanics and their 
    primary manifestation. Many other factors 
    can be found in \citep{2007_Naus_ORNL}.
  \label{Table:Degradation_Mechanisms_Concrete}}
  \begin{tabular}{|c|c|c|}
    \hline
    & Degradation \ factor &Primary \ manifestation\\  \hline  
    \multirow{5}{*}{Physical processes} & cracking & 
    reduced \ durability \\ 
    & vibration & cracking  \\
    & freezing and thawing & cracking/scaling/disintegration  \\
    & abrasion/erosion/cavitation & section loss  \\
    &thermal exposure/thermal cycling & 
    cracking/spalling/strength loss  \\
    \hline
    \multirow{5}{*}{Chemical processes} & 
    efflorescence/leaching & increased porosity \\ 
    & phosphate& surface deposits  \\
    & sulfate attack & volume change/cracking  \\
    & acids/bases & disintegration/spalling/leaching  \\
    & alkali-aggregate reactions & disintegration/cracking  \\
   \hline
  \end{tabular}
\end{table}

The coupling effects between these mechanisms can have 
a significant impact on the rate of deterioration of materials and 
structures. For instance, see Table \ref{Table:Degradation_Mechanisms_Concrete} 
for some important factors that affect the degradation modeling in 
infrastructural materials such as concrete. Therefore, developing 
an appropriate and general model for material degradation is useful
to predict the life span of a given structure. A comprehensive 
understanding of chemo-thermo-mechano degradation not only plays a 
pivotal role in improving the quality and reliability of existing 
infrastructure, but also has a tremendous impact on the economy 
\citep{2013_ASCE_FactSheet}. In this paper, we shall assume that 
predominant degradation mechanisms are moisture and temperature. 
We propose a general three-way strongly coupled degradation model 
based on a thermodynamic framework. This three-way coupling is 
between mechanical, thermal, and transport processes. 

%==========================================================;
% Subsection: Thermodynamics of chemo-mechano degradation  ;
%==========================================================;
\subsection{Thermodynamics of chemo-thermo-mechano degradation}
\label{Eqn:Thermodynamics_ChemoThermoMechano}
Herein, we shall provide a brief review and current status of thermal 
and chemical degradation. In the literature, thermal degradation is 
modelled based on variants of thermoelasticity by incorporating damage 
variables. Some popular research works in this direction are \citep{2005_Willam_Rhee_Xi_JMCE_v17_p276_285} 
for modeling thermo-mechanical damage processes in heterogeneous cementitious 
materials and \citep{2013_Allam_Elbakry_Rabeai_AEJ_v52_p749_761} on the behavior 
of reinforced concrete slabs exposed to fire.
On the other hand, some popular research works for the chemical 
degradation are \citep{2003_Bjork_etal_CBM_v17_p213_p221} on 
the environmental effects of alkalinity and humidity on concrete 
slabs, \citep{2010_Cho_Kim_KSCE_v14_p333_p341} on moisture damage 
mechanisms occurring within asphaltic materials and pavements, 
\citep{1990_Bouadi_Sun_JMS_v25_p499_p505} on thermal and moisture 
effects on structural stiffness and damping of laminated composites,
and \citep{2002_Weitsman_Guo_CST_v62_p889_p908,2006_Weitsman_ASM_v37_p617_p623} 
on fluid-induced damage and absorption in polymeric composites.
However, none of the above mentioned papers on thermal or chemical 
degradation have a proper thermodynamic basis.

There are two popular approaches to construct thermodynamically-consistent 
degradation models. The first approach is based on the theory of the internal 
variable, wherein a scalar (or a tensor) variable is introduced to model the 
degree of damage \citep{1987_Weitsman_IJSS_v23_p1003_p1025,2004_Grasberger_Meschke_MS_v37_p244_p256,
2009_Springman_Bassani_JMPS_v57_p909_p931,2007_Rajagopal_Srinivasa_Wineman_IJP_v23_p1618_p1636}.
For instance, the damage variable may represent the measure of the 
fraction of broken cross-links or micro-cracks in a representative 
volume element of the body \citep{Kachanov,Lemaitre_Desmorat,Voyiadjis_Kattan}.
The main disadvantage of this approach is that it is difficult (or 
sometimes impossible) to measure the internal variables through 
experiments or associate them to physical quantities/parameters.

The second approach is to build a thermodynamic framework by 
modeling all the relevant coupled processes. This achieved by 
taking into account the dependence of material properties on 
the deformation of the solid, temperature, and concentration 
of chemical species. The degradation parameters under this 
approach have physical basis and can be calibrated using 
experiments (for example, see Section \ref{Sec:S5_Model_IBVP} 
of this paper). Herein, we shall employ the second approach 
to develop a thermodynamically consistent degradation model. 
It should be noted that certain research works exist in 
literature wherein the degradation models 
using the second approach. For example, see 
\citep{2009_Muliana_Rajagopal_Subramanian_JCM_v43_p1225_p1249,
2009_Darbha_Rajagopal_IJNM_v44_p478_p485,2012_Karra_Rajagopal_MTDM_v16_p85_p104,
2014_Klepach_Zohdi_CPBE_v56_p413_p423}. However, it appears that
the above cited works suffer from the main drawback that they considered 
thermodynamics of chemo-thermo-mechano degradation in the context of 
a closed system as opposed to an open system, which is the approach 
taken in this paper. Moreover, the models are not as comprehensive as
the one proposed in this paper.

%=================================;
% Subsection: Scope of the paper  ;
%=================================;
\subsection{Scope of the paper}
In this paper we set out to achieve the 
following objectives: 
%-----------------------------------------;
%  Enumeration of the main contributions  ;
%-----------------------------------------;
\begin{enumerate}[(i)]
  \item We derive a general chemo-thermo-mechano 
    degradation model by appealing to the maximization 
    of rate of dissipation. It will also be shown that 
    many popular models are special cases of the proposed 
    mathematical model. For example, we will show that the 
    small-strain moisture degradation model proposed in 
    \citep{Mudunuru_Nakshatrala_IJNME_2011} is a special 
    case of the proposed model. 
%    We also provide the 
%    thermodynamic basis of the degradation model proposed 
%    in \citep{Mudunuru_Nakshatrala_IJNME_2011}, which has 
%    not been addressed earlier.
    %
  \item We will calibrate the proposed degradation 
    model with existing experimental data sets. This 
    calibration study should provide confidence in 
    employing the proposed constitutive model to model 
    degradation of various brittle and quasi-brittle 
    materials like ceramics, glass fibers, and concrete.
  \item A systematic mathematical analysis is presented for 
    the proposed model under large/finite deformations. In 
    particular, we shall show that the unsteady solutions 
    under the proposed degradation model are bounded and 
    are stable in the sense of Lyapunov. 
  \item Last but not the least, semi-analytical 
    solutions to several canonical problems are 
    presented, which provide insights into the 
    behavior of degrading structural members. 
    This will be valuable for developing better 
    design and safety codes.
\end{enumerate}

%===========================;
%  An outline of the paper  ;
%===========================;
The rest of the paper is organized as follows. Section 
\ref{Sec:S2_Degradation_GE} introduces the notation, 
mathematical preliminaries, and the relevant balance 
laws. 
Section \ref{Sec:ChemoMechano_DegradModel} presents 
a mathematical model for degradation of materials due 
to moisture and temperature, which is valid even under 
finite deformations and large strains. The constitutive 
relations are obtained by appealing to the maximization 
of rate of dissipation hypothesis, which ensures that 
the constitutive model satisfies the second law of 
thermodynamics \emph{a prior}. 
In Section \ref{Sec:S4_Model_Calibration}, the proposed 
model is calibrated with an experimental dataset. 
The coupled initial boundary value problem arising 
from the proposed degradation model is presented 
in Section \ref{Sec:S5_Model_IBVP}. We also show 
the solutions of the proposed mathematical model 
are bounded and stable. 
In Section \ref{Sec:S6_Model_Canonical}, solutions to 
several canonical problems are presented to illustrate 
the predictive capabilities of the proposed model, and 
to highlight the effects of degradation on the structural 
behavior. Finally, conclusions are drawn in Section 
\ref{Sec:S7_Model_CR}.

A list of the main symbols used in the paper are 
provided in the Nomenclature.

%% file: Sections_Model/S2_Model_GE.tex
%***************************************;
%                                       ;
%  NAME                                 ;  
%    S2_Model_GE.tex                    ;
%                                       ;
%  WRITTEN BY                           ;
%     Can Xu                            ;
%     Kalyana Babu Nakshatrala          ;
%                                       ; 
%***************************************;
\section{NOTATION, PRELIMINARIES, AND BALANCE LAWS}
\label{Sec:S2_Degradation_GE}
Let us consider a body $\mathfrak{B}$. The body occupies a 
reference configuration $\Omega_{0}(\mathfrak{B}) \subset 
\mathbb{R}^{nd}$, where ``$nd$'' denotes the number of spatial 
dimensions. A point in the reference configuration is denoted 
by $\mathbf{p} \in \Omega_0(\mathfrak{B})$. We shall denote 
the time by $t \in [0,\mathcal{T}]$, where $\mathcal{T}$ is 
the length of the time interval of interest. Due to motion, 
the body occupies different spatial configurations with time. 
We shall denote the configuration occupied by the body at time 
$t$ as $\Omega_{t}(\mathfrak{B}) \subset \mathbb{R}^{nd}$. 
A corresponding spatial point will be denoted as 
$\mathbf{x} \in \Omega_{t}(\mathfrak{B})$. 
The gradient and divergence operators with respect 
to $\mathbf{p}$ are, respectively, denoted by 
$\mathrm{Grad}[\bullet]$ and $\mathrm{Div}[\bullet]$. 
Similarly, the gradient and divergence operators with 
respect to $\mathbf{x}$ are, respectively, denoted by 
$\mathrm{grad}[\bullet]$ and $\mathrm{div}[\bullet]$.

The motion of the body is mathematically described 
by the following invertible mapping:
%--------------------------------;
%  Equation: Motion of the body  ;
%--------------------------------;
\begin{align}
  \label{Eqn:Deform_Config_Motion}
  \mathbf{x} = \boldsymbol{\varphi}(\mathbf{p},t)
\end{align}
The displacement vector field can then be written as:
%--------------------------------;
%  Equation: Displacement field  ;  
%--------------------------------;
\begin{align}
  \label{Eqn:Displacement_Field}
  \mathbf{u} = \mathbf{x} - \mathbf{p} = 
  \boldsymbol{\varphi}(\mathbf{p},t) - \mathbf{p} 
\end{align}
The velocity vector field is defined as:
%----------------------------;
%  Equation: Velocity field  ;  
%----------------------------;
\begin{align}
  \label{Eqn:Velocity_Field}
  \mathbf{v} = \dot{\mathbf{x}} := 
  \frac{\partial \boldsymbol{\varphi}(\mathbf{p},t)}{\partial t} 
\end{align}
where a superposed dot indicates the material/total time 
derivative, which is the derivative with respect to time 
holding the reference coordinates fixed. 
The gradient of motion (which is also referred 
to as the deformation gradient) is defined as: 
%--------------------------------;
%  Equation: Gradient of motion  ;
%--------------------------------;
\begin{align}
  \label{Eqn:Gradient_Of_Motion}
  \mathbf{F} = \mathrm{Grad}[\mathbf{x}] \equiv 
  \frac{\partial \boldsymbol{\varphi}(\mathbf{p},t)}
   {\partial \mathbf{p}} = \mathbf{I} + \mathrm{Grad}
   [\mathbf{u}]
\end{align}
where $\mathbf{I}$ denotes the second-order identity 
tensor. The corresponding right Cauchy-Green tensor 
is denoted by:
%---------------------------------------;
%  Equation: Right Cauchy-Green tensor  ;
%---------------------------------------;
\begin{align}
  \label{Eqn:Right_CauchyGreen_Tensor}
  \mathbf{C} = \mathbf{F}^{\mathrm{T}} \mathbf{F}
\end{align}
where $(\bullet)^{\mathrm{T}}$ denotes the transpose 
of a second-order tensor. The velocity gradient with 
respect to $\mathbf{x}$ and the symmetric part of the 
velocity gradient are, respectively, defined as follows:
%------------------------------------;
%  Equation: Definitions of L and D  ;
%------------------------------------;
\begin{align}
  \label{Eqn:L_Def}
  \mathbf{L} &:= \mathrm{grad}[\mathbf{v}] \equiv 
  \dot{\mathbf{F}} \mathbf{F}^{-1} \\
  \label{Eqn:D_Def}
  \mathbf{D} &:= \frac{1}{2} \left(\mathbf{L} + 
  \mathbf{L}^{\mathrm{T}}\right)
\end{align}	
The Green-St.~Venant strain tensor is defined as:
%--------------------------;
%  Equation: Green strain  ;
%--------------------------;
\begin{align}
  \label{Eqn:GreenVenant_StrainTensor}
  \mathbf{E} =\frac{1}{2}(\mathbf{C}-\mathbf{I}) = 
  \frac{1}{2} \left(\mathrm{Grad}[\mathbf{u}] + 
  \mathrm{Grad}[\mathbf{u}]^{\mathrm{T}} + 
  \mathrm{Grad}[\mathbf{u}]^{\mathrm{T}} 
  \mathrm{Grad}[\mathbf{u}]\right)
\end{align}
In situations the following assumption holds:
%--------------------------------------;
%  Equation: Linear strain assumption  ;
%--------------------------------------; 
\begin{align}
  \label{Eqn:linear}
  \mathop{\mathrm{max}}_{\mathbf{p} \in \Omega_0(\mathfrak{B}), 
  t \in [0,\mathcal{T}]} \sqrt{\| \boldsymbol{\varphi} (\mathbf{p},t) 
  - \mathbf{p}\|^2 + \|\mathrm{Grad}[\mathbf{u}]\|^{2}} \ll 1
\end{align}
one is justified to employ the following linearized strain tensor: 
%-------------------------------;
%  Equation: Linearized strain  ;
%-------------------------------;
\begin{align}
  \label{Eqn:Degradation_El}
  \mathbf{E}_{l} = \frac{1}{2} \left(\mathrm{Grad}[\mathbf{u}] 
  + \mathrm{Grad}[\mathbf{u}]^{\mathrm{T}}\right) 
  \approx \frac{1}{2} \left(\mathrm{grad}[\mathbf{u}] 
  + \mathrm{grad}[\mathbf{u}]^{\mathrm{T}}\right)
\end{align}
where $\|\bullet\|$ denotes the Frobenius norm \citep{Antman}.

Since we will be dealing with processes in addition 
to the mechanical deformation, we need to introduce 
quantities other than the ones that are associated 
with the kinematics. 
We will denote the temperature by $\vartheta$ and the 
specific entropy by $\eta$. The mass fraction of the 
chemical species is denoted by $c$ and the corresponding 
chemical potential is denoted by $\varkappa$. The temperature, 
mass fraction of chemical 
species, entropy, and chemical potential are all scalar fields, while 
the displacement, velocity, and acceleration are vector fields. In some 
situations, it may be needed to explicitly indicate the functional 
dependence of these quantities. We employ a standard notation, which 
will be illustrated through the temperature field. The temperature in 
terms of reference coordinates and spatial coordinates will be 
denoted as follows: 
%---------------------------------------------------------------;
%  Equation: Temperature (Reference and current configuration)  ;
%---------------------------------------------------------------;
\begin{align}
  \label{Eqn:Temp_RefCurr_Config}
  \vartheta = \tilde{\vartheta}(\mathbf{p},t) 
  = \hat{\vartheta}(\mathbf{x},t)
\end{align}

%============================;
%  Subsection: Balance laws  ;
%============================;
\subsection{Balance laws}
\label{Subsec:Balance_Laws_CM_Degradation}
For our study, we shall consider the thermodynamic 
system to be the entire degrading body. Moreover, 
we shall assume this thermodynamic system to 
be an open system. That is, heat and mass transfers 
can occur across the boundary of the system. We now 
present the balance laws that govern the evolution 
of the chosen system. 

The \emph{balance of mass of the solid} in the 
degrading body takes the following form: 
%------------------------------------------------;
%  Equation: Balance of mass in deformed domain  ;
%------------------------------------------------;
\begin{align}
  \label{Eqn:Balance_of_mass_D}
  \dot{\rho} + \rho \, \mathrm{div}[\mathbf{v}] = 0 
\end{align}
where $\rho$ is the density of the solid in the 
deformed configuration $\Omega_{t}(\mathfrak{B})$. 
The \emph{balance of a chemical species}, which is being 
transported in the degrading body, can be mathematically 
written as:
%-------------------------------------------------------------------;
%  Equation: Balance of chemical species in deformed configuration  ;
%-------------------------------------------------------------------;
\begin{align}
  \label{Eqn:Balance_of_species_D}
  \rho \dot{c} + \mathrm{div}[\mathbf{h}] = h
\end{align}
where $\mathbf{h}$ is the mass transfer flux 
vector in the deformed configuration, and $h$ 
is the volumetric source of the chemical species 
in the deformed configuration. We assume that the 
chemical species cannot take partial stresses, 
which is a reasonable assumption in the degradation 
of materials due to small concentrations of moisture. 
One can handle large moisture contents by introducing 
partial stresses and using the theory of interacting 
continua (which is often referred to mixture theory) 
\citep{Bowen}. We do not address such issues, as our 
focus is degradation due to small concentrations of 
moisture or chemicals. 
The \emph{balance of linear momentum of the solid} can be written as:
%-----------------------------------------------------------;
%  Equation: Balance of linear momentum in deformed domain  ;
%-----------------------------------------------------------;
\begin{align}
  \label{Eqn:Balance_of_LM_D}
  &\rho \dot{\mathbf{v}} = \mathrm{div}[\mathbf{T}] 
  + \rho \mathbf{b} 
\end{align}
where $\mathbf{b}$ is the specific body force, 
and $\mathbf{T}$ is the Cauchy stress in the 
solid. Assuming that there is no internal 
couples, the \emph{balance of angular momentum 
  of the solid} reads:
%-----------------------------------------;
%  Equation: Balance of angular momentum  ;
%-----------------------------------------;
\begin{align}
  \label{Eqn:Model_BoAM}
  \mathbf{T} = \mathbf{T}^{\mathrm{T}}
\end{align}
Assuming that the balance of linear momentum 
(i.e., equation \eqref{Eqn:Balance_of_LM_D}) 
holds, the \emph{balance of energy of the system} 
(i.e., the first law of thermodynamics) can be 
written as:
%-------------------------------------------;
%  Equation: BoE in deformed configuration  ;
%-------------------------------------------;
\begin{align}
  \label{Eqn:Balance_of_energy_D}
  \rho \frac{\mathrm{d}}{\mathrm{d}t} 
  \left(A + \vartheta \eta \right) 
  = \mathbf{T} \bullet \mathbf{D} 
  - \mathrm{div}[\varkappa \mathbf{h}] 
  + \varkappa h - \mathrm{div}[\mathbf{q}] 
  + q 
\end{align}
where ${A}$ is the specific Helmholtz potential, 
$\mathbf{q}$ is the heat flux vector in the 
deformed configuration, and $q$ is the volumetric 
heat source in the deformed configuration. 
In our study, we assume that the Helmholtz potential 
$A$ to depend on $\mathbf{F}$, $c$, and $\vartheta$. 
We also have the following relations for the chemical 
potential and specific entropy:
%-----------------------------------------------------;
%  Equation: Chemical potential and specific entropy  ;
%-----------------------------------------------------;
\begin{align}
  \label{Eqn:Chemical_Potential_Definition}
  \varkappa &:= +\frac{\partial A}{\partial c} \\
  \label{Eqn:Specific_Entropy_Definition}
  \eta &:= - \frac{\partial A}{\partial \vartheta}
\end{align}
Assuming the balance of chemical species to hold, 
we then have the following convenient form for the 
balance of energy:
%----------------------------------------------------------;
%  Equation: BoE (reduced form) in deformed configuration  ;
%----------------------------------------------------------;
\begin{align}
  \label{Eqn:Balance_of_energy_D_reduced}
  \rho \left( \frac{\partial A}{\partial 
  \mathbf{F}} \mathbf{F}^{\mathrm{T}} 
  \bullet \mathbf{D} + \vartheta \dot{\eta} 
  \right) = \mathbf{T} \bullet \mathbf{D} - 
  \mathrm{div}[\mathbf{q}] - \mathrm{grad}
  [\varkappa] \bullet \mathbf{h} + q 
\end{align}
The \emph{localized version of the second law of 
thermodynamics in the deformed configuration} (by 
assuming that all the aforementioned balance laws 
to hold) takes the following form:
%--------------------------------------------------;
%  Equation: Second law in deformed configuration  ;
%--------------------------------------------------;
\begin{align}
  \label{Eqn:Second_Law_Thermodynamics_D}
  \rho \left( \frac{\partial A}{\partial \mathbf{F}} 
  \mathbf{F}^{\mathrm{T}} \bullet \mathbf{D} \right) = 
  \mathbf{T} \bullet \mathbf{D} - \frac{1}{\vartheta} 
  \mathrm{grad}[\vartheta] \bullet \mathbf{q} - 
  \mathrm{grad}[\varkappa] \bullet \mathbf{h} - 
  \rho \zeta
\end{align}
where $\zeta$ is the specific rate of dissipation functional, 
which is non-negative. The above equation is a stronger 
version than the second law of thermodynamics, which is 
a global law and not a local one. The second law of 
thermodynamics does \emph{not} assert that the rate 
of entropy production be non-decreasing at \emph{each 
and every point} in the system/body. Strictly speaking, 
equation \eqref{Eqn:Second_Law_Thermodynamics_D} 
should be referred to as the reduced local dissipation equality. 
Another point to highlight is that the second law of thermodynamics, 
in its original form, is in the form of an inequality. The introduction 
of the non-negative dissipation functional, which acts as a slack 
variable, converts the inequality into an equality, as provided 
in equation \eqref{Eqn:Second_Law_Thermodynamics_D}.

%==========================================================;
%  Subsubsection: The maximization of rate of dissipation  ;
%==========================================================;
\subsection{The maximization of rate of dissipation}
\label{SubSubSec:MaxRateEntropyProduction}
Among the various methodologies to derive constitutive 
relations (e.g., see \citep{Maugin}), the axiom of 
maximization of rate of dissipation put-forth by 
Ziegler \citep{Ziegler} is an attractive procedure. 
Herein, we extend this procedure to the open thermodynamic 
system under consideration. We obtain the constitutive 
relations using the maximization of rate of dissipation hypothesis, 
which needs the prescription of two functionals -- the 
Helmholtz potential and the dissipation functional. 
We assume the functional dependence of the Helmholtz 
potential and the dissipation functional to be 
$\hat{A}(\mathbf{F},c,\vartheta)$ and $\hat{\zeta}
(\mathbf{D}, \mathrm{grad}[\vartheta], \mathrm{grad}
[\varkappa];\mathbf{F},\vartheta,c)$. 

The mathematical statement of maximization of 
rate of dissipation can be written as follows:
%------------------;
%  Equation: MREP  ;
%------------------;
\begin{subequations}
  \begin{alignat}{2}
    \label{Eqn:MREP_1}
    &\mathop{\mathrm{maximize}}\limits_{\mathbf{D}, \mathrm{grad}
    [\vartheta], \mathrm{grad}[\varkappa]} \; \; &&\rho \zeta = 
    \rho \hat{\zeta}(\mathbf{D},\mathrm{grad}[\vartheta],\mathrm{grad} 
    [\varkappa];\mathbf{F},\vartheta,c)\\
    \label{Eqn:MREP_2}
    &\mathrm{subject \ to} \ &&\rho\left(\frac{\partial A}{\partial 
    \mathbf{F}} \mathbf{F}^{\mathrm{T}} \bullet \mathbf{D}\right) = 
    \mathbf{T} \bullet \mathbf{D}-\frac{1}{\vartheta} \mathrm{grad}
    [\vartheta] \bullet \mathbf{q}-\mathrm{grad}[\varkappa] \bullet 
    \mathbf{h} -\rho  \zeta
  \end{alignat}
\end{subequations}
Note that $\rho \zeta$ is maximized with
respect to arguments to the left side of `$;$'.
Using the method of Lagrange multipliers, the above 
constrained optimization problem is equivalent to 
the following unconstrained optimization problem: 
%----------------------------------;
%  Equation: Lagrange multipliers  ;
%----------------------------------;
\begin{align}
  \label{Eqn:MLM_MREP}
  \mathop{\mathrm{extremize}} \limits_{\mathbf{D}, \mathrm{grad}
  [\vartheta], \mathrm{grad}[\varkappa], \Lambda_t} \; \; & \rho \hat{\zeta}(
  \mathbf{D},\mathrm{grad}[\vartheta],\mathrm{grad}[\varkappa];
  \mathbf{F}, \vartheta,c) \nonumber\\ 
  + \, &\Lambda_t\left(\rho\left(\frac{\partial A}{\partial \mathbf{F}}
  \mathbf{F}^{\mathrm{T}} \bullet \mathbf{D}\right)-\mathbf{T} \bullet 
  \mathbf{D}+\frac{1} {\vartheta} \mathrm{grad}[\vartheta] \bullet 
  \mathbf{q}+\mathrm{grad}[\varkappa] \bullet \mathbf{h} + \rho \zeta \right)
\end{align}
where $\Lambda_t$ is the Lagrange multiplier enforcing 
the constraint \eqref{Eqn:MREP_2}. The first-order 
optimal conditions give rise to the following relations: 
%--------------------------------------------------------------------;
%  Equation: First-order optimal conditions (current configuration)  ;
%--------------------------------------------------------------------;
\begin{subequations}
  \begin{align}
    \label{Eqn:MREP_CurrConfig_Stress}
    &\mathbf{T} = \rho \frac{\partial A}
    {\partial \mathbf{F}} \mathbf{F}^{\mathrm{T}} + \left(\frac{1 + 
    \Lambda_t}{\Lambda_t} \right) \rho \frac{\partial \zeta}{\partial 
    \mathbf{D}} \\
    \label{Eqn:MREP_CurrConfig_HeatFlux}
    &\frac{1}{\vartheta} 
    \mathbf{q} = -\left(\frac{1 + \Lambda_t}{\Lambda_t} \right) 
    \rho\frac{\partial \zeta}{\partial\mathrm{grad}[\vartheta]} \\
    \label{Eqn:MREP_CurrConfig_DiffFlux}
    &\mathbf{h} =
    -\left(\frac{1 + \Lambda_t}{\Lambda_t} \right) \rho \frac{\partial 
      \zeta}{\partial\mathrm{grad}[\varkappa]} \\
    \label{Eqn:MREP_CurrConfig_ReducedSecondLaw}
    &\rho \left(\frac{\partial A}{\partial 
    \mathbf{F}} \mathbf{F}^{\mathrm{T}} \bullet \mathbf{D} \right) - 
    \mathbf{T} \bullet \mathbf{D} + \frac{1}{\vartheta} \mathrm{grad}
    [\vartheta] \bullet \mathbf{q} + \mathrm{grad}[\varkappa] \bullet
    \mathbf{h} + \rho\zeta = 0 
  \end{align}
\end{subequations}
The above equations can be obtained by taking 
(G\^{a}teaux) variation of the objective function 
in equation \eqref{Eqn:MLM_MREP} with respect 
to $\mathbf{D}$, $\mathrm{grad}[\vartheta]$, 
$\mathrm{grad}[\varkappa]$ and $\Lambda_{t}$, 
respectively. 
By straightforward manipulations on equations 
\eqref{Eqn:MREP_CurrConfig_Stress}--\eqref{Eqn:MREP_CurrConfig_ReducedSecondLaw}, the Lagrange multiplier $\Lambda_{t}$ can be 
explicitly calculated as follows:
%---------------------------------------------------------;
%  Equation: Lagrange multiplier (current configuration)  ;
%---------------------------------------------------------;
\begin{align}
  \label{Eqn:lambda_LM_Curr}
  \Lambda_t = \left[\frac{\zeta}{\frac{\partial \zeta}{\partial \mathbf{D}} 
      \bullet \mathbf{D} 
      + \frac{\partial \zeta}{\partial \mathrm{grad}
        [\vartheta]} \bullet \mathrm{grad}[\vartheta] 
      + \frac{\partial \zeta}{\partial \mathrm{grad}[\varkappa]} 
      \bullet \mathrm{grad}[\varkappa]} - 1 \right]^{-1}
\end{align}
If the rate of dissipation functional $\zeta$ is 
a homogeneous functional of order 2 with respect 
to $\mathbf{D}$, $\mathrm{grad}[\vartheta]$ and 
$\mathrm{grad}[\varkappa]$, we then have 
%----------------------------------------------------------------------;
%  Equation: 2nd order homogeneous functional (\zeta, current config)  ;
%----------------------------------------------------------------------;
\begin{align}
  \label{Eqn:LagrMultiExpr_CurrConfig}
  \frac{\partial \zeta}{\partial \mathbf{D}} 
  \bullet \mathbf{D} + \frac{\partial \zeta}{\partial 
  \mathrm{grad}[\vartheta]} \bullet \mathrm{grad}[\vartheta] 
  + \frac{\partial \zeta}{\partial \mathrm{grad}
  [\varkappa]} \bullet \mathrm{grad}[\varkappa] = 2 \zeta
\end{align}
which further implies that $\Lambda_t = -2$. The 
constitutive relations under $\Lambda_t = -2$ will 
simplify to: 
%-------------------------------------;
%  Equations: constitutive relations  ;
%-------------------------------------;
\begin{subequations}
  \begin{align}
    \label{Eqn:StressGenExpr_CurrConfig}
    &\mathbf{T} = \rho \frac{\partial A}{\partial \mathbf{F}} 
    \mathbf{F}^{\mathrm{T}} + \frac{1}{2} \rho \frac{\partial 
      \zeta}{\partial \mathbf{D}} \\
    \label{Eqn:HeatFluxsGenExpr_CurrConfig}    
    &\mathbf{q} = -\frac{\vartheta}{2} \rho \frac{\partial \zeta}
    {\partial\mathrm{grad}[\vartheta]} \\
    \label{Eqn:DiffFluxGenExpr_CurrConfig}
    &\mathbf{h} = -\frac{1}{2} \rho \frac{\partial \zeta}
    {\partial \mathrm{grad}[\varkappa]}
  \end{align}
\end{subequations}

%-----------------------------------------------;
%  Remark: Remark about homogeneous functional  ;
%-----------------------------------------------;
\begin{remark}
  It should be emphasized that the dissipation functional 
  need not be a homogeneous functional of order two in 
  terms of $\mathbf{F}$, $c$ and $\vartheta$. The 
  maximization of the rate of dissipation certainly 
  does not require such an assumption. However, we 
  shall make such an assumption, as it is convenient 
  and the resulting constitutive relations can still 
  model the desired degradation mechanisms. 
\end{remark}

%==================================================================;
%  Subsection: Governing equations in the reference configuration  ;
%==================================================================;
\subsection{Governing equations in the reference configuration}
\label{Eqn:GovEqn_RefConfig}
Since we are also interested in developing a computational 
framework and obtaining numerical solutions, it will be 
convenient to write the balance laws in the reference 
configuration. To this end, we introduce:
%--------------------------;
%  Equation: J definition  ;
%--------------------------;
\begin{align}
  J \equiv \mathrm{det}[\mathbf{F}]
\end{align}
where $\mathrm{det}[\bullet]$ denotes the determinant. 
The balance of mass in the reference configuration can 
be written as:
%-------------------------------------------------;
%  Equation: Balance of mass in reference domain  ;
%-------------------------------------------------;
\begin{align}
  \label{Eqn:Balance_of_mass_R}
  \rho_0  = J \rho
\end{align}
where $\rho_0$ is the density of the undeformed solid. The 
balance of chemical species in the reference configuration 
can be rewritten as:
%-------------------------------------------------------------;
%  Equation: Balance of chemical species in reference domain  ;
%-------------------------------------------------------------;
\begin{align}
  \label{Eqn:Balance_of_species_R}
  \rho_0 \dot{c} + \mathrm{Div}[\mathbf{h}_0] = h_0
\end{align}
where $\mathbf{h}_0 = J \mathbf{F}^{-1} \mathbf{h}$ is the diffusive 
flux vector in the reference configuration and $h_0 = J h$ is the 
volumetric source in the reference configuration. The balance of 
linear momentum in the reference configuration takes the following 
form:
%------------------------------------------------------------;
%  Equation: Balance of linear momentum in reference domain  ;
%------------------------------------------------------------;
\begin{align}
  \label{Eqn:Balance_of_LM_R}
  &\rho_0 \dot{\mathbf{v}} = \mathrm{Div}[\mathbf{P}] 
  + \rho_0 \mathbf{b} 
\end{align}
where $\mathbf{P} = J \mathbf{T} \mathbf{F}^{-\mathrm{T}}$ is the 
first Piola-Kirchhoff stress. The balance of angular momentum in 
the reference configuration takes the following form:
%-----------------------------------------;
%  Equation: Balance of angular momentum  ;
%-----------------------------------------;
\begin{align}
  \label{Eqn:BoAM_in_R}
  \mathbf{P} \mathbf{F}^{\mathrm{T}} = \mathbf{F} 
  \mathbf{P}^{\mathrm{T}} 
\end{align}
In the reference configuration, the balance of energy can be written as:
%---------------------------------------------------;
%  Equation: Balance of energy in reference domain  ;
%---------------------------------------------------;
\begin{align}
  \label{Eqn:Balance_of_energy_R}
  \rho_0 \left( \frac{\partial A}{\partial \mathbf{F}} \bullet
  \dot{\mathbf{F}} + \vartheta \dot{\eta} \right) = \mathbf{P} 
  \bullet \dot{\mathbf{F}} - \mathrm{Div}[\mathbf{q}_0] - \mathrm{Grad}
  [\varkappa] \bullet \mathbf{h}_0 + q_0
\end{align}
where $\mathbf{q}_0 = J \mathbf{F}^{-1} \mathbf{q}$ is 
the heat flux vector in the reference configuration 
and $q_0 = J q$ is the volumetric heat source in the 
reference configuration.
In the reference configuration, the 
second law can be rewritten as:
%-------------------------------------------------------------;
%  Equation: Second law of thermodynamics in deformed domain  ;
%-------------------------------------------------------------;
\begin{align}
  \label{Eqn:Second_Law_Thermodynamics_R}
  \rho_0 \left( \frac{\partial A}{\partial \mathbf{F}} \bullet
  \dot{\mathbf{F}} \right) = \mathbf{P} \bullet \dot{\mathbf{F}}
  - \frac{1}{\vartheta} \mathrm{Grad}[\vartheta] \bullet \mathbf{q}_0 
  - \mathrm{Grad}[\varkappa] \bullet \mathbf{h}_0 - \rho_0 \zeta_{0}
\end{align}
where $\zeta_{0} = \zeta$ is the non-negative 
rate of dissipation functional in the reference 
configuration.

%=====================================================;
%  Subsubsection: MRD in the reference configuration  ;
%=====================================================;
\subsubsection{Maximization of rate of dissipation in the reference configuration}
\label{SubSubSec:MaxDissp_RefConfig}
The mathematical statement of maximization of 
rate of dissipation can be written as follows:
%-----------------;
%  Equation: MRD  ;
%-----------------;
\begin{subequations}
  \begin{alignat}{2}
    \label{Eqn:MREP_R_1}
    &\mathop{\mathrm{maximize}} \limits_{\dot{\mathbf{F}}, 
    \mathrm{Grad}[\vartheta],\mathrm{Grad}[\varkappa]} \; \; 
    &&\rho_0 \zeta_{0} =\rho_0 \tilde{\zeta}(\dot{\mathbf{F}},\mathrm{Grad}[\vartheta], 
    \mathrm{Grad}[\varkappa];\mathbf{F},
    \vartheta,c)\\
    \label{Eqn:MREP_R_2}
    &\mathrm{subject \ to} \ &&\rho_0 \left(\frac{\partial 
    A}{\partial \mathbf{F}} \bullet \dot{\mathbf{F}}\right) = 
    \mathbf{P} \bullet \dot{\mathbf{F}} - \frac{1}{\vartheta}
    \mathrm{Grad}[\vartheta] \bullet \mathbf{q}_0 - \mathrm{Grad}
    [\varkappa] \bullet \mathbf{h}_0 - \rho_0\zeta_{0}
  \end{alignat}
\end{subequations}
Using the method of Lagrange multipliers, 
one can obtain the following equivalent 
unconstrained optimization problem: 
%----------------------------------;
%  Equation: Lagrange multipliers  ;
%----------------------------------;
\begin{align}
  \label{Eqn:MLM_MREP_R}
  \mathop{\mathrm{extremize}} \limits_{\dot{\mathbf{F}}, 
  \mathrm{Grad}[\vartheta],\mathrm{Grad}[\varkappa],\Lambda_0} 
  \; \; &\rho_0 \tilde{\zeta}(\dot{\mathbf{F}},\mathrm{Grad}[\vartheta],
  \mathrm{Grad}[\varkappa];\mathbf{F},\vartheta,c) \nonumber \\ 
  + \, &\Lambda_0\left(\rho_0 \left(\frac{\partial A}{\partial 
  \mathbf{F}} \bullet \dot{\mathbf{F}}\right) - \mathbf{P} \bullet 
  \dot{\mathbf{F}} + \frac{1}{\vartheta} \mathrm{Grad}[\vartheta] 
  \bullet \mathbf{q}_0+\mathrm{Grad}[\varkappa] \bullet \mathbf{h}_0 + 
 \rho_0\zeta_{0} \right)
\end{align}
where $\Lambda_0$ is the Lagrange multiplier enforcing 
the constraint given by equation \eqref{Eqn:MREP_R_2}. 
The first-order optimality conditions give rise to the 
following constitutive relations: 
%------------------------------------------------------------------------;
% Equation: First-order optimality conditions (Reference configuration)  ;
%------------------------------------------------------------------------;
\begin{subequations}
  \begin{align}
    \label{Eqn:MREP_RefConfig_Stress}
    &\mathbf{P} = \rho_0 
    \frac{\partial A}{\partial \mathbf{F}} + \left(\frac{1 
    + \Lambda_0}{\Lambda_0} \right) \rho_0 \frac{\partial \zeta_0}
    {\partial \dot{\mathbf{F}}} \\
    \label{Eqn:MREP_RefConfig_HeatFlux}
    &\frac{1}{\vartheta} 
    \mathbf{q}_0 = -\left(\frac{1 + \Lambda_0}{\Lambda_0}\right) 
    \rho_0 \frac{\partial \zeta_0}{\partial \mathrm{Grad}[\vartheta]} \\
     \label{Eqn:MREP_RefConfig_DiffFlux}
    &\mathbf{h}_0 = -\left(\frac{1 + \Lambda_0}{\Lambda_0} \right) 
    \rho_0 \frac{\partial \zeta_0}{\partial \mathrm{Grad}[\varkappa]} \\
    \label{Eqn:MREP_RefConfig_ReducedSecondLaw}
    &\rho_0 \left(\frac{\partial A}{\partial 
    \mathbf{F}} \bullet \dot{\mathbf{F}} \right) - \mathbf{P} \bullet 
    \dot{\mathbf{F}} + \frac{1}{\vartheta} \mathrm{Grad}[\vartheta] 
    \bullet \mathbf{q}_0 + \mathrm{Grad}[\varkappa] \bullet \mathbf{h}_0 
    + \rho_0 \zeta_0 = 0 
  \end{align}
\end{subequations}
Similar to the derivation presented earlier in the 
context of current configuration, the Lagrange multiplier 
$\Lambda_{0}$ can be explicitly calculated as follows:
%-----------------------------------------------------------;
%  Equation: Lagrange multiplier (Reference configuration)  ;
%-----------------------------------------------------------;
\begin{align}
  \label{Eqn:lambda_LM_Ref}
  \Lambda_0 = \left[\frac{\zeta_0}{\frac{\partial 
  \zeta_0}{\partial \dot{\mathbf{F}}} \bullet 
  \dot{\mathbf{F}} + \frac{\partial \zeta_0}{\partial 
  \mathrm{Grad}[\vartheta]} \bullet \mathrm{Grad}[\vartheta] 
  + \frac{\partial \zeta_0}{\partial \mathrm{Grad}[\varkappa]}
  \bullet \mathrm{Grad}[\varkappa]} - 1 \right]^{-1}
\end{align}
If the rate of dissipation functional in the reference 
configuration $\zeta_{0}$ is a homogeneous functional 
of order 2, we have 
%--------------------------------------------;
%  Equation: 2nd order homogeneous function  ;
%--------------------------------------------;
\begin{align}
  \label{Eqn:LagrMultiExpr_CurrConfig}
  \frac{\partial \zeta_0}{\partial \dot{\mathbf{F}}}
  \bullet \dot{\mathbf{F}} + \frac{\partial \zeta_0}{\partial 
  \mathrm{Grad}[\vartheta]} \bullet \mathrm{Grad}[\vartheta] 
  + \frac{\partial \zeta_0}{\partial \mathrm{Grad}[\varkappa]}
  \bullet \mathrm{Grad}[\varkappa] = 2 \zeta_0
\end{align}
which further implies that $\Lambda_0=-2$. The constitutive 
relations under $\Lambda_0 = - 2$ take the following form:
%-------------------------------------;
%  Equations: constitutive relations  ;
%-------------------------------------;
\begin{subequations}
  \begin{align}
    \label{Eqn:StressGenExpr_RefConfig}
    &\mathbf{P} = \rho_{0} \frac{\partial A}{\partial \mathbf{F}} 
    + \frac{1}{2} \rho_0 \frac{\partial \zeta_0}{\partial \dot{\mathbf{F}}}\\ 
    \label{Eqn:HeatFluxsGenExpr_RefConfig}   
    &\mathbf{q}_0 = -\frac{\vartheta}{2} \rho_0 \frac{\partial \zeta_0}
    {\partial \mathrm{Grad}[\vartheta]}\\
    \label{Eqn:DiffFluxGenExpr_RefConfig}
    &\mathbf{h}_0 = -\frac{1}{2}\rho_0 \frac{\partial\zeta_0}
    {\partial \mathrm{Grad}[\varkappa]}
  \end{align}
\end{subequations}

\begin{figure}
  \centering
  \includegraphics[scale=0.6]
  {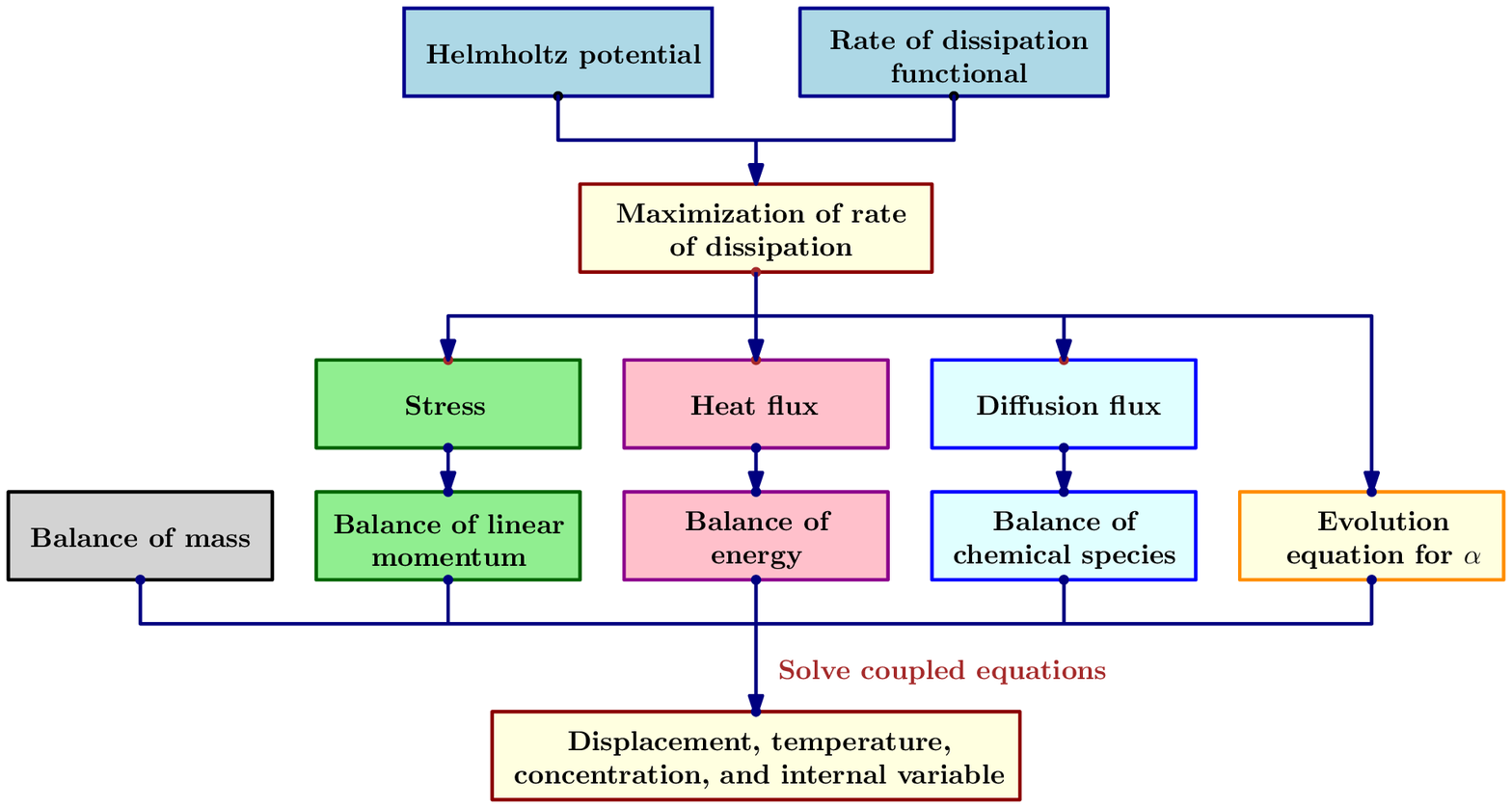}
  \caption{\textsf{Overarching idea of the proposed degradation 
    framework:}~This flowchart shows the overarching idea behind 
    the proposed framework. We shall appeal to the axiom of 
    maximization of rate of dissipation to obtain constitutive 
    relations for stress, heat flux, diffusion flux, and evolution 
    equation for internal variable (if required). Solving the 
    coupled equations, we get the solution for displacement, 
    temperature, concentration, and internal variable (if 
    required).
  \label{Fig:overarching_idea}}
\end{figure}

The overarching idea behind the proposed chemo-thermo-mechano 
degradation model is shown in Figure \ref{Fig:overarching_idea}. 
In the next section, we will develop the proposed constitutive 
model by appealing to the maximization of rate of dissipation. 
Solving the coupled balance laws (even including the evolution 
equation for internal variable), we can get the displacement, 
temperature, and concentration (internal variable, if needed).

%------------------------------------------------------------------;
%  Figure: Overarching idea of the proposed degradation framework  ;
%------------------------------------------------------------------;

%% file: Sections_Model/S3_Model_MODEL.tex
%***************************************;
%                                       ;
%  NAME                                 ;  
%    S3_Model_MODEL.tex                 ;
%                                       ;
%  WRITTEN BY                           ;
%     Can Xu                            ;
%     Kalyana Babu Nakshatrala          ;
%                                       ; 
%***************************************;
\section{A GENERAL CONSTITUTIVE MODEL FOR CHEMO-THERMO-MECHANO DEGRADATION}
\label{Sec:ChemoMechano_DegradModel}
Under the maximization of rate of dissipation hypothesis, 
the constitutive relations can be obtained by prescribing 
two functionals -- the Helmholtz potential and the dissipation 
functional. Philosophically, the Helmholtz potential quantifies 
the way in which the material stores energy, whereas the dissipation 
functional quantifies the way in which the material dissipates energy.
For our proposed chemo-thermo-mechano degradation model, we prescribe 
the following functional forms for the specific Helmholtz potential 
and the rate of dissipation functional:
%-----------------------------------------------------------------;
%  Equation: Specific Helmholtz potential (chemo-thermo-mechano)  ;
%-----------------------------------------------------------------;
\begin{align}
  \label{Eqn:Degradation_Helmholtz}
  A = \hat{A}(\mathbf{F},c,\vartheta) &= \frac{1}{\rho_0}\psi
  - \frac{1}{2} \frac{c_{p}}{\vartheta_{\mathrm{ref}}} \left 
  \lbrace \vartheta - \vartheta_{\mathrm{ref}} \right \rbrace^{2} 
  - \frac{1}{\rho_0} \left\lbrace \vartheta - \vartheta_{\mathrm{ref}} 
  \right \rbrace \mathbf{M}_{\vartheta \mathbf{E}} \bullet \mathbf{E} 
  + d_{\vartheta c} \left \lbrace \vartheta - \vartheta_{\mathrm{ref}} 
  \right \rbrace \left \lbrace c - c_{\mathrm{ref}} \right \rbrace 
  \nonumber \\
  &- \frac{1}{\rho_0} \left \lbrace c - c_{\mathrm{ref}} \right \rbrace 
  \mathbf{M}_{c \mathbf{E}} \bullet \mathbf{E} + \frac{R_s \vartheta_
  {\mathrm{ref}}}{2} \{c - c_{\mathrm{ref}}\}^2
\end{align}
%%
%--------------------------------------------;
%  Equation: Rate of dissipation functional  ;
%--------------------------------------------;
\begin{align}
  \label{Eqn:Dissipation_Functional}
  \zeta = \hat{\zeta}(\mathbf{D},\mathrm{grad}[\vartheta],
  \mathrm{grad}[\varkappa];\mathbf{F},\vartheta,c) 
  %%\nonumber\\
  %%
  &=\frac{c_p}{\vartheta} \mathrm{grad}[\vartheta] \bullet 
  \mathbf{D}_{\vartheta \vartheta} \mathrm{grad}[\vartheta]
  +\frac{1}{\vartheta} \mathrm{grad}[\vartheta] \bullet 
  \mathbf{D}_{\vartheta \varkappa} \mathrm{grad}[\varkappa] 
  \nonumber \\
  &+\frac{1}{\vartheta}\mathrm{grad}[\varkappa] \bullet 
  \mathbf{D}_{\varkappa \vartheta}\mathrm{grad}[\vartheta]
  +\frac{1}{R_s\vartheta_{\mathrm{ref}}} \mathrm{grad}[\varkappa] 
  \bullet \mathbf{D}_{\varkappa \varkappa}\mathrm{grad}[\varkappa]
\end{align}
where $R_s = R/M$. $R_s$ and $R$ denote the specific vapor constant 
and the universal vapor constant respectively, $M$ is the molecular 
mass of chemical species. $\vartheta_{\mathrm{ref}}$ and $c_{\mathrm{ref}}$ 
are the specified reference temperature and reference mass concentration, 
which depend on the underlying boundary value problem. We denote $c_p$ 
as the coefficient of heat capacity, $ d_{\vartheta c}$ as the thermo-chemo 
coupled parameter \citep{Sih_etal}, 
$\mathbf{M}_{\vartheta \mathbf{E}}$ as the anisotropic 
coefficient of thermal expansion (which is assumed to be independent of 
temperature, concentration, and strain), and $\mathbf{M}_{c \mathbf{E}}$ 
as the anisotropic coefficient of chemical expansion due to concentration 
(which is also assumed to be independent of temperature, concentration, 
and strain). Both $\mathbf{M}_{\vartheta \mathbf{E}}$ and $\mathbf{M}_{c 
\mathbf{E}}$ are assumed to be symmetric. $\mathbf{D}_{\vartheta\vartheta}$ 
is the anisotropic thermal conductivity tensor and $\mathbf{D}_{\varkappa 
\varkappa}$ is the anisotropic diffusivity tensor. $\mathbf{D}_{\vartheta 
\varkappa}$ corresponds to the anisotropic Soret effect tensor, which 
characterizes the transport of chemical species caused by temperature 
gradient. Similarly, $\mathbf{D}_{\varkappa\vartheta}$ is the Dufour 
effect tensor, which represents the heat flow caused by a concentration 
gradient.
%%
%---------------------------------------------------------;
%  Remark: Coefficient of thermal and chemical expansion  ;
%---------------------------------------------------------;
\begin{remark}
  In chemo-thermo-elasticity and in modeling degradation of materials 
  due to transport and reaction of chemical species, coefficient 
  of chemical expansion $\mathbf{M}_{c \mathbf{E}}$ and thermo-chemo 
  coupling parameter $d_{\vartheta c}$ play a vital role (see 
  \citep[Chapter-5]{Sih_etal} and references therein). Induced-strains 
  due to chemical expansivity will be significant in harsh environmental 
  conditions and cannot be neglected \citep{Sih_etal}. Considerable 
  inquest has been made in literature to experimentally measure $\mathbf{M}
  _{c \mathbf{E}}$ in ceramics \citep{2001_Adler_JACS_v84_p2117_p2119, 
  2011_Morozovska_PRB_v83_p195313,2008_Blond_Richet_JECS_v28_p793_p801}, 
  laminated and polymer composites \citep{Sih_etal,1989_Bouadi_Sun_JRPC_v8_p40_p54,
  1994_Cai_Weitsman_JCM_v28_p130_p154}, elastomers and biological 
  materials \citep{Harper,1984_Myers_etal_JBE_v106_p151_p158,
  1991_Lai_etal_JBE_v113_p245_p258}, and concrete structures 
  \citep{2000_Ulm_etal_JEM_v126_p233_p242, Cerny_Rovnanikova,
  Swamy}. However, adequate progress has not been made yet 
  to develop constitutive models and computational frameworks 
  for such chemo-thermo-elastic materials or materials undergoing 
  chemical-induced degradation. Herein, we shall take a step 
  forward to address this issue.  
\end{remark}

%--------------------------------------------------------;
%  Remark: Onsager relations (Dufour and Soret tensors)  ;
%--------------------------------------------------------;
\begin{remark}
  It should be noted that in the absence of electrical and 
  magnetic fields, all of the above tensors are symmetric 
  \citep{Bowen,Coussy_poromechanics,2001_Jarkova_etal_EPJE_v5_p583_p588}. 
  Moreover, from the Onsager reciprocal relations (which was 
  put-forth by Onsager in 1930s \citep{1931_Onsager_PR_v37_p405,
  1931_Onsager_PR_v38_p2265}) we have the following relationship 
  between the Soret effect tensor and the Dufour effect tensor.
  %------------------------------------------;
  %  Equation: Onsager Reciprocal Relations  ;
  %------------------------------------------;
  \begin{align}
    \label{Eqn:Onsager_Relations}
    \mathbf{D}_{\vartheta\varkappa} = \mathbf{D}_{\varkappa\vartheta}
  \end{align}
  Additionally, physics demands that the tensors $\mathbf{D}_
  {\vartheta\vartheta}$ and $\mathbf{D}_{\varkappa\varkappa}$ 
  are positive definite.
\end{remark}

%---------------------------------------------------------------------------;
%  Remark: On Helmholtz potential and dissipation functional for diffusion  ;
%---------------------------------------------------------------------------;
\begin{remark}
  Note that the specific Helmholtz potential and correspondingly the 
  dissipation functional for diffusion can also be modelled using the 
  following expressions:
  %---------------------------------------------------------------------;
  %  Equation: Specific Helmholtz potential and dissipation functional  ;
  %---------------------------------------------------------------------; 
  \begin{align}
    \label{Eqn:Degradation_Helmholtz1}
    A_{c} &= R_s \vartheta_{\mathrm{ref}} c \{\mathrm{ln}[c] - 1\} \\
    \label{Eqn:Dissipation_Functional1}
    \zeta_{c} &= \frac{c}{R_s \vartheta_
    {\mathrm{ref}}} \mathrm{grad}[\varkappa] 
    \bullet \mathbf{D}_{\varkappa\varkappa} 
    \mathrm{grad}[\varkappa]
  \end{align}
  Both equations \eqref{Eqn:Degradation_Helmholtz}--\eqref{Eqn:Dissipation_Functional} 
  and \eqref{Eqn:Degradation_Helmholtz1}--\eqref{Eqn:Dissipation_Functional1} 
  result in similar partial differential equation structure for 
  modeling Fickian diffusion.  
\end{remark}

Under the proposed model, the specific entropy and chemical potential 
take the following form:
%-----------------------------------------------------;
%  Equation: Specific entropy and chemical potential  ;
%-----------------------------------------------------;
\begin{align}
  \label{Eqn:SpecificEntropy_DeformedConfig}
  &\eta = -\frac{\partial A}{\partial \vartheta} = 
  -\frac{1}{\rho_0} \frac{\partial \psi}{\partial \vartheta} 
  +\frac{c_{p}}{\vartheta_{\mathrm{ref}}}\{\vartheta - 
  \vartheta_{\mathrm{ref}}\} + \frac{1}{\rho_0}\mathbf{M}_
  {\vartheta\mathbf{E}} \bullet \mathbf{E} - d_{\vartheta 
  c} \{c - c_{\mathrm{ref}}\}\\
  \label{Eqn:ChemicalPotential_DeformedConfig}
  &\varkappa = \frac{\partial A}{\partial c} = \frac{1}{\rho_0} 
  \frac{\partial \psi}{\partial c} + R_s \vartheta_{\mathrm{ref}} 
  \{c - c_{\mathrm{ref}}\} - \frac{1}{\rho_0}\mathbf{M}_{c\mathbf{E}} 
  \bullet \mathbf{E} + d_{\vartheta c} \{\vartheta - \vartheta_
  {\mathrm{ref}}\}
\end{align}
From equations \eqref{Eqn:StressGenExpr_CurrConfig}--\eqref{Eqn:DiffFluxGenExpr_CurrConfig}, 
we have the constitutive relations in deformed configuration as: 
%--------------------------------------------------------------;
%  Equations: Constitutive relations (deformed configuration)  ;
%--------------------------------------------------------------;
\begin{subequations}
  \begin{align}
    \label{Eqn:Stress_DeformedConfig}
    &\mathbf{T} = \rho \frac{\partial A}{\partial \mathbf{F}} 
    \mathbf{F}^{\mathrm{T}} = \frac{1}{J} \frac{\partial \psi}{\partial 
    \mathbf{F}} \mathbf{F}^{\mathrm{T}} - \frac{1}{J} \left\lbrace 
    \vartheta - \vartheta_{\mathrm{ref}} \right \rbrace \mathbf{F} 
    \mathbf{M}_{\vartheta \mathbf{E}} \mathbf{F}^{\mathrm{T}} - 
    \frac{1}{J} \left \lbrace c - c_{\mathrm{ref}} \right \rbrace 
    \mathbf{F} \mathbf{M}_{c\mathbf{E}} \mathbf{F}^{\mathrm{T}} \\  
    \label{Eqn:HeatFlux_DeformedConfig}
    &\mathbf{q} = -\frac{\vartheta}{2} \rho \frac{\partial \hat{\zeta}}
    {\partial \mathrm{grad}[\vartheta]} = -\rho c_p 
    \mathbf{D}_{\vartheta \vartheta}
    \mathrm{grad}[\vartheta] - \frac{\rho}{2}  \mathbf{D}_{\vartheta \varkappa}
    \mathrm{grad}[\varkappa] - \frac{\rho}{2}  \mathbf{D}_{\varkappa \vartheta}
    \mathrm{grad}[\varkappa] \\
    \label{Eqn:DiffFlux_DeformedConfig}
    &\mathbf{h} = -\frac{1}{2} \rho \frac{\partial \hat{\zeta}}{\partial 
    \mathrm{grad}[\varkappa]} = -\frac{\rho}{R_s \vartheta_{\mathrm{ref}}}
    \mathbf{D}_{\varkappa\varkappa}\mathrm{grad}[\varkappa] - \frac{\rho}{2 
    \vartheta} \mathbf{D}_{\vartheta \varkappa} \mathrm{grad}[\vartheta] 
    -\frac{\rho}{2\vartheta}\mathbf{D}_{\varkappa \vartheta}\mathrm{grad}[\vartheta]
  \end{align}
\end{subequations}
The rate of dissipation functional for the degradation 
model in the reference configuration is taken as follows:
%----------------------------------------------------------------------;
%  Equation: Rate of dissipation functional (reference configuration)  ;
%----------------------------------------------------------------------;
\begin{align}
  \label{Eqn:Dissipation_Functional_R}
  \zeta &= \tilde{\zeta}(\dot{\mathbf{F}}, \mathrm{Grad}[\vartheta], 
  \mathrm{Grad}[\varkappa]; \mathbf{F},\vartheta,c) \nonumber \\
  &= \frac{c_p}{\vartheta}\mathrm{Grad}[\vartheta] \bullet \overline{\mathbf{D}}
  _{\vartheta\vartheta} \mathrm{Grad}[\vartheta] + \frac{1}{\vartheta} 
  \mathrm{Grad}[\vartheta] \bullet \overline{\mathbf{D}}_{\vartheta \varkappa}
  \mathrm{Grad}[\varkappa] \nonumber \\
  &+ \frac{1}{\vartheta} \mathrm{Grad}[\varkappa] \bullet \overline{\mathbf{D}}_
  {\varkappa \vartheta} \mathrm{Grad}[\vartheta] + \frac{1}{R_s \vartheta_
  {\mathrm{ref}}}\mathrm{Grad}[\varkappa] \bullet \overline{\mathbf{D}}_ 
  {\varkappa \varkappa} \mathrm{Grad}[\varkappa]
\end{align}
where $\overline{\mathbf{D}}_{\alpha \beta} = \mathbf{F}^{-1} 
\mathbf{D}_{\alpha \beta} \mathbf{F}^{-T}$, $\alpha$ and $\beta$ 
represent $\vartheta$ or $\varkappa$.
Correspondingly, the constitutive relations in the reference configuration 
take the following form: 
%----------------------------------------------------------------;
%  Equations: Constitutive relations (undeformed configuration)  ;
%----------------------------------------------------------------;
\begin{subequations}
  \begin{align}
    \label{Eqn:Stress_UnDeformedConfig}
    \mathbf{P} &= \rho_0 \frac{\partial A}{\partial \mathbf{F}} =
    \frac{\partial \psi} {\partial \mathbf{F}} - \left\lbrace \vartheta 
    - \vartheta_{\mathrm{ref}} \right \rbrace \mathbf{F} \mathbf{M}_
    {\vartheta\mathbf{E}} - \left \lbrace c - c_{\mathrm{ref}} \right 
    \rbrace \mathbf{F} \mathbf{M}_{c\mathbf{E}} \\ 
    \label{Eqn:HeatFlux_DeformedConfig}  
    \mathbf{q}_0 &= -\frac{\vartheta}{2} \rho_0 \frac{\partial 
    \tilde{\zeta}}{\partial \mathrm{Grad}[\vartheta]} = -\rho_0 
    c_p \overline{\mathbf{D}}_{\vartheta \vartheta} \mathrm{Grad}
    [\vartheta] - \frac{\rho_0}{2} \overline{\mathbf{D}}_{\vartheta
    \varkappa} \mathrm{Grad}[\varkappa] - \frac{\rho_0}{2} \overline{
    \mathbf{D}}_{\varkappa \vartheta} \mathrm{Grad}[\varkappa] \\
    \label{Eqn:DiffFlux_DeformedConfig}
    \mathbf{h}_0 &= -\frac{1}{2} \rho_0 \frac{\partial \tilde{\zeta}}
    {\partial \mathrm{Grad}[\varkappa]} = -\frac{\rho_0}{R_s \vartheta
    _{\mathrm{ref}}} \overline{\mathbf{D}}_{\varkappa \varkappa} 
    \mathrm{Grad}[\varkappa] - \frac{\rho_0}{2 \vartheta} \overline{
    \mathbf{D}}_{\vartheta \varkappa} \mathrm{Grad}[\vartheta] - 
    \frac{\rho_0}{2 \vartheta} \overline{\mathbf{D}}_{\varkappa 
    \vartheta}\mathrm{Grad}[\vartheta] 
  \end{align}
\end{subequations}

%=====================================================================;
%  Subsection: Constitutive specifications for the degradation model  ;
%=====================================================================;
\subsection{Coupling terms for the degradation model}
\label{Subsec:Constitutive_specification}
The following hyperelastic material models will be employed in this paper: 
%----------------------------------;
%  Equations: Hyperelastic models  ;
%----------------------------------;
\begin{subequations}
  \begin{alignat}{2}
    \label{Eqn:StVenantKirchhoff}
    \psi &= \frac{\lambda}{2}(\mathrm{tr}[\mathbf{E}])^2 +
    \mu \mathbf{E} \bullet \mathbf{E} \quad 
    &&\mbox{St.~Venant-Kirchhoff model} \\
    \label{Eqn:ModifiedStVenantKirchhoff}
    \psi &= \frac{\kappa}{2}(\mathrm{ln}[J])^2 + \mu 
    \mathbf{E} \bullet \mathbf{E} \quad 
    &&\mbox{Modified St.~Venant-Kirchhoff model} \\
    \label{Eqn:NeoHookean}
    \psi &= \mu \mathrm{tr}[\mathbf{E}] +
    \mu \mathrm{ln}[J] + \frac{\lambda}{2}(\mathrm{ln}[J])^2 
    \quad 
    &&\mbox{Neo-Hookean model}
  \end{alignat}
\end{subequations}
where $\psi$ is the stored strain energy density functional, 
$\lambda$ and $\mu$ are the Lam\'e parameters, and $\kappa = 
\lambda + \frac{2 \mu}{3}$ is the bulk modulus. Recall that 
$J = \mathrm{det}[\mathbf{F}]$. The Lam\'e parameters in the 
degrading model are given by the following expressions:
%------------------------------------------------------;
%  Equation: Lame parameters in CTM degradation model  ;
%------------------------------------------------------;
\begin{subequations}
  \begin{align}
    \label{Eqn:lama_parameters1}
    \lambda(\mathbf{x},c) &= \lambda_0(\mathbf{x}) - 
    \lambda_{1}(\mathbf{x}) \frac{c}{c_{\mathrm{ref}}}
    - \lambda_{2}(\mathbf{x}) \frac{\vartheta}{\vartheta_
    {\mathrm{ref}}} \\
    \label{Eqn:lama_parameters2}
    \mu(\mathbf{x},c) &= \mu_0(\mathbf{x}) - \mu_{1}
    (\mathbf{x}) \frac{c}{c_{\mathrm{ref}}} - \mu_{2}
    (\mathbf{x}) \frac{\vartheta}{\vartheta_{\mathrm{ref}}}
  \end{align}
\end{subequations}
where $\lambda_0$ and $\mu_0$ are the Lam\'e parameters of 
the virgin material. $\lambda_{1}$ and $\mu_{1}$ are the 
parameters that account for the effect of concentration 
of chemical species on degradation of solid. $\lambda_{2}$ 
and $\mu_{2}$ are the parameters that account for the temperature 
effect on the degrading solid. It should be noted that $\lambda_
{1}$, $\mu_{1}$, $\lambda_{2}$, and $\mu_{2}$ are all positive. 
Furthermore, these parameters are constrained such that the bulk 
modulus and shear modulus are strictly positive.

%====================================================;
%  Subsubsection: Deformation dependent diffusivity  ;
%====================================================;
\subsubsection{Deformation dependent diffusivity}
\label{SubSubSec:DeformDiffusivity}
The effect of deformation on diffusivity is modeled as follows:~When 
tensile and shear strains are predominant, we have the following 
constitutive model
%---------------------------------------------------------;
%  Equation: Deformation-dependent diffusivity (Tension)  ;
%---------------------------------------------------------;
\begin{align}
  \label{Eqn:Diffusivity_coeff_t}
  \mathbf{D}_{\varkappa \varkappa} =  \mathbf{D}_{0} &+ 
  \left( \mathbf{D}_{T} - \mathbf{D}_{0} \right)
  \frac{\left(\exp[\eta_T I_{\mathbf{E}}] - 1 \right)}
  {\left(\exp[\eta_T E_{\mathrm{ref}T}] - 1 \right)}
  +\left( \mathbf{D}_{S} - \mathbf{D}_{0} \right)
  \frac{ \left(\exp[\eta_S II_{\mathbf{E}}] - 1 \right)}
  {\left(\exp[\eta_S E_{\mathrm{ref}S}] - 1 \right)} \nonumber 
  \\
  &+ \left( \mathbf{D}_{MS} - \mathbf{D}_{0} \right)
  \frac{ \left(\exp[\eta_{MS} III_{\mathbf{E}}] - 1 \right)}
  {\left(\exp[\eta_{MS} E_{\mathrm{ref}MS}] - 1 \right)}
\end{align}
where $I_{\mathbf{E}}$, $II_{\mathbf{E}}$, and $III_{\mathbf{E}}$ 
are the first, second, and third invariants of the Green-St.~Venant 
strain tensor. These are defined as follows:
%----------------------------------------------;
%  Equations: Invariants of strain tensor (E)  ;
%----------------------------------------------;
\begin{subequations}
  \begin{align}
    \label{Eqn:First_Invariant}
    I_{\mathbf{E}} &:= \mathrm{tr}[\mathbf{E}]\\
    \label{Eqn:Second_Invariant}
    II_{\mathbf{E}} &:= \sqrt{2 \, \mathrm{dev}[\mathbf{E}] 
    \bullet \mathrm{dev}[\mathbf{E}]} = \sqrt{\frac{2}{3}(3 
    \mathrm{tr}[{\mathbf{E}^2}] - (\mathrm{tr}[\mathbf{E}])^2)} \\
    \label{Eqn:Third_Invariant}
    III_{\mathbf{E}} &:= \mathrm{det}\left[\frac{1}{II_{\mathbf{E}}} 
    \mathrm{dev}[\mathbf{E}] \right]
  \end{align}
\end{subequations}
where $\mathrm{dev}[\mathbf{E}] := \mathbf{E} - \frac{1}{3} \mathrm{tr}
[\mathbf{E}] \mathbf{I}$ is the deviatoric part of $\mathbf{E}$. These 
invariants are used to model the effect of dilation, magnitude of distortion,
and mode of distortion on the diffusivity of the solid. $\eta_T$, $\eta_S$, 
and $\eta_{MS}$ are non-negative parameters. $E_{\mathrm{ref}T}$, $E_{\mathrm{ref}S}$,
and $E_{\mathrm{ref}MS}$ are reference measures of the tensile strain, shear 
strain, and mode of shear strain respectively. $\mathbf{D}_0$, $\mathbf{D}_T$,
$\mathbf{D}_S$, and $\mathbf{D}_{MS}$ are, respectively, the reference 
diffusivity tensors under no strain, tensile strain, and shear strain. 

When compression and shear strains are predominant, deformation dependent 
diffusivity is modeled as follows:
%-------------------------------------------------------------;
%  Equation: Deformation-dependent diffusivity (Compression)  ;
%-------------------------------------------------------------;
\begin{align}
  \label{Eqn:Diffusivity_coeff_c}
  \mathbf{D}_{\varkappa \varkappa} =  \mathbf{D}_{0} &+ 
  \left( \mathbf{D}_{0} - \mathbf{D}_{C} \right)
  \frac{\left(\exp[\eta_T I_{\mathbf{E}}] - 1 \right)}
  {\left(\exp[\eta_T E_{\mathrm{ref}T}] - 1 \right)}
  +\left( \mathbf{D}_{S} - \mathbf{D}_{0} \right)
  \frac{ \left(\exp[\eta_S II_{\mathbf{E}}] - 1 \right)}
  {\left(\exp[\eta_S E_{\mathrm{ref}S}] - 1 \right)} \nonumber 
  \\
  &+ \left( \mathbf{D}_{MS} - \mathbf{D}_{0} \right)
  \frac{ \left(\exp[\eta_{MS} III_{\mathbf{E}}] - 1 \right)}
  {\left(\exp[\eta_{MS} E_{\mathrm{ref}MS}] - 1 \right)}
 \end{align}
where $\eta_C$ is a non-negative parameter, $E_{\mathrm{ref}C}$ 
is a reference measure of the compression strain, and $\mathbf{D}_C$ 
is the reference diffusivity tensor under compressive strain. 

%%-------------------------------------------------------------------------;
%%  Remark: Deformation-dependent diffusivity for small and large strains  ;
%%-------------------------------------------------------------------------;
%\begin{remark}
%  In \citep{Mudunuru_Nakshatrala_IJNME_2011}, a constitutive 
%  model has been developed for deformation dependent diffusivity 
%  based on small-strain assumption. However, it should be noted 
%  that the model proposed by the authors is a special case and 
%  is obtained by linearizing the equations \eqref{Eqn:Diffusivity_coeff_t} 
%  and \eqref{Eqn:Diffusivity_coeff_c}. This model is developed 
%  based on the experimental evidence that the relative diffusion 
%  rate varies exponentially with respect to the trace of strain 
%  \citep{McAfee_JoCP_1958_v28_p218,McAfee_JoCP_1958_v28_p226}. 
%  In this paper, we have taken a step further to calibrate 
%  these materials parameters according to the experimental 
%  data for finite strains based on the model given by equations 
%  \eqref{Eqn:Diffusivity_coeff_t} and \eqref{Eqn:Diffusivity_coeff_c}.
%\end{remark}

%--------------------------------;
%  Remark: Choice of invariants  ;
%--------------------------------;
\begin{remark}
  Note that deformation dependent diffusivity given by equations 
  \eqref{Eqn:Diffusivity_coeff_t} and \eqref{Eqn:Diffusivity_coeff_c}
  can be constructed using a different set of invariants of a given 
  strain tensor. This invariants can be either principal or Hencky
  type \citep{Lurie_Elasticity,Plesek_Kruisova_CompStruct_2006_v84_p1141,
  Criscione_Humphrey_Douglas_Hunter_JMPS_2000_v48_p2445} based on the 
  nature of material and associated experimental data. The proposed 
  framework can accommodate such models with minor modifications.
  
  In case of transversely isotropic materials with fibers running 
  along the direction $\mathbf{M}_{tf}$, the following invariants 
  are needed to model deformation dependent diffusivity in addition 
  to the invariant set given by equations \eqref{Eqn:First_Invariant}--\eqref{Eqn:Third_Invariant}
  %---------------------------------------------------------;
  %  Equations: Additional invariants of strain tensor (E)  ;
  %---------------------------------------------------------;
  \begin{subequations}
    \begin{align}
      \label{Eqn:Fourth_Invariant}
      IV_{\mathbf{E}} &:= \mathbf{M}_{tf} \bullet \mathbf{E} 
      \mathbf{M}_{tf} \\
      \label{Eqn:Fifth_Invariant}
      V_{\mathbf{E}} &:= \mathbf{M}_{tf} \bullet \mathbf{E}^2 
      \mathbf{M}_{tf}
    \end{align}
  \end{subequations}
\end{remark}
For more details on selection of invariants for 
transversely isotropic or orthotropic materials 
see \citep{Lurie_Elasticity,Holzapfel,Ogden}.

%-------------------------------------------------------------;
%  Subsubsection: Deformation dependent thermal conductivity  ;
%-------------------------------------------------------------;
\subsubsection{Deformation dependent thermal conductivity}
\label{SubSubSec:DeformThermalConductivity}
The effect of deformation of the solid on thermal conductivity 
is modeled as follows \citep{2006_Bhowmick_Shenoy_JCP_v125}: 
%--------------------------------------;
%  Equation: conductivity coefficient  ;
%--------------------------------------;
\begin{align}
  \label{Eqn:conductivity_coeff_t}
  \mathbf{D}_{\vartheta\vartheta} = \mathbf{K}
  _{0\vartheta} (1 + I_{\mathbf{E}})^{-\delta}
\end{align}
where $\delta$ is a non-negative parameter. $\mathbf{K}_
{0\vartheta}$ is the reference conductivity tensor under 
no strain. Based on molecular dynamics simulations, Bhowmick 
and Shenoy \citep{2006_Bhowmick_Shenoy_JCP_v125} suggested 
$\delta$ to be $9.59$ and $\mathbf{K}_{0\vartheta} = 
4.61 \vartheta^{-1.45}$ (for certain brittle-type materials). 
For various other ductile or brittle-type materials, these 
parameters can be determined by experiments or can be constructed 
using Lennard-Jones potential in molecular dynamics.

%-----------------------------------;
%  Remark: Dufour and soret tensor  ;
%-----------------------------------;
\begin{remark}
  Due to the lack of experimental data, we assume that 
  Dufour and Soret tensors do not depend on the deformation 
  of solid. However, it should be noted that the proposed 
  thermodynamic and computational framework can accommodate 
  deformation dependent Dufour and Soret tensors with minor 
  modifications (whenever such an experimental evidence is 
  available).
\end{remark}

%=============================================================================================;
%  Subsection: Special cases of the general degradation model and their thermodynamic status  ;
%=============================================================================================;
\subsection{Special cases of the general degradation model and their thermodynamic status}
\label{SubSec:Special_Cases_DegradModel}
\begin{figure}
  \centering
  \includegraphics[scale=0.6,clip]{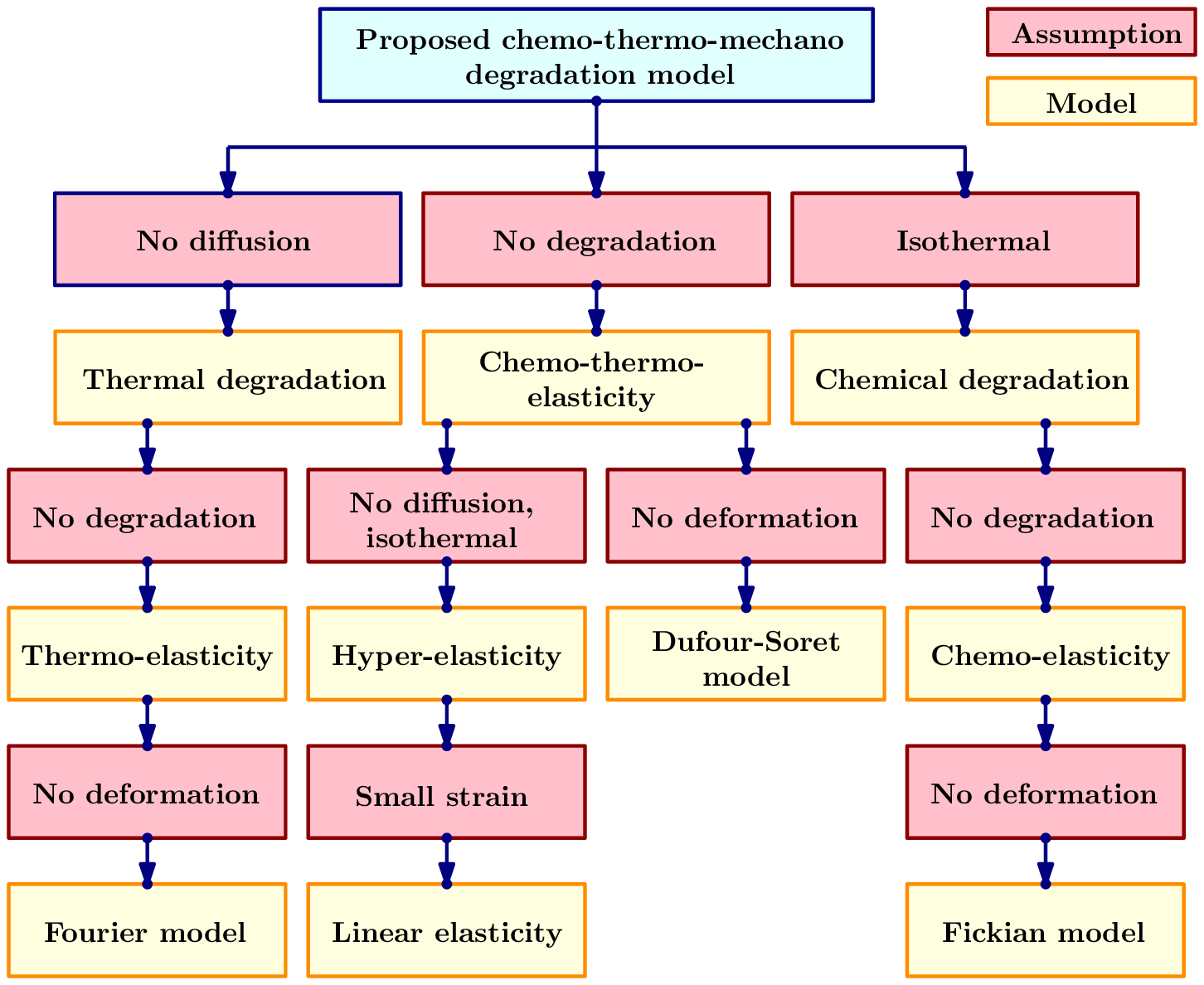}
  \caption{\textsf{Special cases of the proposed chemo-thermo-mechano 
    degradation model:}~Many existing degrading and non-degrading 
    constitutive models are special cases of the proposed hierarchical 
    model, with appropriate assumptions.
  \label{Fig:special_cases}}
\end{figure}

The following popular non-degradation constitutive models 
can be shown as special cases of the proposed degradation 
model, as shown in Figure \ref{Fig:special_cases}, 
when the material parameters are assumed to be independent 
of concentration, temperature, and deformation of the solid.
That is, the Lam\'e parameters and $\mathbf{D}_{\alpha \beta}$ 
($\alpha$ and $\beta$ represent either $\vartheta$ or $\varkappa$) 
are independent of $c$, $\vartheta$, and $\mathbf{E}$.
%---------------------------------------------------;
%  Itemize: Special cases of the degradation model  ;
%---------------------------------------------------;
\begin{enumerate}
  \item \texttt{Fourier and Fickian models:}~The standard heat 
    conduction constitutive model is obtained by assuming the 
    solid to be rigid and mass concentration of diffusing chemical 
    species to be equal to zero. Similarly, to recover the standard 
    Fickian model we assume the solid to be rigid and temperature 
    of the homogenized body to be constant.
  \item \texttt{Dufour-Soret model:}~This model is obtained by 
    assuming the solid to be rigid. Furthermore, the thermo-chemo 
    coupling parameter $d_{\vartheta c}$ is neglected. 
  \item \texttt{Linearized elasticity and hyperelasticity:}~To obtain 
    hyperelastic constitutive models, we assume isothermal conditions 
    and mass concentration of diffusing chemical species to be equal 
    to zero. The linearized elasticity model can be recovered from any 
    given hyperelastic model by assuming that the small strains assumption 
    given by equation \eqref{Eqn:linear} holds.
  \item \texttt{Thermoelasticity:}~The standard thermoelasticity model 
    can be recovered by assuming mass concentration of diffusing chemical 
    species to be equal to zero. The material parameters are assumed to 
    be independent of temperature and deformation.
  \item \texttt{Chemoelasticity:}~ Similarly, the standard chemoelasticity 
    model can be recovered by assuming isothermal conditions. The material 
    parameters are assumed to be independent of concentration and deformation.
  \item \texttt{Chemo-Thermoelasticity:}~Herein, we assume that the material 
    parameters are independent of concentration, temperature, and deformation. 
    In addition, thermo-chemo coupling parameter $d_{\vartheta c}$, Dufour 
    tensor, and Soret tensor are neglected.
\end{enumerate}
One can also derive specialized (thermo-mechano and chemo-mechano) degradation models:
%---------------------------------------------------;
%  Itemize: Special cases of the degradation model  ;
%---------------------------------------------------;
\begin{enumerate}
  \item \texttt{Thermo-mechano degradation model:}~This model is 
    obtained from the thermoelasticity model by relaxing the assumption 
    that material parameters are independent of temperature and deformation.
  \item \texttt{Chemo-mechano degradation model:}~Similar to thermo-mechano 
    degradation model, this degradation model is obtained from the chemoelasticity 
    model by relaxing the assumption that material parameters are independent 
    of concentration and deformation.
\end{enumerate}
%%
%-----------------------------------------------;
%  Figure: Special cases of the proposed model  ;
%-----------------------------------------------;
%==========================================;
%  Subsubsection: Status of MKM-KBN model  ;
%==========================================;
\subsubsection{Status of the degradation model in \citep{Mudunuru_Nakshatrala_IJNME_2011}}
\label{Subsec:Status_Proposed_Model}
The small-strain chemo-mechano degradation model proposed in 
\citep{Mudunuru_Nakshatrala_IJNME_2011} is a special case 
of the proposed chemo-thermo-mechano degradation, and can 
be obtained under a plethora of assumptions. These assumptions 
include steady-state response, small strains, and isothermal 
conditions with negative volumetric heat source in the entire 
degrading body. One also needs to neglect chemo-thermo, 
chemo-mechano, and thermo-mechano couplings. Moreover, 
the functional forms of the specific Helmholtz potential 
and rate of dissipation functional need to be:
%--------------------------------------------------------------------;
%  Equation: Specific Helmholtz potential and dissipation (MKM-KBN)  ;
%--------------------------------------------------------------------;
\begin{align}
  \label{Eqn:Degradation_Helmholtz_MKM_KBN}
  A &= \frac{1}{\rho}\psi + \frac{R_s \vartheta_
  {\mathrm{ref}}}{2} \{c - c_{\mathrm{ref}}\}^2 \\
  \zeta &= \frac{1}{R_s\vartheta_{\mathrm{ref}}} \mathrm{grad}[\varkappa] 
  \bullet \mathbf{D}_{\varkappa \varkappa}\mathrm{grad}[\varkappa]
\end{align}
where the stored strain energy density functional is given by:
%-------------------------------------------------------;
%  Equation: Stored strain potential for MKM-KBN model  ;
%-------------------------------------------------------;
\begin{align}
  \label{Eqn:MKM_KBN_StoredEnergy}
  \psi = \hat{\psi}(\mathbf{E}_l,c) = 
  \frac{\lambda(c)}{2} 
  \left( \mathrm{tr}
  [\mathbf{E}_l]\right)^2 + \mu(c) \mathbf{E}_l 
  \bullet \mathbf{E}_l
\end{align}
%
%Note that balance of mass is trivially satisfied. 
%
Under the small strain assumption given by equation 
\eqref{Eqn:linear}, the Cauchy stress, chemical potential, 
and mass transfer flux vector can be written as:
%--------------------------------------------------;
%  Equation: Cauchy stress and chemical potential  ;
%--------------------------------------------------;
\begin{align}
  \label{Eqn:Cauchy_Stress_MKMKBN}
  &\mathbf{T} = \rho \frac{\partial A}{\partial \mathbf{E}_l} =
  \lambda(c) \mathrm{tr}[\mathbf{E}_l] \mathbf{I} + 2\mu(c) \mathbf{E}_l \\
  \label{Eqn:Chemical_Potential_MKMKBN}
  &\varkappa = \frac{\partial A}{\partial c} 
  = R_s\vartheta_{\mathrm{ref}} \{c - c_{\mathrm{ref}}\} \\
  \label{Eqn:Mass_Transfer_Flux_MKMKBN}
  &\mathbf{h} = -\frac{1}{2} \rho \frac{\partial \hat{\zeta}}{\partial 
  \mathrm{grad}[\varkappa]} = -\frac{\rho}{R_s \vartheta_{\mathrm{ref}}}
  \mathbf{D}_{\varkappa\varkappa}\mathrm{grad}[\varkappa]
\end{align}
The balance of chemical species and the balance 
of linear momentum for the solid are given by 
equations \eqref{Eqn:Balance_of_species_D} and 
\eqref{Eqn:Balance_of_LM_D}. Under the isothermal 
condition, the balance of energy simplifies to the 
following expression:
%-----------------------------------------------------------;
%  Equation: Balance of energy under isothermal conditions  ;
%-----------------------------------------------------------;
\begin{align}
  \label{Eqn:Isot_assumption}
  q = - \frac{\rho}{R_s \vartheta_{\mathrm{ref}}} 
  \mathrm{grad}[\varkappa] \bullet \mathbf{D}_{\varkappa 
  \varkappa}\mathrm{grad}[\varkappa]
\end{align}
which means that $q$ needs to be non-positive in order to maintain 
the isothermal condition. The deformation dependent diffusivity $\mathbf{D}
_{\varkappa \varkappa}$ is based on the small strain assumption, which 
is obtained by linearizing the equations \eqref{Eqn:Diffusivity_coeff_t} 
and \eqref{Eqn:Diffusivity_coeff_c}. Note that this model is developed 
based on the experimental evidence that the relative diffusion 
rate varies exponentially with respect to the trace of strain 
\citep{McAfee_JoCP_1958_v28_p218,McAfee_JoCP_1958_v28_p226}. 
In this paper, we have taken a step further to calibrate these materials 
parameters according to the experimental data for finite strains based on 
the model given by equations \eqref{Eqn:Diffusivity_coeff_t} and 
\eqref{Eqn:Diffusivity_coeff_c}.

%% file: Sections_Model/S4_Model_Calibration.tex
\section{CALIBRATION WITH EXPERIMENTAL DATA}
\label{Sec:S4_Model_Calibration}
In this section, we will calibrate the proposed model for 
the diffusivity using the experimental data set reported 
in \citep{McAfee_JoCP_1958_v28_p218,McAfee_JoCP_1958_v28_p226}. 
These experiments were conducted on spherical shells made of 
glass, which is a brittle material. These papers report the 
variation of diffusivity under various deformation modes: 
tension, compression, and shear.
The calibration study presented below, which also includes 
a statistical analysis of the fit, will be valuable in two 
ways. 
\emph{First}, it demonstrates the predictive capabilities of 
the proposed constitutive model, and provides confidence in 
the model to be able to apply to other brittle materials like 
ceramics and even to quasi-brittle materials like concrete with 
appropriate modifications.
\emph{Second}, it provides order-of-magnitude estimates for 
various parameters in the diffusivity model for realistic 
materials. This will guide in the selection of values for 
these parameters in the subsequent numerical studies.
%------------------------------------------------------;
%  Figure-1: Pictorial description of spherical shell  ;
%------------------------------------------------------;
\begin{figure}
  \centering
  \includegraphics[scale=0.7]{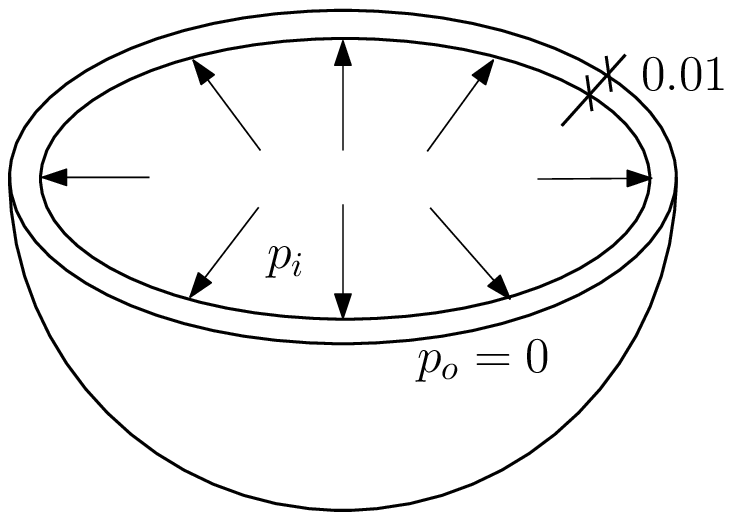}
  \caption{\textsf{Calibration with experimental data:}~A 
    pictorial description of the degrading shell used for 
    calibrating the proposed model with the experimental data.
    \label{Fig:glass_experimental_data_pictorial_description}}
\end{figure}
Figure \ref{Fig:glass_experimental_data_pictorial_description} 
provides the geometry and the loading on a spherical shell. The 
inner and outer radii are, respectively, $a = 0.99$ and $b = 1.0$. 
The boundary conditions for the deformation subproblem is that the 
pressure within the sphere is varied from $p_i = 0$ to $p_i = 0.68947 
\, \mathrm{MPa}$ ($100 \, \mathrm{psi}$) and the external surface 
is traction free. For the diffusion subproblem, $c(a) = 0$ and $c(b) = 
1$. In this scenario, it can be assumed that the tensile strain is 
predominant. The diffusion can be assumed to be isotropic. Hence, 
equation \eqref{Eqn:Diffusivity_coeff_t} is simplified as follows:
%----------------------------------;
%  Equation: Exp Data Diffusivity  ;
%----------------------------------;
\begin{align}
  \label{Eqn:Simplified_Diffusivity_coeff_t}
  D =  D_{0} + \left( D_{T} -  D_{0} \right) 
  \frac{ \left(\exp[\eta_T I_{\mathbf{E}}] 
  - 1 \right)}{ \left(\exp[\eta_T E_{\mathrm{ref}_{T}}] 
  - 1\right)}
\end{align}
The sample size to estimate the parameters in 
the proposed deformation-diffusivity model has 
been taken to be 3. It has been reported that 
$D_0 = 7.26\times 10^{-13} \mathrm{m}^2/\mathrm{sec}^{-1}$ 
for glass fibers by \citep{McAfee_JoCP_1958_v28_p218}. 
Based on the chosen sample size and value of $D_0$, 
the estimated diffusivity parameters are given as 
follows: 
%-----------------------------------------------;
%  Equation: Diffusivity parameters in tension  ;
%-----------------------------------------------;
\begin{align}
  \label{Eqn:ParamSet_3points_Tension}
  \eta_T = 1.43\times 10^4, \ D_T=23.39\times 10^{-13},\ 
  E_{\mathrm{ref}_{T}} = 1.833 \times 10^{-3}
\end{align}
Using the experimental data reported in 
\citep{McAfee_JoCP_1958_v28_p226} under 
compressive and shear strains, and following 
a similar procedure as before, the following 
diffusivity parameters are obtained: 
%-------------------------------------------------------------;
%  Equation: Diffusivity parameters in compression and shear  ;
%-------------------------------------------------------------;
\begin{subequations}
  \begin{align}
    \label{Eqn:ParamSet_3points_Compression}
    &\eta_C = 401.19, \ D_C=8.66\times 10^{-13},\ 
    E_{\mathrm{ref}_{C}} = 1.0 \times 10^{-3} \\
    \label{Eqn:ParamSet_3points_Shear}
    &\eta_S = -239.61, \ D_S=8.65\times 10^{-13},\ 
    E_{\mathrm{ref}_{S}} = 3.0 \times 10^{-3}
  \end{align}
\end{subequations}
%
%-----------------------------------------------------------------;
%  Figure-2: Comparison of experimental data with proposed model  ;
%-----------------------------------------------------------------;
\begin{figure}
  \centering
  \includegraphics[scale=0.4,clip]{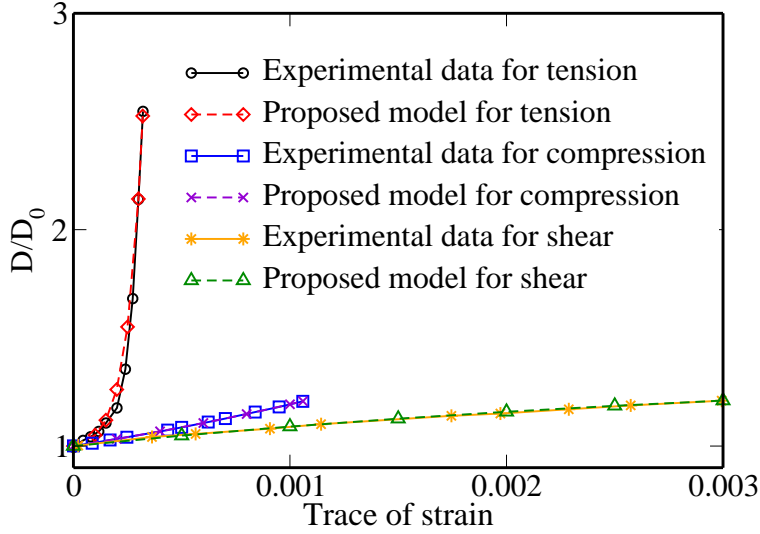}
  \caption{\textsf{Calibration with experimental data:}~This 
    figure compares the experimental data reported in 
    \citep{McAfee_JoCP_1958_v28_p218,McAfee_JoCP_1958_v28_p226} 
    with the proposed constitutive model. The sample size 
    is taken to be 3. The strain invariants are given by equations 
    \eqref{Eqn:First_Invariant}--\eqref{Eqn:Second_Invariant}. 
    A good agreement has been observed between the experimental 
    data and the proposed constitutive model for the diffusivity 
    under tensile, compressive, and shear strains. 
    \label{Fig:glass_experimental_data_comparison}}
\end{figure}
We then compared the proposed model (which is obtained based on sample 
size of 3 points) with the experimental data set of 10 points. Figure 
\ref{Fig:glass_experimental_data_comparison} shows the relation between 
the relative diffusion coefficient $D/D_0$ and various strain invariants. 
From this figure, it is evident that the proposed model is in a good 
agreement with the experimental data. Table \ref{Table:StatAnalysis_DataFit_TenCompShear} 
provides a statistical analysis on the fit of the experimental data with 
the proposed model. The coefficient of determination is close to 1. This 
means that the proposed model based on parameter set given by equations 
\eqref{Eqn:ParamSet_3points_Tension}--\eqref{Eqn:ParamSet_3points_Shear} 
is a good fit to the set of experimental data of various sample sizes.
To calibrate $\mathbf{D}_{MS}$, $\eta_{MS}$, and $E_{\mathrm{ref}MS}$, 
we need additional experimental data related to mode of shear. However, 
such a data set to calibrate the effect of distortion due to shear on 
the diffusivity of glass is not available in literature yet. Hence, we 
could not calibrate $\mathbf{D}_{MS}$, $\eta_{MS}$, and $E_{\mathrm{ref}MS}$.
Whenever such an experimental data is available, one can easily calibrate 
these parameters in a similar fashion.

%----------------------------------------------------------------------------;
%  Table: Statistical analysis of the fit (Tension, compression, and shear)  ;
%----------------------------------------------------------------------------;
\begin{table}
  \centering
  \caption{\textsf{A statistical analysis of the fit}:~This 
    table provides the goodness of fit of the proposed 
    model with the experimental data set reported in 
    \citep{McAfee_JoCP_1958_v28_p218,McAfee_JoCP_1958_v28_p226}. 
    Analysis is performed for various extracted sample sizes, 
    and under tension, compression and shear strains. It is 
    observed that the coefficient of determination is close 
    to 1. \label{Table:StatAnalysis_DataFit_TenCompShear}}
  \begin{tabular}{|c|c|c|c|c|c|c|c|c|c|} \hline
    \multirow{2}{*}{{\small Sample size}} & \multicolumn{3}{|c|}{{\small Mean data}} 
    & \multicolumn{3}{|c|}{{\small Standard deviation data}} & \multicolumn{3}{|c|}{{\small 
    Coefficient of determination}} \\ \cline{2-3} \cline{3-4} \cline{5-6} \cline{6-7} 
    \cline{8-9} \cline{9-10}
    & {\scriptsize Tension} & {\scriptsize Compression} & {\scriptsize Shear} & 
    {\scriptsize Tension} & {\scriptsize Compression} & {\scriptsize Shear} & 
    {\scriptsize Tension} & {\scriptsize Compression} & {\scriptsize Shear}
    \\ \hline
    $10$ & 1.507 & 1.093 & 1.114 & 0.526 & 0.073 & 0.068 & 0.988 & 0.999 & 0.997 \\
    \hline
    $25$ & 1.505 & 1.094 & 1.108 & 0.511 & 0.062 & 0.065 & 0.986 & 0.999 & 0.997 \\
    \hline
    $50$ & 1.521 & 1.095 & 1.107 & 0.515 & 0.062 & 0.062 & 0.987 & 0.999 & 0.997 \\
    \hline
    $75$ & 1.391 & 1.097 & 1.115 & 0.468 & 0.062 & 0.059 & 0.984 & 0.999 & 0.997 \\
    \hline
  \end{tabular}
\end{table}

%% file: Sections_Model/S5_Model_IBVP.tex
%***************************************;
%                                       ;
%  NAME                                 ;  
%    S5_Model_IBVP.tex                  ;
%                                       ;
%  WRITTEN BY                           ;
%     Can Xu                            ;
%     Kalyana Babu Nakshatrala          ;
%                                       ; 
%***************************************;
\section{INITIAL BOUNDARY VALUE PROBLEM AND MATHEMATICAL ANALYSIS}
\label{Sec:S5_Model_IBVP}
From the above statements, the governing equations for the 
proposed chemo-thermo-mechano degrading model are stated 
as follows. Let the boundary of $\Omega_0(\mathfrak{B})$ 
be denoted as $\partial \Omega_0$ and the corresponding 
unit outward normal to this boundary be denoted by $\widehat{\mathbf{n}}_0
(\mathbf{p})$. 
Similarly, $\partial \Omega_{t}$ denotes the boundary of 
$\Omega_{t}(\mathfrak{B})$ and the corresponding unit outward 
normal to this boundary is denoted by $\widehat{\mathbf{n}}
(\mathbf{x},t)$. For the deformation subproblem, the boundary 
is divided into two complementary parts: $\Gamma^{\mathrm{D}}_u$ 
and $\Gamma^{\mathrm{N}}_u$ such that $\Gamma^{\mathrm{D}}_u 
\cup \Gamma^{\mathrm{N}}_u = \partial \Omega_0$ and $\Gamma^
{\mathrm{D}}_u \cap \Gamma^{\mathrm{N}}_u = \emptyset$.
$\Gamma^{\mathrm{D}}_u$ is the part of the boundary on which 
displacement is prescribed and $\Gamma^{\mathrm{N}}_u$ is the 
part of the boundary on which traction is prescribed. 

Similarly, for the transport and thermal subproblem, the boundary 
is divided into complementary parts: $\Gamma^{\mathrm{D}}_c$ and 
$\Gamma^{\mathrm{N}}_c$ and $\Gamma^{\mathrm{D}}_{\vartheta}$ and 
$\Gamma^{\mathrm{N}}_{\vartheta}$ such that $\Gamma^{\mathrm{D}}_c 
\cup \Gamma^{\mathrm{N}}_c = \partial \Omega_0$, $\Gamma^{\mathrm{D}}
_{\vartheta} \cup \Gamma^{\mathrm{N}}_{\vartheta} = \partial \Omega$, 
$\Gamma^{\mathrm{D}}_c \cap \Gamma^{\mathrm{N}}_c = \emptyset$, and 
$\Gamma^{\mathrm{D}}_{\vartheta} \cap \Gamma^{\mathrm{N}}_{\vartheta} 
= \emptyset$. $\Gamma^{\mathrm{D}}_c$ is the part of the boundary on 
which concentration is prescribed. $\Gamma^{\mathrm{N}}_c$ is the part 
of the boundary on which total/diffusive flux is prescribed. $\Gamma^
{\mathrm{D}}_{\vartheta}$ is the part of the boundary on which temperature 
is prescribed. $\Gamma^{\mathrm{N}}_{\vartheta}$ is the part of the boundary 
on which thermal flux is prescribed. In case of steady-state analysis, it 
should be noted that the $\mathrm{meas}\left(\Gamma^{\mathrm{D}}_c \right) 
> 0$, $\mathrm{meas}\left(\Gamma^{\mathrm{D}}_{\vartheta} \right) > 0$, 
and $\mathrm{meas}\left(\Gamma^{\mathrm{D}}_u \right) > 0$. However, such 
a condition is not required for studying transient problems.

%=========================================================;
%  Subsection: Governing equations of the proposed model  ;
%=========================================================;
\subsection{Governing equations of the proposed model}
\label{SubSec:Governing_equations}
The governing equations for the \emph{deformation sub-problem} 
take the following form:
%------------------------------------;
%  Equation: Deformation subproblem  ;
%------------------------------------;
\begin{subequations}
  \label{Eqn:Deformation_GE}
  \begin{alignat}{2}
    \label{Eqn:Deformation_GE_1}
    &\rho_0 \dot{\mathbf{v}}(\mathbf{p},t) 
    = \mathrm{Div}[\mathbf{P}] + \rho_0 
    \mathbf{b}(\mathbf{p},t) \quad && 
    \mathrm{in} \; \Omega_{0} \times ]0, 
    \mathcal{I}[ \\
    \label{Eqn:Deformation_GE_2}
    &\mathbf{u}(\mathbf{p},t) = \mathbf{u}
    ^{\mathrm{p}} (\mathbf{p},t) \quad &&
    \mathrm{on} \; \Gamma^{\mathrm{D}}_{u} 
    \times ]0, \mathcal{I}[ \\
    \label{Eqn:Deformation_GE_3}
    &\mathbf{P} \widehat{\mathbf{n}}_{0}(\mathbf{p}) 
    = \mathbf{t}^{\mathrm{p}}(\mathbf{p},t) \quad 
    && \mathrm{on} \; \Gamma^{\mathrm{N}}_{u} \times 
    ]0, \mathcal{I}[ \\
    &\mathbf{u}(\mathbf{p},t=0) = \mathbf{u}^{\mathrm{i}} 
    (\mathbf{p}) \quad &&\mathrm{in} \; \Omega_{0} \\
     &\mathbf{v}(\mathbf{p},t=0) = \mathbf{v}^{\mathrm{i}} 
    (\mathbf{p}) \quad &&\mathrm{in} \; \Omega_{0}
      \end{alignat}
\end{subequations}
where $\mathbf{u}^{\mathrm{p}}(\mathbf{p},t)$ denotes the 
prescribed displacement on the boundary and $\mathbf{t}^
{\mathrm{p}}(\mathbf{p},t)$ is the prescribed traction on 
the boundary. $\mathbf{u}^{\mathrm{i}}(\mathbf{p})$ and 
$\mathbf{v}^{\mathrm{i}}(\mathbf{p})$ are the 
initial conditions for the displacement and 
velocity, respectively.

The governing equations for the \emph{transport sub-problem} 
take the following form:
%----------------------------------;
%  Equation: Transport subproblem  ;
%----------------------------------;
\begin{subequations}
  \label{Eqn:Transport_GE}
  \begin{alignat}{2}
    \label{Eqn:Transport_GE_1}
    &\rho_0 \dot{c}(\mathbf{p},t)
    + \mathrm{Div}[\mathbf{h}_0] 
    = h_{0}(\mathbf{p},t) \quad 
    && \mathrm{in} \; \Omega_{0} 
    \times ]0, \mathcal{I}[ \\
    \label{Eqn:Transport_GE_2}
    &c(\mathbf{p},t) = c^{\mathrm{p}}(\mathbf{p},t)
    \quad &&\mathrm{on} \; \Gamma^{\mathrm{D}}_{c}  
    \times ]0, \mathcal{I}[ \\
    \label{Eqn:Transport_GE_3}
    &\mathbf{h}_{0} \bullet \widehat{\mathbf{n}}_{0}
    (\mathbf{p}) = h^{\mathrm{p}} (\mathbf{p},t) \quad 
    &&\mathrm{on} \; \Gamma^{\mathrm{N}}_{c} \times 
    ]0, \mathcal{I}[\\
    &c(\mathbf{p},t=0) = c^{\mathrm{i}} (\mathbf{p}) \quad 
    && \mathrm{in} \; \Omega_{0}
  \end{alignat}
\end{subequations}
where $c^{\mathrm{p}}(\mathbf{p},t)$ denotes the prescribed 
concentration on the boundary, $h^{\mathrm{p}}(\mathbf{p},t)$ 
is the prescribed total/diffusive flux on the boundary, and 
$c^{\mathrm{i}}(\mathbf{p})$ is the initial condition for 
the concentration field. 

The governing equations for the \emph{thermal sub-problem} 
take the following form:
%----------------------------------;
%  Equation: Transport subproblem  ;
%----------------------------------;
\begin{subequations}
  \label{Eqn:Heat_GE}
  \begin{alignat}{2}
    \label{Eqn:Heat_GE_1}
    &\rho_0 \vartheta(\mathbf{p},t) \dot{\eta} = 
    -\mathrm{Div}[\mathbf{q}_0] -\mathrm{Grad}
    [\varkappa] \bullet \mathbf{h}_0 + q_0(\mathbf{p},t)  
    && \quad \mathrm{in} \; \Omega_{0} \times 
    ]0, \mathcal{I}[ \\
    \label{Eqn:Heat_GE_2}
    &\vartheta(\mathbf{p},t)= \vartheta^{\mathrm{p}}
    (\mathbf{p},t) && \quad \mathrm{on} \; \Gamma^
    {\mathrm{D}}_{\vartheta} \times ]0, \mathcal{I}[ \\
    \label{Eqn:Heat_GE_3}
    &\mathbf{q}_{0} \bullet \widehat{\mathbf{n}}_{0}
    (\mathbf{p}) = q^{\mathrm{p}} (\mathbf{p},t) 
    &&\quad \mathrm{on} \; \Gamma^{\mathrm{N}}_{c} 
    \times ]0, \mathcal{I}[ \\
    &\vartheta(\mathbf{p},t=0) = \vartheta^{\mathrm{i}} 
    (\mathbf{p}) && \quad \mathrm{in} \; \Omega_{0}
  \end{alignat}
\end{subequations}
where $\vartheta^{\mathrm{p}}(\mathbf{p},t)$ denotes the 
prescribed temperature on the boundary, $q^{\mathrm{p}}
(\mathbf{p},t)$ is the prescribed heat flux on the 
boundary, and $\vartheta^{\mathrm{i}}(\mathbf{p})$ is 
the initial condition for the temperature field. 

%======================================================;
%  Subsection: On the stability of unsteady solutions  ;
%======================================================;
\subsection{On the stability of unsteady solutions}
\label{SubSec:Degradation_Bounded}
We now show that the unsteady solutions under the proposed 
mathematical model for degradation are stable in the sense 
of a dynamical system. There are different notions of stability, 
and herein we shall establish the stability in the sense of 
Lyapunov \citep{Dym_stability}. For the entire analysis presented 
in this section, we assume homogeneous Dirichlet boundary conditions 
on the entire boundary for the diffusion and thermal sub-problems. Let 
%-------------------------------;
%  Equation: Definition of chi  ;
%-------------------------------;
\begin{align}
  \label{Eqn:Chi_Lyapunov}
  \boldsymbol{\chi} : = \left\{\begin{array}{c}
  \boldsymbol{\varphi}  \\
  \mathbf{v}  \\
  \vartheta   \\
  c 
  \end{array}\right\}
\end{align}
Consider the following functional, which 
is defined on the reference configuration:
%--------------------------;
%  Equation: Functional V  ;
%--------------------------;
\begin{align}
  \label{Eqn:Functional_Lyapunov}
  \mathbb{V}(\boldsymbol{\chi}) := 
  \displaystyle \int \limits_{\Omega_0
  (\mathfrak{B})} \rho_{0} \left(A + 
  \vartheta \eta + \frac{1}{2} \mathbf{v} 
  \bullet \mathbf{v} \right) \; \mathrm{d} 
  \Omega_{0} + \Pi_{\mathrm{mech}, 
  \mathrm{ext}}(\boldsymbol{\varphi})
\end{align}
where $\Pi_{\mathrm{mech},\mathrm{ext}}(\boldsymbol{\varphi})$ 
is the potential energy due to external mechanical 
loading, which is assumed to be conservative. This 
implies the following 
\begin{align}
  \frac{\mathrm{d}}{\mathrm{d}t} 
  \Pi_{\mathrm{mech},\mathrm{ext}}(\boldsymbol{\varphi}) 
  = - \int_{\Omega_{0}(\mathfrak{B})} \rho_{0} \mathbf{b} 
  \bullet \mathbf{v} \; \mathrm{d} \Omega_{0} 
  - \int_{\Gamma_{u}^{\mathrm{N}}} \mathbf{t}^{\mathrm{p}} 
  \bullet \mathbf{v} \; \mathrm{d} \Gamma_{0} 
\end{align}

In the literature, the above functional has been shown to 
be a Lyapunov functional for linearized thermoelasticity 
and for themo-hyperelasticity. For example, see 
\citep{ericksen1966thermoelastic,Coleman_Dill,
Gurtin_ARMA_1975_v59_p53} and references therein. 
Herein, we shall show that the above functional is 
a legitimate Lyapunov functional for the proposed 
degradation model, and specifically use the Lyapunov's 
second method for stability (which is a classical result 
in the theory of dynamical systems; for example, see 
\citep{Hale_Kocak,Strogatz,Wiggins}) to establish the 
stability of the solutions under the proposed degradation 
model. 

To this end, we shall take the reference or equilibrium state as: 
%-----------------------------;
%  Equation: chi_equilibrium  ;
%-----------------------------;
\begin{align}
  \boldsymbol{\chi}_{\mathrm{eq}} : = \left\{\begin{array}{c}
  \boldsymbol{\varphi}_{\mathrm{eq}}  \\
  \mathbf{0}  \\
  0   \\
  0 
  \end{array}\right\}
\end{align}
where $\boldsymbol{\varphi}_{\mathrm{eq}}$ is the static 
equilibrium deformation. The above functional is a 
candidate for Lyapunov functional, as it satisfies: 
%-----------------------------;
%  Equation: Properties of V  ;
%-----------------------------;
\begin{align}
  \label{Eqn:Prop_Lyapunov_Functional}
  \mathbb{V}(\boldsymbol{\chi} = \boldsymbol{\chi}_{\mathrm{eq}}) = 0 
  \quad \mathrm{and} \quad 
  \mathbb{V}(\boldsymbol{\chi} \neq \boldsymbol{\chi}_{\mathrm{eq}}) > 0 
\end{align}
We now show that 
%------------------------------------------------;
%  Equation: Lyapunov functional (final result)  ;
%------------------------------------------------;
\begin{align}
  \label{Eqn:dV_dt}
  \frac{\mathrm{d} \mathbb{V}}{\mathrm{d}t} 
  \leq 0
\end{align}
%------------------------------------------------;
%  Equation: Lyapunov functional (final result)  ;
%------------------------------------------------;
\begin{align}
  \label{Eqn:Proof_dv_dt_lessThanZero}
  \frac{\mathrm{d} \mathbb{V}}{\mathrm{d}t} 
  &= \displaystyle \int \limits_{\Omega_{0}
  (\mathfrak{B})} \rho_{0} \left( \frac{\partial 
  A}{\partial \mathbf{F}} \bullet \dot{\mathbf{F}} 
  + \frac{\partial A}{\partial \vartheta} \dot{\vartheta} 
  + \frac{\partial A}{\partial c} \dot{c} + \dot{\vartheta} 
  \eta + \vartheta \dot{\eta} + \mathbf{v} \bullet \dot{\mathbf{v}} 
  \right) \; \mathrm{d} \Omega_{0} + \frac{\mathrm{d}}{\mathrm{d}t} 
  \Pi_{\mathrm{mech,ext}}(\boldsymbol{\varphi}) \nonumber \\
  &= \displaystyle \int \limits_{\Omega_{0}(\mathfrak{B})} 
  \rho_{0} \left(\varkappa \dot{c} + \vartheta \dot{\eta} 
  \right) \; \mathrm{d} \Omega_{0} + \displaystyle \int 
  \limits_{\Omega_{0}(\mathfrak{B})} \rho_{0} \left( 
  \frac{\partial A}{\partial \vartheta} + \eta\right) 
  \dot{\vartheta} \; \mathrm{d} \Omega_{0} + \displaystyle 
  \int \limits_{\Omega_{0}(\mathfrak{B})} \left(\rho_0 
  \mathbf{v} \bullet \dot{\mathbf{v}}+ \mathbf{P} \bullet 
  \dot{\mathbf{F}} \right) \; \mathrm{d} \Omega_{0}
  + \frac{\mathrm{d}}{\mathrm{d}t} \Pi_{\mathrm{mech,ext}}
  (\boldsymbol{\varphi}) \nonumber \\
  &= \displaystyle \int \limits_{\Omega_{0}(\mathfrak{B})} 
  \rho_{0} \left( \varkappa \dot{c} + \vartheta \dot{\eta} 
  \right) \; \mathrm{d} \Omega_{0} \nonumber \\
  &= -\displaystyle \int \limits_{\Omega_{0}(\mathfrak{B})} 
  \varkappa \mathrm{Div}[\mathbf{h}_0] \; \mathrm{d} \Omega_{0}
  -\displaystyle \int \limits_{\Omega_{0}(\mathfrak{B})} 
  \frac{\vartheta - \vartheta_{\mathrm{ref}}}{\vartheta} 
  \mathrm{Div}[\mathbf{q}_{0}] \; \mathrm{d} \Omega_{0} 
  - \displaystyle \int \limits_{\Omega_{0}(\mathfrak{B})} 
  \frac{\vartheta - \vartheta_{\mathrm{ref}}}{\vartheta} 
  \mathrm{Grad}[\varkappa] \bullet \mathbf{h}_{0} \; 
  \mathrm{d} \Omega_{0} \nonumber \\
  &= \displaystyle \int \limits_{\Omega_{0}(\mathfrak{B})} 
  \mathrm{Grad}[\varkappa] \bullet \mathbf{h}_{0} \; 
  \mathrm{d} \Omega_{0} - \displaystyle \int \limits_
  {\Omega_{0}(\mathfrak{B})} \left(1 - \frac{\vartheta_
  {\mathrm{ref}}}{\vartheta}\right) \mathrm{Div}[\mathbf{q}_0] 
  \; \mathrm{d} \Omega_{0} - \displaystyle \int \limits_
  {\Omega_{0}(\mathfrak{B})} \left(1 - \frac{\vartheta_
  {\mathrm{ref}}}{\vartheta}\right) \mathrm{Grad}[\varkappa] 
  \bullet \mathbf{h}_{0} \; \mathrm{d} \Omega_{0} \nonumber \\
  &= \displaystyle \int \limits_{\Omega_{0}(\mathfrak{B})} 
  \frac{\vartheta_{\mathrm{ref}}}{\vartheta} \left(\frac{1}{
  \vartheta} \mathrm{Grad}[\vartheta] \bullet \mathbf{q}_{0}
  + \mathrm{Grad}[\varkappa] \bullet \mathbf{h}_{0} \right) \; 
  \mathrm{d} \Omega_{0} 
  = -\displaystyle \int \limits_{\Omega_{0}(\mathfrak{B})} 
  \frac{\vartheta_{\mathrm{ref}}}{\vartheta} \zeta_0 \; 
  \mathrm{d} \Omega_{0} 
\end{align}
Since $\zeta_0 > 0$ if  $\boldsymbol{\chi} \neq \boldsymbol{\chi}_{\mathrm{eq}}$, 
$\vartheta, \, \vartheta_{\mathrm{ref}} > 0$, one 
can conclude that 
%---------------------------------;
%  Equation: V is non-increasing  ;
%---------------------------------;
\begin{align}
  \label{Eqn:dVdt_FinalStep}
  \frac{\mathrm{d}\mathbb{V}}{\mathrm{d}t} < 0 
\end{align}
From the Lyapunov stability of continuous systems 
\citep{Dym_stability,Hale_Kocak}, one can conclude 
that $\boldsymbol{\chi} = \boldsymbol{\chi}_{\mathrm{eq}}$ 
is asymptotically stable.

%% file: Sections_Model/S6_Model_Canonical.tex
%***************************************;
%                                       ;
%  NAME                                 ;
%    S6_Model_Canonical.tex             ;
%                                       ;
%  WRITTEN BY                           ;
%    Can Xu                             ;
%    Maruti Kumar Mudunuru              ;
%    Kalyana Babu Nakshatrala           ;
%                                       ;
%***************************************;
\section{SEMI-ANALYTICAL SOLUTIONS TO CANONICAL PROBLEMS}
\label{Sec:S6_Model_Canonical}
In this section, we shall appeal to semi-inverse methods to obtain
solutions to some popular canonical boundary value problems \citep{Ogden}.
Incompressible neo--Hookean chemo-thermo-mechano
degradation model is considered here. Similar analysis can be performed 
for other compressible and incompressible chemo-mechano, thermo-mechano, 
and chemo-thermo-mechano degradation models. Coordinate system under
consideration is either spherical or cylindrical. In all the problems 
discussed below, we assume concentration and temperature to be 
functions of time $t$ and radius $r$ (which is a current configuration 
variable). This assumption is often made because the underlying problem 
has either cylindrical or spherical symmetry. We also assume that the
volumetric sources corresponding to temperature and concentration are 
equal to zero. In this paper, as we are mainly interested in degradation 
of solid due to temperature and transport of chemical species, we shall 
neglect Dufour effect, Soret effect, thermo-chemo coupling parameter 
$d_{\vartheta c}$, and anisotropic coefficient of thermal and chemical 
expansions. In order to reduce the complexity of finding solutions based 
on semi-inverse method for deformation sub-problem, we shall neglect the 
inertial effects and body forces.

Based on the assumptions provided here, the governing equations for the 
transport sub-problem in cylindrical coordinates reduce to:
%----------------------------------------------------------------------------;
%  Equation: Reduced GE for diffusion sub-problem (cylindrical coordinates)  ;
%----------------------------------------------------------------------------;
\begin{align}
  \label{Eqn:ReducedGE_DiffSubProb}
  \rho \frac{\partial c}{\partial t} + \frac{1}{r}
  \frac{\partial r h_r}{\partial r} = 0, \; 
  c(r = r_i,t) = c_i, \; 
  c(r = r_o,t) = c_o, \; 
  c(r,t=0) &= c_0
\end{align}
where $h_r$ is the mass transfer flux in the radial 
direction. Similarly, the governing equations for 
the thermal sub-problem in cylindrical coordinates 
can be written as:
%--------------------------------------------------------------------------;
%  Equation: Reduced GE for thermal sub-problem (cylindrical coordinates)  ;
%--------------------------------------------------------------------------;
\begin{align}
  \label{Eqn:ReducedGE_ThermalSubProb}
    \rho \vartheta \frac{\partial \eta}{\partial 
    t} + \frac{1}{r} \frac{\partial r q_r}{\partial r} 
    = - \frac{\partial \varkappa}{\partial r} h_{\mathrm{r}}, \;
    \vartheta(r = r_i,t) = \vartheta_i, \;
    \vartheta(r = r_o,t) = \vartheta_o, \;
    \vartheta(r,t=0) = \vartheta_0
  \end{align}
where $q_r$ is the heat flux in the radial direction.
In spherical coordinates, the governing equations 
for the transport sub-problem are:
%--------------------------------------------------------------------------;
%  Equation: Reduced GE for diffusion sub-problem (spherical coordinates)  ;
%--------------------------------------------------------------------------;
\begin{align}
  \label{Eqn:ReducedGE_DiffSubProb_Sphr}
  \rho \frac{\partial c}{\partial t} + \frac{1}{r^2}
  \frac{\partial r^2 h_r}{\partial r} 
  = 0, \; 
  c(r = r_i,t) = c_i, \; 
  c(r = r_o,t) = c_o, \;
  c(r,t=0) = c_0
\end{align}
The governing equations for the thermal sub-problem 
in spherical coordinates are:
%------------------------------------------------------------------------;
%  Equation: Reduced GE for thermal sub-problem (spherical coordinates)  ;
%------------------------------------------------------------------------;
\begin{align}
  \label{Eqn:ReducedGE_ThermalSubProb_Sphr}
  \rho \vartheta \frac{\partial \eta}{\partial 
    t} + \frac{1}{r^2} \frac{\partial r^2 q_r}{\partial r} 
  = - \frac{\partial \varkappa}{\partial r} h_{\mathrm{r}}, \;
  \vartheta(r = r_i,t) = \vartheta_i, \;
  \vartheta(r = r_o,t) = \vartheta_o, \;
  \vartheta(r,t=0) = \vartheta_0
\end{align}

Another quantity of interest in material degradation is the extent 
of damage at a particular location or along the cross-section of 
the degrading body. In case of incompressible neo-Hookean 
chemo-thermo-mechano degradation model, this quantity can 
be defined as follows:
%------------------------------;
%  Equation: Extent of damage  ;
%------------------------------;
\begin{align}
  \label{Eqn:Extent_Of_Damage}
  \mathcal{D}_{\mu}(\mathbf{x},t) := \frac{\mu}{
  \mu_0} = 1 - \left(\frac{\mu_1 c}{\mu_0 c_{
  \mathrm{ref}}} \right) - \left(\frac{\mu_2 
  \vartheta}{\mu_0 \vartheta_{\mathrm{ref}}}
  \right)
\end{align}
For virgin material, $\mathcal{D}_{\mu} = 1$. If $\mathcal{D}_{\mu}$ 
approaches zero, then the material has degraded the most. In addition, 
equation \eqref{Eqn:Extent_Of_Damage} also provides the following 
information:
%-----------------------------------------------------;
%  Itemize: Information provided by extent of damage  ;
%-----------------------------------------------------;
\begin{itemize}
  \item [$\blacktriangleright$] Amount of degradation at a given location and time, 
  \item [$\blacktriangleright$] The parts of the body that suffered extensive damage, and
  \item [$\blacktriangleright$] The effect of temperature and moisture (or concentration of 
    chemical species) on the mechanical properties of materials.
\end{itemize}

%===========================;
%  Include all subsections  ;
%===========================;
\input{Sections_Model/S6_Model_Shell_Inflation}

\input{Sections_Model/S6_Model_Beam_Bending}

\input{Sections_Model/S6_Model_Cylinder_Torsion}

%% file: Sections_Model/S6_Model_Shell_Inflation.tex
%***************************************;
%                                       ;
%  NAME                                 ;
%    S6_Model_Shell_Inflation.tex       ;
%                                       ;
%  WRITTEN BY                           ;
%    Can Xu                             ;
%    Maruti Kumar Mudunuru              ;
%    Kalyana Babu Nakshatrala           ;
%                                       ;
%***************************************;
\subsection{Inflation of a degrading spherical shell}
\label{Sec:Spherical_Shell}
We now study the behavior of a degrading (thick) 
spherical shell subjected to pressure loading. 
Figure \ref{Fig:Spherical_Shell_Pictorial} provides 
a pictorial description of the boundary value problem. 
In addition to the obvious theoretical significance, this 
problem has relevance to safety, reliability and defect 
monitoring of degrading spherical structures (such as 
a tank shell and a bearing structure) due to pressure 
loading.

Due to the spherical symmetric associated with the 
problem, spherical coordinates are used to analyze 
the inflation of degrading spherical shell. Consider 
a spherical body of inner radius $R_i$ and outer 
radius $R_o$ defined in the reference configuration 
as follows:
%----------------------------------------------------------;
%  Equation: Spherical shell in a reference configuration  ;
%----------------------------------------------------------;
\begin{align}
  \label{Eqn:RefConfig_Spherical_Shell}
  R_i \leq R \leq R_o, \quad 
  0 \leq \Theta \leq \pi, \quad 
  0 \leq \Phi \leq 2\pi 
\end{align}
where $(R,\Theta,\Phi)$ are the spherical polar 
coordinates in the reference configuration. The 
inner and outer surfaces $R = R_i$ and $R = R_o$ 
are, respectively, subjected to pressures $p_i$ 
and $p_o$ with $p_i \geq p_o$. That is, the thick 
cylinder is inflated with pressure. The deformation 
in the current configuration can be described as 
follows:
%--------------------------------------------------------;
%  Equation: Spherical shell in a current configuration  ;
%--------------------------------------------------------;
\begin{align}
  \label{Eqn:CurrConfig_Inflation_Shell}
   r_i \leq r = m(R) \leq r_o, \quad \theta = \Theta, \quad 
   \phi = \Phi 
\end{align}
where $(r,\theta,\phi)$ are the spherical polar 
coordinates in the current configuration, and 
$r_i$ and $r_o$ are, respectively, the inner 
and outer radii of the shell in the current 
(deformed) configuration. 
The deformation gradient, the left Cauchy-Green 
tensor, and the right Cauchy-Green tensor have 
the following matrix representations:
%---------------------------------------------------------------;
%  Equation: Deformation gradient for degrading spherical shell ;
%---------------------------------------------------------------;
\begin{align}
  \label{Eqn:DefGrad_Spherical_Shell}
  \left\{\mathbf{F}\right\} = \left(\begin{array}{ccc}
    \frac{dm}{dR} & 0 & 0 \\
    0 & \frac{m}{R} & 0 \\
    0 & 0 & \frac{m}{R} \\
  \end{array} \right), \quad
    \left\{\mathbf{C}\right\} = \left\{\mathbf{B}\right\} 
      = \left(\begin{array}{ccc}
    \left(\frac{dm}{dR} \right)^2 & 0 & 0 \\
    0 & \frac{m^2}{R^2} & 0 \\
    0 & 0 & \frac{m^2}{R^2} \\
  \end{array} \right)
\end{align}
Incompressibility implies that 
%-------------------------------;
%  Equation: Incompressibility  ;
%-------------------------------;
\begin{align}
  r = m(R) = \sqrt[3]{R^3 + r^3_i - R^3_i} 
  \qquad r_i \leq r \leq r_o
\end{align}
where $r_o = \sqrt[3]{R_o^3 + r^3_i - R^3_i}$. 
The non-zero components of the Cauchy stress are:
%------------------------------------------------------------------;
%  Equation: Balance of linear momentum (spherical shell problem)  ;
%------------------------------------------------------------------;
\begin{align}
  \label{Eqn:Stress_SphericalShell}
  T_{rr} = -p + \mu(c,\vartheta)\left( \frac{\mathrm{d}m}{\mathrm{d}R}\right)^2
  = -p + \mu(c,\vartheta) \frac{R^4}{r^4}, \quad 
  T_{\theta \theta} = T_{\phi\phi} =-p + \mu(c,\vartheta) 
  \frac{r^2}{R^2}
\end{align}
The governing equations for the balance of linear 
momentum in the spherical polar coordinates (e.g., 
see \citep{sadd_elasticity}) reduce to: 
%------------------------------------------------------------------;
%  Equation: Balance of linear momentum (spherical shell problem)  ;
%------------------------------------------------------------------;
\begin{align}
  \label{Eqn:BoLM_ReducedSphr_SphericalShell1}
  \frac{\partial T_{rr}}{\partial r} + \frac{2T_{rr} 
  - T_{\theta \theta}- T_{\phi \phi}}{r} = 0, \; 
  \frac{\partial p}{\partial \theta} = 0, \; 
  \frac{\partial p}{\partial \phi} = 0
\end{align}
The above equations imply that $p$ is independent 
of $\theta$ and $\phi$. That is, 
\begin{align}
  p = p(r,t)
\end{align}
From equation \eqref{Eqn:SpecificEntropy_DeformedConfig} and 
\eqref{Eqn:ChemicalPotential_DeformedConfig}, the specific chemical potential 
and specific entropy for the degrading spherical shell are given as follows:
%--------------------------------------------------------------;
%  Equations: Chemical potential and specific entropy (shell)  ;
%--------------------------------------------------------------;
\begin{subequations}
  \begin{align}
    \label{Eqn:Chemical_Potential_SphShell}
    \varkappa &= \frac{1}{\rho_0} \frac{\partial 
    \psi}{\partial c} + R_s \vartheta_{\mathrm{ref}} 
    \{c - c_{\mathrm{ref}}\} = -\frac{\mu_1}{2 \rho_0 
    c_{\mathrm{ref}}} \left(\frac{R^4}{r^4} + 
    2\frac{r^2}{R^2} - 3\right) + R_s \vartheta_
    {\mathrm{ref}} \{c - c_{\mathrm{ref}}\} \\
    \label{Eqn:Specific_Entropy_SphShell}
    \eta &= -\frac{1}{\rho_0} \frac{\partial \psi}{\partial 
    \vartheta} + \frac{c_p}{\vartheta_{\mathrm{ref}}} \{\vartheta
    - \vartheta_{\mathrm{ref}}\} = \frac{\mu_1}{2 \rho_0 
    \vartheta_{\mathrm{ref}}} \left(\frac{R^4}{r^4} + 
    2\frac{r^2}{R^2} - 3 \right) + \frac{c_p}{ \vartheta_
    {\mathrm{ref}}} \{\vartheta - \vartheta_{\mathrm{ref}}\} 
    \end{align}
    \end{subequations}
Before deriving the governing equations for the degrading shell 
problem, we shall do the non-dimensionalization by choosing primary 
variables and associated reference quantities that are convenient for 
studying this problem. To distinguish, we shall denote all the 
non-dimensional quantities using a superposed bar. 
We shall take $\mu_0$, $R_o$, $\vartheta_{\mathrm{ref}}$, 
$c_{\mathrm{ref}}$, and $D_0$ as the reference quantities, which 
give rise to the following non-dimensional quantities:
%----------------------------------------;
%  Equation: Non-dimensional quantities  ;
%----------------------------------------;
\begin{align}
  \label{Eqn:NonDimQuantities1_DegradShell}
  &\overline{r} = \frac{r}{R_o}, \ \overline{R} = \frac{R}{R_o}, 
  \ \overline{D}
  _{\varkappa \varkappa} = \frac{D_{\varkappa \varkappa}
  }{D_0},\ \overline{D}_{\vartheta \vartheta} 
  = \frac{D_{\vartheta \vartheta}}{D_0} \\
  \label{Eqn:NonDimQuantities2_DegradShell}
 & \overline{\mu}_1 = \frac{\mu_1}{\mu_0}, 
  \ \overline{\mu}_2 =  \frac{\mu_2}{\mu_0}, 
  \ \overline{c} = \frac{c}{c_{\mathrm{ref}}}, 
  \ \overline{\vartheta} = \frac{\vartheta}{\vartheta_{\mathrm{ref}}},
 \ \overline{t} = \frac{D_0 t}{R_o^2}  
\end{align}
With the stress field in equation \eqref{Eqn:Stress_SphericalShell}, 
we shall integrate equation \eqref{Eqn:BoLM_ReducedSphr_SphericalShell1} 
and then have the following non-linear equation in deformation sub-problem 
after non-dimensionalization:
%---------------------------------------------------;
%  Equation: Non-linear equation for determing r_i  ;
%---------------------------------------------------;
\begin{align}
  \label{Eqn:NonLin_SphrShell_Gamma_Nondim}
   \overline{T}_{rr}(\overline{R} = \overline{R}_i, \overline{t}) 
   - \overline{T}_{rr}(\overline{R} = \overline{R}_o, \overline{t}) 
   = \overline{p}_o-\overline{p}_i = \displaystyle \int 
  \limits_{\overline{R}_i}^{\overline{R}_o} \frac{2 
  \overline{\mu}(\overline{c},
  \overline{\vartheta}) \left(
  \overline{R}^6 - \left(\overline{R}^3 + \overline{r}_i^3
  - \overline{R}_i^3 \right)^2 \right)}{ \left(\overline{R}^3 
  + \overline{r}_i^3-\overline{R}_i^3 \right)^{\frac{7}{3}}} 
  \mathrm{d}\overline{R}
\end{align}
In order to reduce the complexity in finding semi-analytical solutions, 
we shall assume $\frac{\partial r}{\partial t} \ll 
\frac{\partial \vartheta}{\partial t}$.  
Substituting equation \eqref{Eqn:Chemical_Potential_SphShell} and 
\eqref{Eqn:Specific_Entropy_SphShell} into the constitutive relations of the 
proposed model, the governing equations 
of these two sub-problems \eqref{Eqn:ReducedGE_DiffSubProb_Sphr}, 
\eqref{Eqn:ReducedGE_ThermalSubProb_Sphr} can be written  
as follows after non-dimensionalization:
%------------------------------------------------;
%  Equation: Non-dim thermal and transport PDEs  ;
%------------------------------------------------;
\begin{align}
  \label{Eqn:NonDimFinalForm_Transport_DegradShell}
 & \frac{\partial \overline{c}}{\partial \overline{t}} - \left(\frac{ 
  2\overline{D}_{\varkappa \varkappa}}{\overline{r}} + \frac{\partial 
  \overline{D}_{\varkappa \varkappa}}{\partial \overline{r}} \right) 
  \frac{\partial \overline{c}}{\partial \overline{r}} - \overline{D}_
  {\varkappa \varkappa} \frac{\partial^2 \overline{c}}{\partial 
  \overline{r}^2} = 2\overline{\omega} \frac{\partial \overline{D}_
  {\varkappa \varkappa}}{\partial \overline{r}} \left(\frac{\overline{R}
  ^4}{\overline{r}^5} - \frac{\overline{r}}{\overline{R}^2} 
  \right) 
  - 6\overline{\omega} \overline{D}_{\varkappa \varkappa} \left(\frac{1}{
  \overline{R}^2} + \frac{\overline{R}^4}{\overline{r}^6} \right)\\
  \label{Eqn:NonDimFinalForm_Thermal_DegradShell}
     & \overline{\vartheta} \frac{\partial \overline{\vartheta}}{\partial 
        \overline{t}} - \left(2\frac{\overline{D}_{\vartheta \vartheta}}{
        \overline{r}} + \frac{\partial \overline{D}_{\vartheta \vartheta}}{
        \partial \overline{r}} \right) \frac{\partial \overline{\vartheta}}{
        \partial \overline{r}} - \overline{D}_{\vartheta\vartheta} \frac{\partial
        ^2 \overline{\vartheta}}{\partial \overline{r}^2} 
        = \overline{\tau} \overline{D}_{\varkappa \varkappa}
         \left(\frac{\partial \overline{c}}{\partial 
        \overline{r}}-2\overline{\omega}\left(\frac{\overline{r}}{
                 \overline{R}^2} - \frac{\overline{R}^4}{\overline{r}^5} \right) \right)^2
\end{align}
where $\overline{\omega}$ and $\overline{\tau}$ are two non-dimensional 
parameters, which have the following expressions:
%$\omega = \frac{\mu_1}{R_s \vartheta_{\mathrm{ref}} 
%c_{\mathrm{ref}}}$. In deriving \eqref{Eqn:FinalForm_Thermal_DegradShell},
%we have assumed that $\frac{\partial r}{\partial t} \ll 
%\frac{\partial \vartheta}{\partial t}$ (in order to reduce 
%the complexity in finding semi-analytical solutions). 
%%%
%The above non-dimensionalization introduces the following 
%non-dimensional parameters:
%----------------------------------------;
%  Equation: Non-dimensional parameters  ;
%----------------------------------------;
\begin{align}
  \label{Eqn:NonDimParam1_DegradShell}
  \overline{\omega} = \frac{\mu_1}{\rho_0 R_s \vartheta_{\mathrm{ref}} c^2_{\mathrm{ref}}}, 
 \quad \overline{\tau} = \frac{R_s c^2_{\mathrm{ref}}}{c_p}
\end{align}
The non-linear equation \eqref{Eqn:NonLin_SphrShell_Gamma_Nondim} 
enables us to find $\overline{r}_i$ at various $\overline{t}$ for given 
$\overline{c}(\overline{R},\overline{t})$ and $\overline{\vartheta}(\overline{R},\overline{t})$. 
However, it should be noted that $\overline{c}(\overline{R},
\overline{t})$ and $\overline{\vartheta}(\overline{R},\overline{t})$ are also a 
function of $\overline{r}_i$ in case of strong coupling. This 
is because diffusivity and thermal conductivity depend on the 
invariants of strain $\overline{\mathbf{E}}$. Hence, the integral 
equation \eqref{Eqn:NonLin_SphrShell_Gamma_Nondim} and partial 
differential equations \eqref{Eqn:NonDimFinalForm_Transport_DegradShell} 
and \eqref{Eqn:NonDimFinalForm_Thermal_DegradShell} are strongly 
coupled. 
%===========================================================================;
%  Subsection: Steady-state and quasi-static analysis for shell degradation  ;
%===========================================================================;
\subsubsection{Steady-state analysis for shell degradation}
\label{Subsec:SteadyState_DegradShell}
For steady-state, we have $h_r r^2 = C_1$ and $q_r r^2 + \varkappa 
h_r r^2 = C_2$, where $C_1$ and $C_2$ are integration constants. 
This implies that $\overline{c}$ and $\overline{\vartheta}$ are 
the solutions of the following ODEs:
%-----------------------------------------------------;
%  Equation: Steady-state thermal and transport ODEs  ;
%-----------------------------------------------------;
\begin{subequations}
  \begin{align}
    \label{Eqn:SteadyState_Thermal_DegradShell}
    &\overline{D}_{\varkappa \varkappa} \overline{r}^2 
    \frac{d \overline{c}}{d \overline{r}} - 2 \overline{D}
    _{\varkappa \varkappa} \overline{\omega} \left( \frac{
    \overline{r}^3}{\overline{R}^2} - \frac{\overline{R}^
    4}{\overline{r}^3}\right) + \overline{C}_1 = 0\\
    \label{Eqn:SteadyState_Transport_DegradShell}    
    &\overline{D}_{\vartheta \vartheta}\overline{r}^2 
    \frac{d \overline{\vartheta}}{d \overline{r}} + 
    \overline{\tau} \left(\frac{\overline{w}}{2} \left( 
    \frac{\overline{R}^4}{\overline{r}^4} + 2 \frac{
    \overline{r}^2}{\overline{R}^2} - 3 \right) - 
    \overline{c} + 1\right) \overline{C}_1 + \overline{C}
    _2 = 0
  \end{align}
\end{subequations}
where the integration constants $\overline{C}_1$ and 
$\overline{C}_2$ are determined from the boundary 
conditions for the transport and thermal sub-problems. 
Under weak coupling (i.e., $D_{\vartheta \vartheta}$ and $D_{\varkappa 
\varkappa}$ are constants), a simplified form of the analytical solutions
for $\overline{c}$ and $\overline{\vartheta}$ can be obtained as 
follows:
%---------------------------------------------------------------------------;
%  Equation: Analytical solutions for Conc and Temperature (weak coupling)  ;
%---------------------------------------------------------------------------;
\begin{align}
  \label{Eqn:AnalyticSoln_Thermal_DegradShell_WC}
  \overline{c} = \overline{\omega} \left(\frac{
    \overline{r}^2}{\overline{R}^2} + \frac{\overline{
      R}^4}{2\overline{r}^4}\right) + \frac{B_1}{\overline{
      r}} + A_1, \quad 
  \overline{\vartheta} =- \frac{\overline{\tau} B_1^2
    \overline{D}_{\varkappa \varkappa}}{2 
    \overline{D}_{\vartheta \vartheta}\overline{r}^2} 
  +\frac{Z_1}{\overline{r}}
  + Y_1
\end{align}
where $A_1$, $B_1$, $Y_1$, and $Z_1$ are constants, 
which are given in terms of the boundary conditions 
$\overline{c}_i$, $\overline{c}_o$, $\overline{
\vartheta}_i$, and $\overline{\vartheta}_o$ as follows:
%---------------------------------------------------;
%  Equation: Integration constants (weak coupling)  ;
%---------------------------------------------------;
\begin{subequations}
  \begin{align}
    &A_1 = \overline{c}_i - \frac{B_1}{\overline{r}_i} - 
     \overline{\omega} \left(\frac{
    \overline{r}_i^2}{\overline{R}^2} + \frac{8\overline{R}^4}{
    \overline{r}_i^4}\right) \\
    &B_1 = \frac{\overline{r}_i \overline{r}_o}{\overline{r}_i - \overline{r}_o}
    \left(\overline{c}_o-\overline{c}_i-\overline{
    \omega}\left(\frac{
    \overline{r}_o^2}{\overline{R}^2} + \frac{8\overline{R}^4}{
    \overline{r}_o^4}-\frac{
    \overline{r}_i^2}{\overline{R}^2} - \frac{8\overline{R}^4}{
    \overline{r}_i^4}
    \right)\right)\\
    &Y_1 = \overline{\vartheta}_i + \frac{\overline{\tau} B_1^2
    \overline{D}_{\varkappa \varkappa}}{2 \overline{D}
    _{\vartheta \vartheta}\overline{r}_i^2}
    - \frac{Z_1}{\overline{r}_i} \\
    &Z_1 = \frac{\overline{r}_i-\overline{r}_o}{\overline{r}_i\overline{r}_o}
    \left(\overline{\vartheta}_o - \overline{\vartheta}_i 
    - \frac{\overline{\tau} B_1^2 \overline{D}_{\varkappa \varkappa}}{2 
    \overline{D}_{\vartheta \vartheta}} \left(\frac{1}{\overline{r}_i^2}
   -\frac{1}{\overline{r}_o^2} \right) \right)
  \end{align}
\end{subequations}

\subsubsection{Unsteady analysis for shell degradation}
\label{Subsec:Quasistatic_Degradshell}
Herein, we shall use the method of horizontal lines 
\citep{rothe1930zweidimensionale,picardleis1980} 
and shooting method \citep{Heath_Numerical} to 
obtain numerical solutions to equations \eqref{Eqn:NonDimFinalForm_Transport_DegradShell} 
and \eqref{Eqn:NonDimFinalForm_Thermal_DegradShell}. In the method 
of horizontal lines, the time is discretized first followed by spatial 
discretization. The time interval of interest $[0,\overline{\mathcal{I}}]$ 
is divided into $N$ non-overlapping subintervals such that $\Delta 
\overline{t} = \frac{\overline{\mathcal{I}}}{N}$ and $\overline{t}_n 
= n \Delta \overline{t}$. $\overline{t}_n$ is called the integral 
time level, where $n = 0, \cdots, N$. $\Delta \overline{t}$ 
is the time-step, which is assumed to be uniform. Employing the 
method of horizontal lines with backward Euler time-stepping scheme, 
we obtain the following ODEs at each time-level for equations 
\eqref{Eqn:NonDimFinalForm_Transport_DegradShell} and \eqref{Eqn:NonDimFinalForm_Thermal_DegradShell}:
%------------------------------------------------;
%  Equation: Non-dim thermal and transport PDEs  ;
%------------------------------------------------;
\begin{align}
  \label{Eqn:NonDimFinalForm_Transport_DegradShell_NumSoln}
  \frac{d^2 \overline{c}^{(n+1)}}{d \overline{r}^2} &+ \left(\frac{ 
  2}{\overline{r}^{(n)}} + \left( \frac{1}{\overline{D}^{(n)}_{\varkappa 
  \varkappa}} \right) \frac{d \overline{D}_{\varkappa \varkappa}}{d 
  \overline{r}} \Bigg|_{\overline{t} = \overline{t}_n} \right) \frac{d \overline{c}^{(n+1)}}{d 
  \overline{r}} - \frac{\overline{c}^{(n+1)}}{\overline{D}^{(n)}_{\varkappa 
  \varkappa} \Delta \overline{t}} = 6\overline{\omega} \left(\frac{1}{\left(\overline{R}^{(n)}
  \right)^2} + \frac{\left(\overline{R}^{(n)}\right)^4}{
  \left(\overline{r}^{(n)}\right)^6} \right) \nonumber \\
  &- \frac{\overline{c}^{(n)}}{\overline{D}^{(n)}_{\varkappa \varkappa} 
  \Delta \overline{t}} - \left(\frac{2\overline{\omega}}{\overline{D}^{(n)}_{\varkappa 
  \varkappa}} \right) \left( \frac{d \overline{D}_{\varkappa \varkappa}}{d 
  \overline{r}} \Bigg|_{\overline{t} = \overline{t}_n} \right) \left(\frac{\left(\overline{R}
  ^{(n)} \right)^4}{\left(\overline{r}^{(n)} \right)^5} - \frac{\overline{r}
  ^{(n)}}{\left(\overline{R}^{(n)} \right)^2} \right)
\end{align}
%------------------------------------------------;
%  Equation: Non-dim thermal and transport PDEs  ;
%------------------------------------------------;
\begin{align}
  \label{Eqn:NonDimFinalForm_Thermal_DegradShell_NumSoln}
  \frac{d^2 \overline{\vartheta}^{(n+1)}}{d \overline{r}^2} &+ \left(\frac{ 
  2}{\overline{r}^{(n)}} + \left( \frac{1}{\overline{D}^{(n)}_{\vartheta 
  \vartheta}} \right) \frac{d \overline{D}_{\vartheta \vartheta}}{d 
  \overline{r}} \Bigg|_{\overline{t} = \overline{t}_n} \right) \frac{d \overline{\vartheta}^{(n+1)}}{d 
  \overline{r}} - \frac{\overline{\vartheta}^{(n)} \overline{\vartheta}^{(n+1)}}{
  \overline{D}^{(n)}_{\vartheta \vartheta} \Delta \overline{t}} = 
  - \frac{\left(\overline{\vartheta}^{(n)} \right)^2}{\overline{D}^{(n)}_
  {\vartheta \vartheta} \Delta \overline{t}} \nonumber \\
  &- \frac{\overline{\tau} \overline{D}^{(n)}_{\varkappa \varkappa}}
  {\overline{D}^{(n)}_{\vartheta \vartheta}} 
  \left(\frac{d \overline{c}}{d \overline{r}} \Bigg|_{\overline{t} = \overline{t}_n} 
  -2\overline{\omega} \left(\frac{\left(\overline{r}^{(n)} \right)
  }{\left(\overline{R}^{(n)}\right)^2} - \frac{\left(\overline{R}^{(n)}
  \right)^4}{\left(\overline{r}^{(n)} \right)^5} \right)
  \right)^2 
\end{align}
where $\overline{c}^{(n)} = \overline{c}(\overline{r},\overline{t} = \overline{t}_n)$
and $\overline{\vartheta}^{(n)} = \overline{\vartheta}(\overline{r}, \overline{t} = 
\overline{t}_n)$. Algorithm \ref{Algo:DegradShell_NumSolnMethod} 
describes a procedure to determine $\overline{c}(\overline{r},\overline{t})$, 
$\overline{\vartheta}(\overline{r},\overline{t})$, and $r_i$ at
various times using an iterative non-linear numerical solution 
strategy. 
The following values are assumed for the non-dimensional 
parameters in the strong coupling simulations:
%---------------------------------------------------------;
%  Equation: Parameters used in the numerical simulation  ;
%---------------------------------------------------------;
\begin{align}
  \label{Eqn:Shell_Degradation_Params}
 &\overline{R}_o = 1, \ \overline{R}_i = 0.5, \ \Delta \overline{t} = 0.01, \ \overline{t} = 2, \ 
\overline{\omega} = 0.05, \ \overline{\tau} = 0.2,
  \ \overline{c}_i = 0, \overline{\vartheta}_0= 0.5 \nonumber \\
  &\overline{c}_o = 1, \ \overline{\vartheta}_i = 0.5, \ \overline{\vartheta}
  _o = 1, \ \overline{\mu}_0 = 1, \ \overline{\mu}_1 = 0.3, \ \overline{\mu}_2 = 0.4,
  \ \overline{D}_0 = 1, \ \overline{D}_T = 1.5, \nonumber \\
  &\overline{D}_S = 1.2, \ \eta_T = \eta_S = 1, \ E_{\mathrm{ref}T} 
  = E_{\mathrm{ref}S} = 1, \ \overline{K}_0 = 1,\ \delta = 10 
\end{align}
In weakly coupling problem, we use $\overline{D}_0$, $\overline{K}_0$ 
as $\overline{D}^{(n)}_{\varkappa \varkappa}$ and 
$\overline{D}^{(n)}_{\vartheta \vartheta}$, respectively.
It should be noted that these values 
are constructed based on the (brittle-type) material parameters 
such as glass, ceramics, and concrete.

The numerical results are shown in figures 
\ref{Fig:Spherical_Shell_T_theta_theta_given_t}--\ref{Fig:Spherical_Shell_ExtOfDamage_given_t}, which reveal the following conclusions 
on the overall behavior of degrading spherical shells 
under inflation:
%------------------------------------------------------------------;
%  Summary of conclusions: Inflation of degrading spherical shell  ;
%------------------------------------------------------------------;
\begin{enumerate}[(i)]
  \item \textsf{Degradation vs. non-degradation:} After degradation, 
  a spherical shell which is initially homogeneous is not homogeneous anymore.
  \item Due to degradation, creep-like behavior is observed. 
  Therefore, as time progresses, hoop stresses increase. 
  We need to note that the shell ceases to creep 
  after a certain period of time, which is the moment when the 
  transport of chemical species and heat conduction are close to 
  steady-states.
  \item As the pressure loading increases, the hoop stress is 
  increasing in a non-linear fashion, which is 
  significantly different from the non-degradation shell.
\item For non-degrading shell, the chemical potential is 
  unchanged with respect to pressure loading. However, 
  for strong coupling, it increases with $\overline{p}_i$ 
  in a non-linear fashion when $\overline{\omega}$ is small 
  enough. This is because for small $\overline{\omega}$, 
  diffusion takes the dominance in the coupling effect. 
  When pressure loading increases, the diffusivity 
  is increasing due to the growing strain. For large 
  $\overline{\omega}$, the deformation is dominant in the 
  coupling, which is $-\overline{I}_E$ term in chemical potential.
  Since the first invariant $\overline{I}_E$ is always positive in 
  this problem, chemical potential is decreasing when the 
  pressure loading increases.
  \item  \textsf{Thermo-dominated vs. chemo-dominated degradation:} 
  Weak coupling over-predicts the amount of degradation compared 
  to the full (or strong) coupling when thermal degradation dominates. 
  This is because when the thermal degradation dominants, the thermal 
  conductivity decreases due to the increase in strain (note that the 
  first invariant of strain is always positive in this problem). However, in 
  chemo-dominated degradation, weak coupling under-predicts the 
  amount of degradation compared to the strong coupling case.
  \item In case of strong coupling, healing-like behavior is observed at 
  early time steps in thermo-dominated degradation (but still remains below 
  that of the virgin material). This is because variable heat sinks exist 
  in the entire body (due to which temperature gets lower than the initial 
  condition). Hence, the material damage is less than that of at $\overline{t} 
  = 0$. However, this heal-like behavior becomes less distinct (or even doesn't 
  exist) when the chemo-degradation achieves the dominance.    
  \item \textsf{Strong vs. weak coupling:} Quantitatively and qualitatively, 
  extent of damage for both strong and weak coupling are considerably 
  different. 
\end{enumerate}

%-------------------------------------------------------------;
%  Algorithm: Degrading shell numerical solution methodology  ;
%-------------------------------------------------------------;
\begin{algorithm} 
  \caption{Inflation of a degrading spherical shell (numerical methodology 
    to find $\overline{r}_i$, $\overline{c}$, and $\overline{\vartheta}$)}
  \label{Algo:DegradShell_NumSolnMethod}
  \begin{algorithmic}[1]
    %----------------------;
    %  STEP-0: Input data  ;
    %----------------------; 
    \STATE INPUT:~Non-dimensional material parameters, non-dimensional 
      boundary conditions, and non-dimensional initial conditions, 
      \texttt{MaxIters}, tolerances $\epsilon_{\mathrm{tol}}^{(r)}$, 
      $\epsilon_{\mathrm{tol}}^{(c)}$, and $\epsilon_{\mathrm{tol}}^
      {(\vartheta)}$.
    %%
    %------------------------;
    %  STEP-1: Guess \gamma  ;
    %------------------------;
    \STATE Evaluate $\overline{r}_i$ at $\overline{t} = 0$ based on equation 
     \eqref{Eqn:NonLin_SphrShell_Gamma_Nondim}. 
    %%
    %-----------------------------------;
    %  STEP-2: Iterative strategy loop  ;
    %-----------------------------------;
    \FOR{$n = 1, 2, \cdots, N$}
    \FOR{$j = 1, 2, \cdots$}
    %%
    %----------------------------------;
    %  STEP-2a: Stopping criteria # 1  ;
    %----------------------------------; 
    \IF {$j >$ \texttt{MaxIters}}
      \STATE Solution did not converge in specified maximum number of 
        iterations. EXIT.
    \ENDIF
    %%
    %----------------------------------;
    %  STEP-2b: Diffusion sub-problem  ;
    %----------------------------------; 
    \STATE  \texttt{\underline{Diffusion sub-problem}}:~Given $\overline{r}^{(j)}_i$, 
      solve equation \eqref{Eqn:NonDimFinalForm_Transport_DegradShell_NumSoln} 
      to obtain $\overline{c}^{(j+1)}$. Herein, we use shooting method to 
      solve the ODEs.
    %%
    %----------------------------------------;
    %  STEP-2c: Heat conduction sub-problem  ;
    %----------------------------------------; 
    \STATE \texttt{\underline{Heat conduction sub-problem}}:~Given 
      $\overline{r}^{(j)}_i$ and $\overline{c}^{(j+1)}$, solve equation 
      \eqref{Eqn:NonDimFinalForm_Thermal_DegradShell_NumSoln} to obtain 
      $\overline{\vartheta}^{(j+1)}$. Similarly, we use shooting method 
      to solve the non-linear ODEs.
    %%
    %------------------------------------;
    %  STEP-2d: Deformation sub-problem  ;
    %------------------------------------;
    \STATE  \texttt{\underline{Deformation sub-problem}}:~Given 
      $\overline{c}^{(j+1)}$ and $\overline{\vartheta}^{(j+1)}$, 
      solve for $\overline{r}^{(j+1)}_i$ given by equation 
      \eqref{Eqn:NonLin_SphrShell_Gamma_Nondim} using bisection 
      method.
    %%
    %----------------------------------;
    %  STEP-2e: Stopping criteria # 2  ;
    %----------------------------------;
    \IF {$\|\overline{r}^{(j+1)}_i - \overline{r}^{(j)}_i \| < 
      \epsilon_{\mathrm{tol}}^{(r)}$, $\|\overline{c}^{(j+1)} - 
      \overline{c}^{(j)} \| < \epsilon_{\mathrm{tol}}^{(c)}$, 
      and $\|\overline{\vartheta}^{(j+1)} - \overline{\vartheta}
      ^{(j)} \| < \epsilon_{\mathrm{tol}}^{(\vartheta)}$}
      \STATE OUTPUT:~$\overline{r}^{(j+1)}_i$, $\overline{c}^{(j+1)}$, 
      and $\overline{\vartheta}^{(j+1)}$. EXIT.
    \ELSE 
      \STATE Update the guess:~$\overline{r}^{(j)}_i \gets \overline{r}^{(j+1)}_i$.
    \ENDIF
   \ENDFOR
   \ENDFOR
  \end{algorithmic}
\end{algorithm}

%% file: Sections_Model/S6_Model_Beam_Bending.tex
%***************************************;
%                                       ;
%  NAME                                 ;
%    S6_Model_Beam_Bending.tex          ;
%                                       ;
%  WRITTEN BY                           ;
%    Can Xu                             ;
%    Maruti Kumar Mudunuru              ;
%    Kalyana Babu Nakshatrala           ;
%                                       ;
%***************************************;
\subsection{Bending of a degrading beam}
\label{Subsec:Beam_Bending}
Herein, we shall consider pure bending of a degrading beam. 
At time $t = 0$, a finite degrading beam is suddenly bent 
by an action of pure end moments. For $t > 0$, the centerline 
of the beam becomes a sector of a circle of radius $r_c$. This 
centerline is held fixed for all the time. Subsequently, the 
stresses in the degrading beam are allowed to relax. In addition, 
it is assumed that the material remains isotropic with respect 
to the reference configuration throughout the degradation process.
These assumptions enable us to employ the counterpart of universal 
deformations (also known as semi-inverse method) \citep{Ogden} 
to study such degrading beams.

A pictorial description of the initial boundary value problem 
is shown in Figure \ref{Fig:Beam_Bending_PicDescription}. The 
degrading beam is defined as follows:
%-----------------------------------------------;
%  Equation: Beam in a reference configuration  ;
%-----------------------------------------------;
\begin{align}
  \label{Eqn:RefConfig_Beam_Bending}
  -L \leq X \leq L, \quad 
  -W \leq Y \leq W, \quad 
  -H \leq Z \leq H 
\end{align}
where $(X,Y,Z)$ are the Cartesian coordinates in 
the reference configuration. We assume that the 
deformation to be as follows: 
%---------------------------------------------;
%  Equation: Beam in a current configuration  ;
%---------------------------------------------;
\begin{align}
  \label{Eqn:CurrConfig_Beam_Bending}
  r = \sqrt{\frac{2X}{\alpha} + \beta}, \quad 
  \theta = \frac{Y}{\gamma}, \quad z = Z 
\end{align}
where $(r,\theta,z)$ are the cylindrical polar coordinates in 
the current configuration. When $X = 0$, we have $\beta = r^2_c$. 
It should be noted 
that $\alpha$ and $\gamma$ are all unknown time-dependent parameters. 
These unknowns are evaluated from the incompressibility constraint, traction 
boundary conditions, and pure end moments. To reduce the complexity in finding 
semi-analytical solutions, we shall assume $r_c$ is given. The faces $X = -L$ 
and $X = L$ are subjected to ambient atmospheric pressure `$p_{\mathrm{atm}}$'. 
Upon deformation, the corresponding deformed faces $r_i$ and $r_o$ are maintained 
at $p_{\mathrm{atm}}$, where $r_{i} = \sqrt{r^2_c - 2 \gamma L}$ and $r_{o} = 
\sqrt{r^2_c + 2 \gamma L}$ are the inner and outer radius of the degrading beam. 
This gives the following traction boundary conditions:
%-----------------------------;
%  Equation: Atmospheric BCs  ;
%-----------------------------;
\begin{align}
  \label{Eqn:Traction_BCs_Beam_Bending}
  T_{rr}(X = -L, t) = T_{rr}(X = L, t) = p_{\mathrm{atm}}
\end{align}
The deformation gradient $\mathbf{F}$, right Cauchy-Green tensor 
$\mathbf{C}$, and left Cauchy-Green tensor $\mathbf{B}$ for the 
degrading beam are given as follows:
%-----------------------------------------------------;
%  Equation: Deformation gradient for degrading beam  ;
%-----------------------------------------------------;
\begin{align}
  \label{Eqn:DefGrad_BeamBending}
  \{\mathbf{F}\} = \left(\begin{array}{ccc}
    \frac{1}{\alpha r} & 0 & 0 \\
    0 & \frac{r}{\gamma} & 0 \\
    0 & 0 & 1 \\
  \end{array} \right) \qquad
  \{\mathbf{C}\} = \{\mathbf{B}\} = \left(\begin{array}{ccc}
    \frac{1}{\alpha^2 r^2} & 0 & 0 \\
    0 & \frac{r^2}{\gamma^2} & 0 \\
    0 & 0 & 1 \\
  \end{array} \right)
\end{align}
For incompressible degrading neo-Hookean material, we 
have $\alpha \gamma = 1$ and the non-zero components 
of the Cauchy stress tensor are given as follows:
%---------------------------------------------------------------;
%  Equation: Balance of linear momentum (beam bending problem)  ;
%---------------------------------------------------------------;
  \begin{align}
    \label{Eqn:Stress_BeamBending}
    T_{rr} = -p + \frac{\mu(c,\vartheta) \gamma^2}{2 \gamma X + r^2_c}, \quad
    T_{\theta \theta} = -p + \frac{\mu(c,\vartheta) \left(2 
    \gamma X + r^2_c \right)}{\gamma^2}, \quad
    T_{zz} = -p + \mu(c,\vartheta)
  \end{align}
The balance of linear momentum in the cylindrical 
polar coordinates reduces to the following: 
%---------------------------------------------------------------;
%  Equation: Balance of linear momentum (beam bending problem)  ;
%---------------------------------------------------------------;
\begin{align}
  \label{Eqn:BoLM_ReducedCylin_BeamBending}
  \frac{\partial T_{rr}}{\partial r} + \frac{T_{rr} 
  - T_{\theta \theta}}{r} = 0, \quad
  \frac{\partial p}{\partial \theta} = 0, \quad
  \frac{\partial p}{\partial z} = 0
\end{align}
The bending moment in the deformation sub-problem 
can be evaluated based on the following formula:
%---------------------------------;
%  Equation: Pure bending moment  ;
%---------------------------------;
\begin{align}
  \label{Eqn:Bending_Moment_Beam}
  M_{\mathrm{beam}}(t) &= \displaystyle \int 
  \limits_{A_{\mathrm{cross}}} T_{\theta \theta} 
  (r - r_{\mathrm{neu}}) \mathrm{d}A \nonumber \\
  &= 2 H \displaystyle \int \limits_{-L}^{L} T_{\theta 
  \theta} (- \sqrt{r^2_c + 2 \gamma X_{\mathrm{neu}}} + 
  \sqrt{r^2_c + 2 \gamma X}) \frac{\gamma}{\sqrt{r^2_c 
  + 2 \gamma X}} \mathrm{d}X
\end{align}
where $\mathrm{d}A = 2H \mathrm{d}r$, $r_{\mathrm{neu}} = 
\sqrt{r^2_c + 2 \gamma X_{\mathrm{neu}}}$ is the neutral 
axis location, and $X_{\mathrm{neu}}$ is the value at which 
$T_{\theta \theta} = 0$. 
The chemical potential, specific entropy for the degrading beam 
are given as follows:
%-------------------------------------------------------------;
%  Equations: Chemical potential and specific entropy (beam)  ;
%-------------------------------------------------------------;
\begin{subequations}
  \begin{align}
    \label{Eqn:Chemical_Potential_DegradBeam}
    \varkappa &= \frac{1}{\rho_0} \frac{\partial 
    \psi}{\partial c} + R_s \vartheta_{\mathrm{ref}} 
    \{c - c_{\mathrm{ref}}\} = -\frac{\mu_1}{2 \rho_0 
    c_{\mathrm{ref}}} \left(\frac{\gamma^2}{r^2} + 
    \frac{r^2}{\gamma^2} - 2\right) + R_s \vartheta_
    {\mathrm{ref}} \{c - c_{\mathrm{ref}}\} \\
    \label{Eqn:Specific_Entropy_DegradBeam}
    \eta &= -\frac{1}{\rho_0} \frac{\partial \psi}{\partial 
    \vartheta} + \frac{c_p}{\vartheta_{\mathrm{ref}}} \{\vartheta
    - \vartheta_{\mathrm{ref}}\} = \frac{\mu_1}{2 \rho_0 
    \vartheta_{\mathrm{ref}}} \left(\frac{\gamma^2}{r^2} + 
    \frac{r^2}{\gamma^2} - 2 \right) + \frac{c_p}{ \vartheta_
    {\mathrm{ref}}} \{\vartheta - \vartheta_{\mathrm{ref}}\}
  \end{align}
\end{subequations}
Most of the non-dimensional quantities are same as that of the 
degrading shell problem except for the following:
%----------------------------------------;
%  Equation: Non-dimensional quantities  ;
%----------------------------------------;
\begin{align}
  \label{Eqn:NonDimQuantities1_DegradBeam}
   \overline{r} = \frac{r}{r_c}, \ \overline{X} = \frac{X}{r_c}, 
  \ \overline{\gamma} = \frac{\gamma}{r_c},
   \ \overline{t} = \frac{D_0 t}{r_c^2}  
\end{align}
Using equations 
\eqref{Eqn:CurrConfig_Beam_Bending}--\eqref{Eqn:BoLM_ReducedCylin_BeamBending}, we have the following non-linear equation in $\overline{\gamma}$
%------------------------------------------------------;
%  Equation: Non-linear equation for determing \gamma  ;
%------------------------------------------------------;
\begin{align}
  \label{Eqn:NonLin_BeamBend_Gamma_Nondim_1}
  \displaystyle \int \limits_{-L/r_c}^{L/r_c} \frac{
  \overline{\mu}(\overline{c}(\overline{X},\overline{t}), 
  \overline{\vartheta}(\overline{X},\overline{t})) \left(
  \overline{\gamma}^4 - \left(2 \overline{\gamma} \overline{
  X} + 1 \right)^2 \right)}{\overline{\gamma} \left(2 \overline{
  \gamma} \overline{X} + 1 \right)^2} d \overline{X} 
  = 0
\end{align}
From \eqref{Eqn:NonLin_BeamBend_Gamma_Nondim_1}, 
$\overline{\gamma}|_{\overline{t}=0}$ is given as follows:
%-----------------------------;
%  Equation: \gamma at t = 0  ;
%-----------------------------;
\begin{align}
  \label{Eqn:Initial_Gamma_t0}
  \overline{\gamma}|_{\overline{t}=0} = \frac{1}{r_c}\sqrt{-2L^2 + \sqrt{4L^4 + r^4_c}}
\end{align}
which is the case for homogeneous neo-Hookean material. 
%Once $c$, $\vartheta$, and $\gamma$ are known, the Lagrange multiplier $p 
%= p(r,t)$ enforcing the incompressibility constraint can evaluated 
%from \eqref{Eqn:BoLM_ReducedCylin_BeamBending1} and \eqref{Eqn:Traction_BCs_Beam_Bending} 
%as follows:
%%-------------------------------------;
%%  Equation: Determing 'p' from T_rr  ;
%%-------------------------------------;
%\begin{align}
%  \label{Eqn:LagMult_BeamBending}
%  p(r,t) = p_{\mathrm{atm}} + \frac{\mu(c,\vartheta) 
%  \gamma^2}{r^2} - \displaystyle \int \limits_
%  {r_{i}}^{r} \frac{\mu(c,\vartheta) 
%  \left(\gamma^4 - r^4 \right)}{\gamma^2 r^3} dr 
%\end{align}
%%%
%where $r_{i} = \sqrt{r^2_c -2 \gamma L}$. 
As $r_c$ is given, the parameter 
$\overline{\gamma}$ is bounded above and below as follows:
%------------------------------------------------;
%  Equation: Upper and lower bound for $\gamma$  ;
%------------------------------------------------;
\begin{align}
  \label{Eqn:UppLowBound_Gamma}
  \frac{-r_c}{2L} < \overline{\gamma} < \frac{r_c}{2L}
\end{align}
which can be used in finding the solution for the non-linear equation 
given by \eqref{Eqn:NonLin_BeamBend_Gamma_Nondim_1}. 
It should be noted that $\overline{\gamma}|_{t=0}$ satisfies 
the inequality given by \eqref{Eqn:UppLowBound_Gamma}.

From equations \eqref{Eqn:ReducedGE_DiffSubProb} and 
\eqref{Eqn:ReducedGE_ThermalSubProb}, 
the final form for the governing equations for transport and thermal 
sub-problems for degrading beam are given as follows:
%------------------------------------------------;
%  Equation: Non-dim thermal and transport PDEs  ;
%------------------------------------------------;
\begin{align}
  \label{Eqn:NonDimFinalForm_Transport_DegradBeam}
 &\frac{\partial \overline{c}}{\partial \overline{t}} - \left(\frac{ 
  \overline{D}_{\varkappa \varkappa}}{\overline{r}} + \frac{\partial 
  \overline{D}_{\varkappa \varkappa}}{\partial \overline{r}} \right) 
  \frac{\partial \overline{c}}{\partial \overline{r}} - \overline{D}_
  {\varkappa \varkappa} \frac{\partial^2 \overline{c}}{\partial 
  \overline{r}^2} = \overline{\omega} \frac{\partial \overline{D}
  _{\varkappa \varkappa}}{\partial \overline{r}} \left(\frac{\overline{
  \gamma}^2}{\overline{r}^3} - \frac{\overline{r}}{\overline{\gamma}^2} 
  \right) 
  - 2\overline{\omega} \overline{D}_{\varkappa \varkappa} \left(\frac{1}{
  \overline{\gamma}^2} + \frac{\overline{\gamma}^2}{\overline{r}^4} \right)\\
  \label{Eqn:NonDimFinalForm_Thermal_DegradBeam}
  &\overline{\vartheta} \frac{\partial \overline{\vartheta}}{\partial 
  \overline{t}} - \left(\frac{\overline{D}_{\vartheta \vartheta}}{
  \overline{r}} + \frac{\partial \overline{D}_{\vartheta \vartheta}}{
  \partial \overline{r}} \right) \frac{\partial \overline{\vartheta}}{
  \partial \overline{r}} - \overline{D}_{\vartheta\vartheta} \frac{\partial
  ^2 \overline{\vartheta}}{\partial \overline{r}^2} = 
  \overline{\tau} \overline{D}_{\varkappa \varkappa} \left(\frac{\partial \overline{c}}{
  \partial \overline{r}}+\overline{\omega}
  \left(\frac{\overline{\gamma}^2}{\overline{r}^3} - \frac{\overline{r}}{
  \overline{\gamma}^2} \right)\right)^2 
\end{align}
%%
%-------------------------------------------------------------;
%  Subsubsection: Steady-state analysis for beam degradation  ;
%-------------------------------------------------------------;
\subsubsection{Steady-state and unsteady analysis for beam degradation}
\label{Subsec:SteadyState_DegradBeam}
In case of steady-state, we have $h_r r = C_1$ and $q_r r + 
\varkappa h_r r = C_2$, where $C_1$ and $C_2$ are integration 
constants. Equations \eqref{Eqn:NonDimFinalForm_Transport_DegradBeam} 
and \eqref{Eqn:NonDimFinalForm_Thermal_DegradBeam} imply that 
$\overline{c}$ and $\overline{\vartheta}$ are the solutions of the following 
ODEs:
%-----------------------------------------------------;
%  Equation: Steady-state thermal and transport ODEs  ;
%-----------------------------------------------------;
\begin{subequations}
  \begin{align}
    \label{Eqn:SteadyState_Thermal_DegradBeam}
    &\overline{D}_{\varkappa \varkappa} \overline{r} 
    \frac{d \overline{c}}{d \overline{r}} - \overline{D}_
    {\varkappa \varkappa}\overline{\omega} \left( 
    \frac{\overline{\gamma}^2}{\overline{r}^2} - \frac{
    \overline{r}^2}{\overline{\gamma}^2}\right) + \overline{C}
    _1 = 0 \\
    \label{Eqn:SteadyState_Transport_DegradBeam}    
    &\overline{D}_{\vartheta \vartheta} \overline{r} 
    \frac{d \overline{\vartheta}}{d \overline{r}} + \overline{\tau} 
    \left(\frac{\overline{w}}{2} \left( \frac{\overline{\gamma}
    ^2}{\overline{r}^2} + \frac{\overline{r}^2}{\overline{\gamma}^2} 
    - 2\right) - \overline{c}+1\right) \overline{C}_1 + 
    \overline{C}_2 = 0
  \end{align}
\end{subequations}
In case of weak coupling (where $D_{\vartheta \vartheta}$ 
and $D_{\varkappa \varkappa}$ are constants), the solutions 
for $\overline{c}$ and $\overline{\vartheta}$ take the 
following simplified form:
%---------------------------------------------------------------------------;
%  Equation: Analytical solutions for Conc and Temperature (weak coupling)  ;
%---------------------------------------------------------------------------;
\begin{align}
  \label{Eqn:AnalyticSoln_Thermal_DegradBeam_WC}
  \overline{c} = -\frac{\overline{\omega}}{2} \left(\frac{
    \overline{\gamma}^2}{\overline{r}^2} + \frac{\overline{r}^2}{
    \overline{\gamma}^2}\right) + B_2\mathrm{ln}[\overline{r}] 
  + A_2, \quad 
  \overline{\vartheta} =- \frac{\overline{\tau} B_2^2
    \overline{D}_{\varkappa \varkappa}}{2 
    \overline{D}_{\vartheta \vartheta}} 
  \mathrm{ln}[\overline{r}]^2+Z_2\mathrm{ln}[\overline{r}] 
  + Y_2
\end{align}
where the constants $A_2$, $B_2$, $Y_2$, and $Z_2$ 
(which depend on the boundary conditions) are as 
follows:
%---------------------------------------------------;
%  Equation: Integration constants (weak coupling)  ;
%---------------------------------------------------;
\begin{subequations}
  \begin{align}
    &A_2 = \overline{c}_i - B_2\mathrm{ln}[\overline{r}_i] + 
    \frac{\overline{\omega}}{2} \left( \frac{\overline{\gamma}^2}
    {\overline{r}_i^2} + \frac{\overline{r}_i^2}{\overline{\gamma}^2}\right) \\
    &B_2 = \frac{1}{\mathrm{ln}[\overline{r}_o] - \mathrm{ln}
    [\overline{r}_i]}\left(\overline{c}_o-\overline{c}_i-\frac{\overline{
    \omega}}{2}\left( \frac{\overline{\gamma}^2}{\overline{r}_i^2} +
    \frac{\overline{r}_i^2}{\overline{\gamma}^2}-\frac{\overline{\gamma}
    ^2}{\overline{r}_o^2} - \frac{\overline{r}_o^2}{\overline{\gamma}^2} 
    \right)\right)\\
    &Y_2 = \overline{\vartheta}_i + \frac{\overline{\tau} B_2^2
    \overline{D}_{\varkappa \varkappa}}{2 \overline{D}
    _{\vartheta \vartheta}} \mathrm{ln}[\overline{r}_i]^2 
    - Z_2\mathrm{ln}[\overline{r}_i] \\
    &Z_2 = \frac{1}{\mathrm{ln}[\overline{r}_o] - \mathrm{ln}
    [\overline{r}_i]}\left(\overline{\vartheta}_o - \overline{\vartheta}_i 
    - \frac{\overline{\tau} B_2^2 \overline{D}_{\varkappa \varkappa}}{2 
    \overline{D}_{\vartheta \vartheta}} \left( \mathrm{ln}
    [\overline{r}_i]^2 -\mathrm{ln}[\overline{r}_o]^2 \right) \right)
  \end{align}
\end{subequations}
For unsteady analysis, we employ method of horizontal lines 
with backward Euler time-stepping scheme. This gives the following 
ODEs at each time-level for equations \eqref{Eqn:NonDimFinalForm_Transport_DegradBeam} 
and \eqref{Eqn:NonDimFinalForm_Thermal_DegradBeam}:
%------------------------------------------------;
%  Equation: Non-dim thermal and transport PDEs  ;
%------------------------------------------------;
\begin{align}
  \label{Eqn:NonDimFinalForm_Transport_DegradBeam_NumSoln}
  \frac{d^2 \overline{c}^{(n+1)}}{d \overline{r}^2} &+ \left(\frac{ 
  1}{\overline{r}^{(n)}} + \left( \frac{1}{\overline{D}^{(n)}_{\varkappa 
  \varkappa}} \right) \frac{d \overline{D}_{\varkappa \varkappa}}{d 
  \overline{r}} \Bigg|_{\overline{t} = \overline{t}_n} \right) \frac{d \overline{c}^{(n+1)}}{d 
  \overline{r}} - \frac{\overline{c}^{(n+1)}}{\overline{D}^{(n)}_{\varkappa 
  \varkappa} \Delta \overline{t}} = 2\overline{\omega} \left(\frac{1}{\left(\overline{\gamma}^{(n)}
  \right)^2} + \frac{\left(\overline{\gamma}^{(n)}\right)^2}{
  \left(\overline{r}^{(n)}\right)^4} \right) \nonumber \\
  &- \frac{\overline{c}^{(n)}}{\overline{D}^{(n)}_{\varkappa \varkappa} 
  \Delta \overline{t}} - \left(\frac{\overline{\omega}}{\overline{D}^{(n)}_{\varkappa 
  \varkappa}} \right) \left( \frac{d \overline{D}_{\varkappa \varkappa}}{d 
  \overline{r}} \Bigg|_{\overline{t} = \overline{t}_n} \right) \left(\frac{\left(\overline{\gamma}
  ^{(n)} \right)^2}{\left(\overline{r}^{(n)} \right)^3} - \frac{\overline{r}
  ^{(n)}}{\left(\overline{\gamma}^{(n)} \right)^2} \right)
\end{align}
%------------------------------------------------;
%  Equation: Non-dim thermal and transport PDEs  ;
%------------------------------------------------;
\begin{align}
  \label{Eqn:NonDimFinalForm_Thermal_DegradBeam_NumSoln}
  \frac{d^2 \overline{\vartheta}^{(n+1)}}{d \overline{r}^2} &+ \left(\frac{ 
  1}{\overline{r}^{(n)}} + \left( \frac{1}{\overline{D}^{(n)}_{\vartheta 
  \vartheta}} \right) \frac{d \overline{D}_{\vartheta \vartheta}}{d 
  \overline{r}} \Bigg|_{\overline{t} = \overline{t}_n} \right) \frac{d \overline{\vartheta}^{(n+1)}}{d 
  \overline{r}} - \frac{\overline{\vartheta}^{(n)} \overline{\vartheta}^{(n+1)}}{
  \overline{D}^{(n)}_{\vartheta \vartheta} \Delta \overline{t}} = - \frac{\overline{\tau} 
  \overline{D}^{(n)}_{\varkappa \varkappa}}{\overline{D}^{(n)}_{\vartheta \vartheta}} 
  \left(\frac{d \overline{c}}{d \overline{r}} \Bigg|_{\overline{t} = \overline{t}_n} \right)^2 
  - \frac{\left(\overline{\vartheta}^{(n)} \right)^2}{\overline{D}^{(n)}_
  {\vartheta \vartheta} \Delta\overline{t}} \nonumber \\
  &- \frac{2 \overline{\tau} \overline{\omega} \overline{D}^{(n)}_{\varkappa 
  \varkappa}}{\overline{D}^{(n)}_{\vartheta \vartheta}} \left(
  \frac{\left(\overline{\gamma}^{(n)} \right)^2}{\left(\overline{r}^{(n)} 
  \right)^3} - \frac{\overline{r}^{(n)}}{\left(\overline{\gamma}^{(n)} 
  \right)^2} \right) \frac{d \overline{c}}{d \overline{r}} \Bigg |_{\overline{t} 
  = \overline{t}_n} - \frac{\overline{\tau} \overline{D}^{(n)}_{\varkappa \varkappa} 
  \overline{\omega}^2}{\overline{D}^{(n)}_{\vartheta \vartheta}} \left(\frac{\left(
  \overline{\gamma}^{(n)} \right)^2}{\left(\overline{r}^{(n)}\right)^3} 
  - \frac{\left(\overline{r}^{(n)} \right)}{\left(\overline{\gamma}^{(n)} 
  \right)^2} \right)^2
\end{align}
Algorithm \ref{Algo:DegradBeam_NumSolnMethod} describes a procedure 
to determine $\overline{c}(\overline{r},\overline{t})$, 
$\overline{\vartheta}(\overline{r},\overline{t})$, and $\overline{\gamma}$ at 
various times using an iterative non-linear numerical solution 
strategy. 
The boundary conditions for diffusion and thermal 
subproblems are the same as the degrading shell 
problem. The other parameters are assumed in the 
strongly coupling simulations as follows:
\begin{align}
  \label{Eqn:Beam_Degradation_Params}
  &\overline{L} = 1, \ \overline{r}_c = 1, \ \Delta \overline{t} = 0.1, 
  \ \overline{t} = 2, \ \overline{\omega} = 0.05, \ \overline{\tau} = 0.5, 
   \ \overline{\mu}_0 = 1, \ \overline{\mu}_1 = \overline{\mu}_2 = 0.4, 
  \ \overline{D}_0 = 1, \nonumber \\
  &\overline{D}_T = 2.0, \ \overline{D}_S = 1.5, \ \eta_T = \eta_S = 1, \ E_{\mathrm{ref}T} 
  = E_{\mathrm{ref}S} = 1, \ \overline{K}_0 = 1,\ \delta = 10 
\end{align}
In case of weak coupling, we have $\overline{D}_0$ as $\overline{D}^{(n)}
_{\varkappa \varkappa}$ and $\overline{K}_0$ as $\overline{D}^{(n)}_{\vartheta 
\vartheta}$, respectively. 

The numerical results are shown in figures 
\ref{Fig:Beam_Bending_WeakStrongCoupling_NeutralAxis}--\ref{Fig:Beam_Bending_ExtOfDamage}, which reveal the following conclusions 
on the overall behavior of bending of degrading 
beams:
\begin{enumerate}[(i)]
  \item \textsf{Degradation vs. non-degradation:} \emph{The 
    main observation is that the neutral axis shifts further to the 
    left}, similar to the phenomenon observed in viscoelastic solids 
    \citep{1997_Kolberg_Wineman_IJNM_v32_p863_p883}. Moreover, in case 
    of weak coupling for some instants of time \emph{the maximum stress 
    does not occur at either tensile or compressive sides of the beam 
    after the onset of degradation}. This is of primal importance in 
    regards to the calculation of failure loads/moments due to material 
    damage. Hence, a simple approach based on strength of materials or 
    a more complex finite elasticity theory to calculate stresses without 
    accounting for degradation will lead to erroneous results.
  \item Initially at $\overline{t} = 0$ and when there is no degradation, 
    the response is that of a homogeneous neo-Hookean material. \emph{On 
    the onset of degradation, the material ceases to be homogeneous}. 
  \item Moment relaxation is observed for weak and strong coupling 
  degradation. Note that the moment is a constant without degradation.
  Moreover, although diffusion is dominant in the coupling effect for 
  chemical potential, one can still observe the deformation effect 
  on $\overline{\varkappa}$ as compared with no degradation case.
  \item  \textsf{Strong vs. weak coupling:} One can see that 
    $\overline{T}_{\theta \theta}$ for strong coupling is 
    considerably different from the weak coupling. This is 
    because the degradation progress is dependent on the 
    deformation, concentration of the diffusing chemical 
    species, and temperature of the body.
  \item The extent of damage is monotonic for weak coupling, 
    which is not the case for strong coupling (which helps 
    in identifying regions that need retrofitting). 
\end{enumerate}

%-----------------------------------------------------;
%  Remark: On neutral axis  using internal variables  ;
%-----------------------------------------------------;
\begin{remark}
  Based on a semi-inverse approach, under degradation, 
  \citep{2007_Rajagopal_Srinivasa_Wineman_IJP_v23_p1618_p1636}
  have shown that there is a shift in the neutral axis for 
  pure bending of a polymer beam. However, their model is 
  based on internal variables, which is difficult to calibrate 
  experimentally. On the other hand, the proposed (and calibrated) 
  chemo-thermo-mechano degradation model is able to predict 
  the shift of neutral axis without appealing to internal 
  variable framework.
\end{remark}

%----------------------------------------------------------------------------;
%  Algorithm: Pure bending of degrading beam numerical solution methodology  ;
%----------------------------------------------------------------------------;
\begin{algorithm} 
  \caption{Pure bending of degrading beam (numerical methodology 
    to find $\overline{\gamma}$, $\overline{c}$, and $\overline{\vartheta}$)}
  \label{Algo:DegradBeam_NumSolnMethod}
  \begin{algorithmic}[1]
    %----------------------;
    %  STEP-0: Input data  ;
    %----------------------; 
    \STATE INPUT:~Non-dimensional material parameters, non-dimensional 
      boundary conditions, and non-dimensional initial conditions, 
      \texttt{MaxIters}, tolerances $\epsilon_{\mathrm{tol}}^{(\gamma)}$, 
      $\epsilon_{\mathrm{tol}}^{(c)}$, and $\epsilon_{\mathrm{tol}}^
      {(\vartheta)}$.
    %%
    %------------------------;
    %  STEP-1: Guess \gamma  ;
    %------------------------;
    \STATE Evaluate $\overline{\gamma}$ at $\overline{t} = 0$ based on equation 
    \eqref{Eqn:Initial_Gamma_t0}. Use this as an initial guess for solving 
    nonlinear equation given by \eqref{Eqn:NonLin_BeamBend_Gamma_Nondim_1} 
    or guess $\overline{\gamma}$ based on equation \eqref{Eqn:UppLowBound_Gamma}.
    %%
    %-----------------------------------;
    %  STEP-2: Iterative strategy loop  ;
    %-----------------------------------;
    \FOR{$n = 1, 2, \cdots, N$}
    \FOR{$i = 1, 2, \cdots$}
    %%
    %----------------------------------;
    %  STEP-2a: Stopping criteria # 1  ;
    %----------------------------------; 
    \IF {$i >$ \texttt{MaxIters}}
      \STATE Solution did not converge in specified maximum number of 
        iterations. EXIT.
    \ENDIF
    %%
    %----------------------------------;
    %  STEP-2b: Diffusion sub-problem  ;
    %----------------------------------; 
    \STATE \texttt{\underline{Diffusion sub-problem}}:~Given $\overline{\gamma}^{(i)}$, 
      solve equation \eqref{Eqn:NonDimFinalForm_Transport_DegradBeam_NumSoln} 
      to obtain $\overline{c}^{(i+1)}$. Herein, we use shooting method to 
      solve the ODEs.
    %%
    %----------------------------------------;
    %  STEP-2c: Heat conduction sub-problem  ;
    %----------------------------------------; 
    \STATE \texttt{\underline{Heat conduction sub-problem}}:~Given $\overline{\gamma}^{(i)}$ 
      and $\overline{c}^{(i+1)}$, solve equation \eqref{Eqn:NonDimFinalForm_Thermal_DegradBeam_NumSoln}
      to obtain $\overline{\vartheta}^{(i+1)}$. Similarly, we use shooting method 
      to solve the ODEs.
    %%
    %------------------------------------;
    %  STEP-2d: Deformation sub-problem  ;
    %------------------------------------;
    \STATE \texttt{\underline{Deformation sub-problem}}:~Given $\overline{c}^{(i+1)}$ 
      and $\overline{\vartheta}^{(i+1)}$, solve for $\overline{\gamma}^{(i+1)}$ given 
      by equation \eqref{Eqn:NonLin_BeamBend_Gamma_Nondim_1} 
      using bisection method.
    %%
    %----------------------------------;
    %  STEP-2e: Stopping criteria # 2  ;
    %----------------------------------;
    \IF {$\|\overline{\gamma}^{(i+1)} - \overline{\gamma}^{(i)} \| < 
      \epsilon_{\mathrm{tol}}^{(\gamma)}$, 
      $\|\overline{c}^{(i+1)} - \overline{c}^{(i)} \| < \epsilon_
      {\mathrm{tol}}^{(c)}$, and $\|\overline{\vartheta}
      ^{(i+1)} - \overline{\vartheta}^{(i)} \| < \epsilon
      _{\mathrm{tol}}^{(\vartheta)}$}
      \STATE OUTPUT:~$\overline{\gamma}^{(i+1)}$, $\overline{c}^{(i+1)}$, 
      and $\overline{\vartheta}^{(i+1)}$. EXIT the inner loop.
    \ELSE 
      \STATE Update the guess:~$\overline{\gamma}^{(i)} \gets \overline{\gamma}^{(i+1)}$.
    \ENDIF
   \ENDFOR
   \ENDFOR
  \end{algorithmic}
\end{algorithm}

%% file: Sections_Model/S6_Model_Cylinder_Torsion.tex
%***************************************;
%                                       ;
%  NAME                                 ;
%    S6_Model_Cylinder_Torsion.tex      ;
%                                       ;
%  WRITTEN BY                           ;
%    Can Xu                             ;
%    Maruti Kumar Mudunuru              ;
%    Kalyana Babu Nakshatrala           ;
%                                       ;
%***************************************;
\subsection{Torsional shear of a degrading cylinder}
\label{Subsec:Torsion_Cylinder}
A pictorial description of the degrading cylindrical annulus 
of finite length is shown in Figure \ref{Fig:torsion_pictorial_description}.
The bottom of the cylinder is fixed and just after time $t = 0$, 
a twisting moment is applied. We analyze the material degradation 
and corresponding structural response due to the torsional shear 
for a prescribed angle of twist. Initially, the body is a homogeneous 
neo-Hookean material and there is no transport of chemical 
species in the body. For time $t > 0$, the outer boundary of the 
cylinder is always exposed to moisture (or a diffusing chemical 
species). 
The inner surface of the degrading annular cylinder is held at zero 
concentration. This can be achieved by constructing a mechanism which 
continuously removes the moisture (or diffusing chemical species) 
from the inner boundary of the degrading cylinder. Hence, one can 
control the concentration of the moisture at both inner and outer 
surfaces. Similar type of initial and boundary conditions are 
enforced for the thermal counter part.

Consider a closed cylindrical body of inner radius 
$R_i$, outer radius $R_o$, and height $L$ defined 
as follows:
%-----------------------------------------------------------;
%  Equation: Torsion cylinder in a reference configuration  ;
%-----------------------------------------------------------;
\begin{align}
  \label{Eqn:RefConfig_Cylinder}
  R_i \leq R \leq R_o, \quad 0 \leq \Theta 
  \leq 2\pi, \quad 0 \leq Z \leq L 
\end{align}
where $(R,\Theta,Z)$ are the cylindrical polar coordinates 
in the reference configuration. Under torsional shear, the 
deformation can be described as follows:
%---------------------------------------------------------;
%  Equation: Torsion cylinder in a current configuration  ;
%---------------------------------------------------------;
\begin{align}
  \label{Eqn:CurrConfig_Torsion_Cylinder}
   r = R, \quad \theta = \Theta + g(Z,t), 
   \quad z = \Lambda Z
\end{align}
The components of the deformation gradient $\mathbf{F}$ can be written as:
%---------------------------------------------------------;
%  Equation: Deformation gradient for degrading cylinder  ;
%---------------------------------------------------------;
\begin{align}
  \label{Eqn:DefGrad_Cylinder}
  \{\mathbf{F}\} = \left(\begin{array}{ccc}
    1 & 0 & 0 \\
    0 & 1 & rg^{\prime} \\
    0 & 0 & \Lambda \\
  \end{array} \right) \quad 
  \mathrm{where} \; g^{\prime} := \frac{\partial g(Z,t)}{\partial Z}
\end{align}
Incompressibility implied that $\Lambda = 1$. The components 
of the right Cauchy-Green tensor $\mathbf{C}$ and the left 
Cauchy-Green tensor $\mathbf{B}$ can be written as:
%------------------------------------------------------------------;
%  Equation: Left and right Cauchy tensors for degrading cylinder  ;
%------------------------------------------------------------------;
\begin{align}
  \label{Eqn:CBtensors_Cylinder}
  \{\mathbf{C}\} = \left(\begin{array}{ccc}
    1 & 0 & 0 \\
    0 & 1 & rg^{\prime} \\
    0 & rg^{\prime} & 1 + \left( rg^{\prime} \right)^2 \\
  \end{array} \right) \quad
  \{\mathbf{B}\} = \left(\begin{array}{ccc}
    1 & 0 & 0 \\
    0 & 1 + \left(rg^{\prime} \right)^2 
    & r g^{\prime} \\
    0 & r g^{\prime} & 1 \\
  \end{array} \right) 
\end{align}
The non-zero components of the Cauchy stress 
$\mathbf{T}$ are given as follows:
%----------------------------------------------------------;
%  Equation: Balance of linear momentum (torsion problem)  ;
%----------------------------------------------------------;
  \begin{align}
    \label{Eqn:T_Torsion}
    &T_{rr} = -p + \mu(c,\vartheta), \quad
    T_{\theta \theta} = -p + \mu(c,\vartheta) 
    \left(1 + \left(rg^{\prime} \right)^2 \right) \nonumber\\
    &T_{zz} = -p + \mu(c,\vartheta), \quad
    T_{\theta z} = T_{z\theta} = \mu(c,\vartheta)
    rg^{\prime}
  \end{align}
The balance of linear momentum in the cylindrical 
polar coordinates reduces to the following:
%----------------------------------------------------------;
%  Equation: Balance of linear momentum (torsion problem)  ;
%----------------------------------------------------------;
\begin{align}
  \label{Eqn:BoLM_ReducedCylin_Torsion}
  -\frac{\partial p}{\partial r} + \mu(c,\vartheta) r \left(g^{\prime}
  \right)^2 = 0, \quad
  -\frac{1}{r} \frac{\partial p}{\partial \theta} + 
  \mu(c,\vartheta) r g^{\prime\prime} = 0, \quad
  -\frac{\partial p}{\partial z} = 0
\end{align}
Symmetry in the problem implies that $\frac{\partial p}
{\partial \theta} = 0$, which further implies that 
$g^{\prime\prime} = 0$. Hence, $g(Z,t)$ takes the 
following form:
%---------------------------------------------------------------;
%  Equation: Angle of twist (analytical expression for g(Z,t))  ;
%---------------------------------------------------------------;
\begin{align}
  \label{Eqn:Analytic_Expr_gZt}
  g(Z,t) = \Psi_1(t) Z + \Psi_2(t)
\end{align}
where $\Psi_1$ and $\Psi_2$ are evaluated based 
on the input data. As the bottom of the cylinder 
is fixed, we have $g(Z = 0,t) = 0$, which implies 
$\Psi_2(t) = 0$. 
 
The chemical potential and specific entropy are given as follows:
%------------------------------------------------------------------------;
%  Equations: Chemical potential, specific entropy, mass transfer flux,  ;
%             and heat transfer flux in radial direction (torsion)       ;
%------------------------------------------------------------------------;
\begin{subequations}
  \begin{align}
    \label{Eqn:Chemical_Potential_Torsion}
    \varkappa &= \frac{1}{\rho_0} \frac{\partial 
    \psi}{\partial c} + R_s \vartheta_{\mathrm{ref}} 
    \{c - c_{\mathrm{ref}}\} = -\frac{\mu_1 r^2 \Psi_1
    ^2}{2 \rho_0 c_{\mathrm{ref}}} + R_s \vartheta_
    {\mathrm{ref}} \{c - c_{\mathrm{ref}}\} \\
    \label{Eqn:Specific_Entropy_Torsion}
    \eta &= -\frac{1}{\rho_0} \frac{\partial \psi}{\partial 
    \vartheta} + \frac{c_p}{\vartheta_{\mathrm{ref}}} \{\vartheta
    - \vartheta_{\mathrm{ref}}\} = \frac{\mu_1 r^2 \Psi_1^2}{2 \rho_0 
    \vartheta_{\mathrm{ref}}} + \frac{c_p}{ \vartheta_
    {\mathrm{ref}}} \{\vartheta - \vartheta_{\mathrm{ref}}\}
  \end{align}
\end{subequations}
Most of the non-dimensional quantities remain the 
same as that of the previous initial boundary value 
problems except for the following:
%----------------------------------------;
%  Equation: Non-dimensional quantities  ;
%----------------------------------------;
\begin{align}
  \label{Eqn:NonDimQuantities1_DegradCylinder}
  \overline{R} = \frac{R}{R_o}, \ \overline{\psi} 
  = \psi R_o, \ \overline{t} = \frac{D_0 t}{R_o^2}  
\end{align}
The non-dimensional twisting moment 
$\overline{M}(\overline{t})$ satisfies:
%-------------------------------------------------;
%  Equation: Non-linear equation for determing M  ;
%-------------------------------------------------;
\begin{align}
  \label{Eqn:NonLin_Cylinder_Moment_Nondim}
  \overline{M}(\overline{t}) = 2 \pi \displaystyle \int 
  \limits_{\overline{R}_i}^{\overline{R}_o} \overline{\mu}
  (\overline{c} (\overline{R},\overline{t}), \overline{
  \vartheta}(\overline{R},\overline{t})) \overline{\Psi}_1 
  \overline{R}^3 \mathrm{d}\overline{R}
\end{align}
The Poynting effect for hyperelastic materials shall be also 
studied. It implies the axial length change for a cylinder under 
shear. The non-dimensional normal force required to keep the 
length unchanged can be written as follows:
%-------------------------------------------------;
%  Equation: Non-linear equation for determing M  ;
%-------------------------------------------------;
\begin{align}
  \label{Eqn:NonLin_Cylinder_Moment_Nondim}
  \overline{N}(\overline{t}) = \pi \displaystyle \int 
  \limits_{\overline{R}_i}^{\overline{R}_o} \overline{\mu}
  (\overline{c} (\overline{R},\overline{t}), \overline{
  \vartheta}(\overline{R},\overline{t})) \overline{\Psi}_1^2 
  \overline{R}^3 \mathrm{d}\overline{R}
\end{align}
From equations \eqref{Eqn:ReducedGE_DiffSubProb} 
and \eqref{Eqn:ReducedGE_ThermalSubProb}, the 
final form of the governing equations for transport 
and thermal sub-problems can be written as:
%------------------------------------------------;
%  Equation: Non-dim thermal and transport PDEs  ;
%------------------------------------------------;
\begin{align}
  \label{Eqn:NonDimFinalForm_Transport_Torsion}
  &\frac{\partial \overline{c}}{\partial \overline{t}} - \left(\frac{ 
  \overline{D}_{\varkappa \varkappa}}{\overline{r}} + \frac{\partial 
  \overline{D}_{\varkappa \varkappa}}{\partial \overline{r}} \right) 
  \frac{\partial \overline{c}}{\partial \overline{r}} - \overline{D}_
  {\varkappa \varkappa} \frac{\partial^2 \overline{c}}{\partial 
  \overline{r}^2} = -\overline{\omega} \overline{\Psi}_1^2 \left(2 
  \overline{D}_{\varkappa\varkappa} +\overline{r} \frac{\partial 
    \overline{D}_{\varkappa \varkappa}}{\partial \overline{r}} \right) \\
  \label{Eqn:NonDimFinalForm_Thermal_Torsion}
  &\overline{\vartheta} \frac{\partial \overline{\vartheta}}{\partial 
  \overline{t}} - \left(\frac{\overline{D}_{\vartheta \vartheta}}{
  \overline{r}} + \frac{\partial \overline{D}_{\vartheta \vartheta}}{
  \partial \overline{r}} \right) \frac{\partial \overline{\vartheta}}{
  \partial \overline{r}} - \overline{D}_{\vartheta\vartheta} \frac{\partial
  ^2 \overline{\vartheta}}{\partial \overline{r}^2} = \overline{\tau} 
  \overline{D}_{\varkappa \varkappa} \left(\frac{\partial \overline{c}}{
  \partial \overline{r}} \right)^2 - 2 \overline{\tau} \overline{\omega} 
  \overline{D}_{\varkappa \varkappa} \overline{r}\overline{\Psi}_1^2 \frac{
  \partial \overline{c}}{\partial \overline{r}} + \overline{\tau} \overline{D}
  _{\varkappa \varkappa} \overline{\omega}^2 \overline{r}^2 \overline{\Psi}_1^4
\end{align}
One needs to solve equations 
\eqref{Eqn:NonLin_Cylinder_Moment_Nondim}--\eqref{Eqn:NonDimFinalForm_Thermal_Torsion} to obtain $\overline{c}(\overline{r},\overline{t})$, 
$\overline{\vartheta}(\overline{r},\overline{t})$, 
and $\overline{M}(\overline{t})$. Algorithm \ref{Algo:DegradCylinder_NumSolnMethod} describes a numerical solution procedure to solve 
these equations at various times for a given angle of 
twist per unit length. 

%========================================================;
%  Subsubsection: Steady-state and quasistatic analysis  ;
%========================================================;
\subsubsection{Steady-state and unsteady response 
  of degrading cylinder under torsional shear}
\label{Subsec:SteadyState_DegradCylinder}
In the case of steady-state, $\overline{c}$ and 
$\overline{\vartheta}$ are the solutions of the 
following ODEs:
%----------------------------------------------------;
% Equation: Steady-state thermal and transport ODEs  ;
%----------------------------------------------------;
\begin{subequations}
  \begin{align}
    \label{Eqn:SteadyState_Thermal_DegradCylinder}
    &\overline{D}_{\varkappa \varkappa} \overline{r}^2 
    \frac{d \overline{c}}{d \overline{r}} - \overline{D}
    _{\varkappa \varkappa} \overline{\omega} \overline{r} 
    \overline{\Psi}_1^2 + \overline{C}_1 = 0 \\
    \label{Eqn:SteadyState_Transport_DegradCylinder}
    &\overline{D}_{\vartheta \vartheta} \overline{r} 
    \frac{d \overline{\vartheta}}{d \overline{r}} + 
    \overline{\tau} \left(\frac{\overline{\omega}}{2} 
    \overline{r}^2 \overline{\Psi}_1^2 - \overline{c} 
    + 1\right) \overline{C}_1 + \overline{C}_2 = 0
  \end{align}
\end{subequations}
where $\overline{C}_1$ and $\overline{C}_2$ 
are integration constants. 
Under weak coupling (where $D_{\vartheta \vartheta}$ and 
$D_{\varkappa \varkappa}$ are constants), a simplified 
form of the analytical solutions for $\overline{c}$ 
and $\overline{\vartheta}$ is given as follows:
%---------------------------------------------------------------------------;
%  Equation: Analytical solutions for Conc and Temperature (weak coupling)  ;
%---------------------------------------------------------------------------;
\begin{align}
  \label{Eqn:AnalyticSoln_Thermal_DegradCylinder_WC}
  \overline{c} = \frac{\overline{\omega}}{2} \overline{r}
  ^2 \overline{\Psi}_1^2 + B_3 \mathrm{ln}[\overline{r}] 
  + A_3, \quad  
  %%
  % \label{Eqn:AnalyticSoln_Transport_DegradCylinder_WC}
  \overline{\vartheta} = -\frac{\overline{\tau} B_3^2
    \overline{D}_{\varkappa \varkappa}}{2 \overline{D}_
    {\vartheta \vartheta}} \mathrm{ln}[\overline{r}]^2
  + Z_3\mathrm{ln}[\overline{r}] + Y_3
\end{align}
where $A_3$, $B_3$, $Y_3$, and $Z_3$ are constants, which are obtained 
by the corresponding boundary conditions for thermal and diffusion 
sub-problem. These are given as follows:
%---------------------------------------------------;
%  Equation: Integration constants (weak coupling)  ;
%---------------------------------------------------;
\begin{subequations}
  \begin{align}
    &A_3 = \overline{c}_i - B_3\mathrm{ln}[\overline{r}_i] - 
    \frac{\overline{\omega}}{2} \overline{r}_i^2 
    \overline{\Psi}_1^2 \\
    &B_3 = \frac{1}{\mathrm{ln}[\overline{r}_o] - \mathrm{ln}
    [\overline{r}_i]} \left(\overline{c}_o - \overline{c}_i 
    - \frac{\overline{\omega}}{2} \left( \overline{r}_o^2 
    \overline{\Psi}_1^2 - \overline{r}_i^2 \overline{\Psi}_1
    ^2 \right) \right)\\
    &Y_3 = \overline{\vartheta}_i + \frac{\overline{\tau} 
    B_3^2 \overline{D}_{\varkappa \varkappa}}{2 \overline{D}
    _{\vartheta \vartheta}} \mathrm{ln}[\overline{r}_i]^2 
    - Z_3 \mathrm{ln}[\overline{r}_i] \\
    &Z_3 = \frac{1}{\mathrm{ln}[\overline{r}_o] - \mathrm{ln}
    [\overline{r}_i]} \left(\overline{\vartheta}_o - \overline{
    \vartheta}_i - \frac{\overline{\tau} B_3^2 \overline{D}_{
    \varkappa \varkappa}}{2 \overline{D}_{\vartheta \vartheta}} 
    \left( \mathrm{ln}[\overline{r}_i]^2 -\mathrm{ln}[\overline{r}
    _o]^2 \right) \right)
  \end{align}
\end{subequations}

For unsteady analysis, method of horizontal lines with backward Euler 
time-stepping scheme is employed. This gives the following ODEs at each 
time-level:
%------------------------------------------------;
%  Equation: Non-dim thermal and transport PDEs  ;
%------------------------------------------------;
\begin{align}
  \label{Eqn:NonDimFinalForm_Transport_DegradCylinder_NumSoln}
  \frac{d^2 \overline{c}^{(n+1)}}{d \overline{r}^2} &+ \left(\frac{ 
  1}{\overline{r}^{(n)}} + \left( \frac{1}{\overline{D}^{(n)}_{\varkappa 
  \varkappa}} \right) \frac{d \overline{D}_{\varkappa \varkappa}}{d 
  \overline{r}} \Bigg|_{t = t_n} \right) \frac{d \overline{c}^{(n+1)}}{d 
  \overline{r}} - \frac{\overline{c}^{(n+1)}}{\overline{D}^{(n)}_{\varkappa 
  \varkappa} \Delta t} = 2\overline{\omega} \left(\overline{\Psi}_1^{(n)} 
  \right)^2 \nonumber \\
  &+\overline{\omega} \left(\overline{\Psi}_1^{(n)}\right)^2 
  \frac{\overline{r}^{(n)}}{\overline{D}^{(n)}_{\varkappa 
  \varkappa}} \left( \frac{d \overline{D}_{\varkappa \varkappa}}{d 
  \overline{r}} \Bigg|_{t = t_n} \right) - \frac{\overline{c}
  ^{(n)}}{\overline{D}^{(n)}_{\varkappa \varkappa} \Delta t}
\end{align}
%------------------------------------------------;
%  Equation: Non-dim thermal and transport PDEs  ;
%------------------------------------------------;
\begin{align}
  \label{Eqn:NonDimFinalForm_Thermal_DegradCylinder_NumSoln}
  \frac{d^2 \overline{\vartheta}^{(n+1)}}{d \overline{r}^2} &+ \left(\frac{ 
  1}{\overline{r}^{(n)}} + \left( \frac{1}{\overline{D}^{(n)}_{\vartheta 
  \vartheta}} \right) \frac{d \overline{D}_{\vartheta \vartheta}}{d 
  \overline{r}} \Bigg|_{t = t_n} \right) \frac{d \overline{\vartheta}^{(n+1)}}{d 
  \overline{r}} - \frac{\overline{\vartheta}^{(n)} \overline{\vartheta}^{(n+1)}}{
  \overline{D}^{(n)}_{\vartheta \vartheta} \Delta t} = - \frac{\overline{\tau} 
  \overline{D}^{(n)}_{\varkappa \varkappa}}{\overline{D}^{(n)}_{\vartheta \vartheta}} 
  \left(\frac{d \overline{c}}{d \overline{r}} \Bigg|_{t = t_n} \right)^2 
  - \frac{\left(\overline{\vartheta}^{(n)} \right)^2}{\overline{D}^{(n)}_
  {\vartheta \vartheta} \Delta t} \nonumber \\
  &+ \frac{2 \overline{\tau} \overline{\omega} \overline{D}^{(n)}
  _{\varkappa \varkappa}}{\overline{D}^{(n)}_{\vartheta \vartheta}} 
  \overline{r}^{(n)} \left(\overline{\Psi}_1^{(n)} \right)^2 \frac{d 
  \overline{c}}{d \overline{r}} \Bigg |_{t = t_n} - \frac{\overline{
  \tau} \overline{D}^{(n)}_{\varkappa \varkappa} \overline{\omega}
  ^2}{\overline{D}^{(n)}_{\vartheta \vartheta}} \left( \overline{r}^
  {(n)}\right)^2 \left(\overline{\Psi}_1^{(n)} \right)^4
\end{align}
The boundary conditions for diffusion and thermal 
sub-problems are the same as that of the previous 
boundary value problems. 

The following non-dimensional parameters are assumed 
in the numerical numerical simulations: 
%------------------------------------------------------;
%  Equation: Non-recurring non-dimensional parameters  ;
%------------------------------------------------------;
\begin{align}
  \label{Eqn:NonDimParams_TorsionProblem}
  &\overline{R}_o = 1, \ \overline{R}_i = 0.5, \ 
  \Delta \overline{t} = 0.1, \ \overline{t} = 2, \ 
  \overline{\omega} = 0.05, \ \overline{\tau} = 0.8, \  
  \overline{\mu}_0 = 1, \ \overline{\mu}_1 = 0.5, \ 
  \overline{\mu}_2 = 0.2, \nonumber \\
  &\overline{D}_0 = 1, \ \overline{D}_T = 1.5, \
  \overline{D}_S = 1.2, \ \eta_T = \eta_S = 0.1, \ 
  E_{\mathrm{ref}T} = E_{\mathrm{ref}S} = 1, \ 
  \overline{K}_0 = 1,\ \delta = 10 
\end{align}
The numerical results are shown in Figure 
\ref{Fig:Torsion_Moment_vs_t} and \ref{Fig:Torsion_normal_force_vs_t}, 
which reveals the following important conclusions on the overall 
behavior of degrading structural members under torsional shear:
%--------------------------------------------------;
%  Enumerate: Discussion on the canonical problem  ;
%--------------------------------------------------;
\begin{enumerate}[(i)]
  \item The numerical results reveal that there 
    is relaxation of moment for fixed deformation. 
    In addition, the twisting moment required to 
    maintain a fixed angle of twist decreases with 
    increase in $\overline{\mu}_{1}$. 
    Similar type of behavior is observed 
    when $\overline{\mu}_1$ is kept constant and 
    $\overline{\mu}_2$ is varied.
  \item We observe moment relaxation 
    due to material degradation when both the transport 
    and thermal sub-problems are close to steady states. 
    Moreover, one can see that 
    moment relaxation depends on the geometry of the specimen. 
    These aspects differentiate the stress relaxation due 
    to degradation from the stress relaxation due to 
    viscoelasticity. 
    \item We observe that the normal force due to 
      Poynting effect is decreasing over time as a 
      result of degradation. Without degradation, 
      the normal force is a constant (which is the 
      case for hyperelastic materials).
  %
%  \item Since thermal degradation dominates in this 
%    problem, so the extent of damage is increasing as the strain 
%    increases.
    % 
\end{enumerate}

%---------------------------------------------------------------------------------------;
%  Algorithm: Torsional shear of a degrading cylinder (numerical solution methodology)  ;
%---------------------------------------------------------------------------------------;
\begin{algorithm} 
  \caption{Torsional shear of a degrading cylinder (numerical methodology 
    to find $\overline{M}$, $\overline{c}$, and $\overline{\vartheta}$)}
  \label{Algo:DegradCylinder_NumSolnMethod}
  \begin{algorithmic}[1]
    %----------------------;
    %  STEP-0: Input data  ;
    %----------------------; 
    \STATE INPUT:~Non-dimensional material parameters, non-dimensional 
      boundary conditions, and non-dimensional initial conditions.
    %%
    %-----------------------------------;
    %  STEP-1: Iterative strategy loop  ;
    %-----------------------------------;
    \FOR{$n = 1, 2, \cdots, N$}
    %%
    %----------------------------------;
    %  STEP-1a: Diffusion sub-problem  ;
    %----------------------------------; 
    \STATE \texttt{\underline{Diffusion sub-problem}}:~Given $\overline{\Psi}_1$, 
      solve equation \eqref{Eqn:NonDimFinalForm_Transport_DegradCylinder_NumSoln} 
      to obtain $\overline{c}^{(n)}$. Herein, we use shooting method to 
      solve the ODEs.
    %%
    %----------------------------------------;
    %  STEP-2c: Heat conduction sub-problem  ;
    %----------------------------------------; 
    \STATE \texttt{\underline{Heat conduction sub-problem}}:~Given 
      $\overline{\Psi}_1$ and $\overline{c}^{(n)}$, solve equation 
      \eqref{Eqn:NonDimFinalForm_Thermal_DegradCylinder_NumSoln}
      to obtain $\overline{\vartheta}^{(n)}$. Similar to diffusion 
      sub-problem, we use shooting method to solve the non-linear 
      ODEs.
    %%
    %------------------------------------;
    %  STEP-2d: Deformation sub-problem  ;
    %------------------------------------;
    \STATE \texttt{\underline{Deformation sub-problem}}:~Given $\overline{c}
      ^{(n)}$ and $\overline{\vartheta}^{(n)}$, solve for $\overline{M}^{(n)}$ 
      given by equation \eqref{Eqn:NonLin_Cylinder_Moment_Nondim}.
   \ENDFOR
  \end{algorithmic}
\end{algorithm}

%% file: Sections_Model/S7_Model_CR.tex
%***************************************;
%                                       ;
%  NAME                                 ;
%    S7_Model_CR.tex                    ;
%                                       ;
%  WRITTEN BY                           ;
%    Kalyana Babu Nakshatrala           ;
%                                       ;
%***************************************;
\section{CONCLUDING REMARKS}
\label{Sec:S7_Model_CR}
This paper has made several contributions to the modeling of 
degradation of materials due to the presence of an adverse 
chemical species and temperature. \emph{First}, a consistent 
mathematical model has been derived that has firm continuum 
thermodynamics underpinning. The constitutive relations, which 
give rise to coupled deformation-thermal-transport equations, 
have been derived by appealing to the maximization of the rate 
of dissipation, which is a stronger version of the second law 
of thermodynamics. The proposed model is hierarchical in the 
sense that it recovers many existing models as special cases. 
\emph{Second}, the materials parameters have been 
calibrated with an experimental dataset available 
in the literature. 
\emph{Third},  it has been shown that the unsteady 
solutions to the proposed degradation model are 
bounded and stable in the sense of Lyapunov even 
under large deformations and large strains. 
\emph{Last but not the least}, using several canonical problems 
in degradation mechanics, we illustrated the effects of chemical 
degradation and thermal degradation on the response of a body 
that is initially hyperelastic. Some of the main features of 
degradation and of the proposed model can be summarized as 
follows: 
%-----------------------------------------;
% Conclusions: Shell, beam, and cylinder  ;
%-----------------------------------------;
\begin{enumerate}[(C1)]
\item Degradation introduces spatial inhomogeneity. 
  That is, a material which is originally homogeneous 
  may cease to be homogeneous due to degradation.
\item The proposed mathematical model can provide 
  the variation of important quantities like chemical 
  potential within the body, which is essential in 
  incorporating chemical reactions into the modeling.
\item The extent of damage in a structural member 
  can be both qualitatively and quantitatively 
  different under strong and weak couplings 
  between mechanical, thermal and transport 
  processes.
  More importantly, weak coupling may over-predict 
  the material degradation in some cases while in 
  other cases it may under-predict the degradation. 
  It is, therefore, of paramount importance to select 
  the extent of coupling between the mechanical, 
  thermal and chemical processes. 
  %
%  The non-dimensional quantities provided in the paper 
%  (which depend on material properties) can collectively 
%  serve as a guideline on the nature of the coupling. Of 
%  course, boundary conditions and geometry will also affect 
%  the nature of coupling.
  %%
\item The usual assumptions on either kinematics or stresses, 
  which may be justified for non-degrading members, may no 
  longer hold under degradation. For example, assumptions 
  on the location of neutral axis or the location of the 
  maximum stress on the outer fibers in beam bending will 
  not hold under degradation. 
\item  Degrading structural members may exhibit some responses 
  that are typically associated with viscoelasticity.  
  In particular, we have shown that degradation can induce 
  stress relaxation and creep in the response of the materials 
  even in the case of finite-sized bodies. In contrast 
  to a viscoelastic body (which creeps continuously upon the 
  application of a load) the body undergoing chemical degradation 
  ceases to creep for practical purposes after a certain period 
  of time. This the moment when the transport of chemical 
  species is close to a steady-state, if there is no volumetric 
  source and the boundary conditions are unchanged over time. 
  A similar trend holds even in the case of thermal degradation. 
  This characteristic behavior of degrading solids can be used to 
  differentiate the creep associated with viscoelasticity and degradation. 
  Moreover, stress relaxation due to degradation depends on the geometry 
  of the specimen, which is also different from the case due to 
  viscoelasticity. 
\end{enumerate}

A possible future research work can be towards incorporating 
fatigue and fracture into the degradation modeling. A related 
scientific question can be towards addressing the effect of 
material degradation on the crack initiation and its propagation.

%% file: Sections_Model/Model_Figures.tex
\begin{figure}[h]
  \centering
  \includegraphics[scale=0.7,clip]{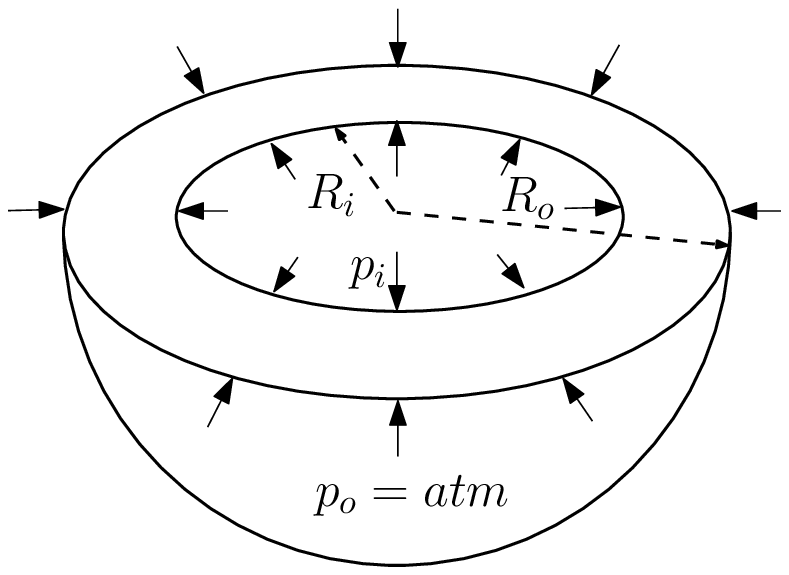}
  \caption{\textsf{Inflation of a degrading spherical shell:}~A 
    pictorial description of degrading shell in the reference 
    configuration. The shell is subjected to an inner pressure 
    $p_i$ and an outer pressure $p_o$.
    \label{Fig:Spherical_Shell_Pictorial}}
\end{figure}

%-------------------------------------------------------------------------;
%  Figure-4: T_theta_theta across the cross section of the shell given p  ;
%-------------------------------------------------------------------------;
\begin{figure}[h]
  \centering
 \includegraphics[scale=0.45,clip]{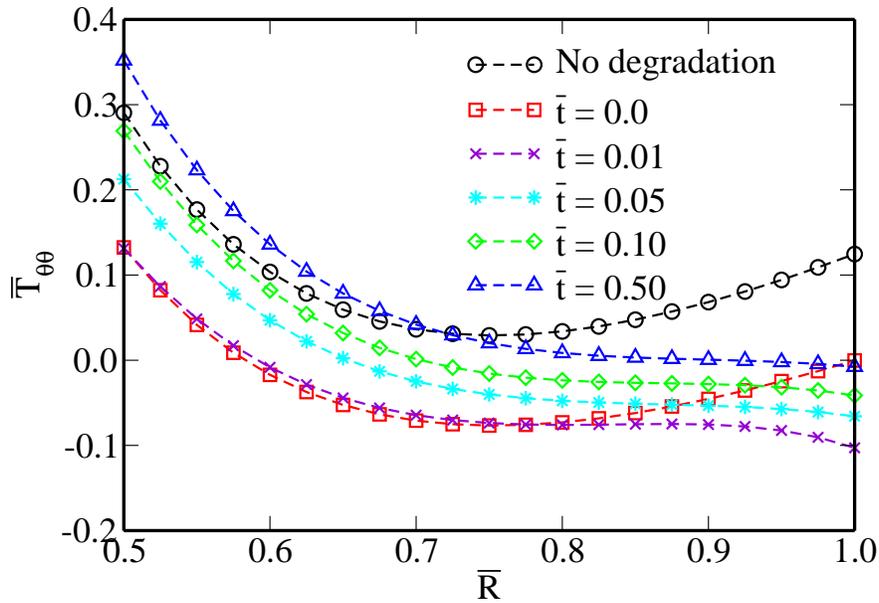}
  \caption{\textsf{Inflation of a degrading spherical shell:}~This 
    figure shows the hoop stress $\overline{T}_{\theta\theta}$ as a 
    function of $\overline{R}$ at various instants of time due to 
    an inner pressure of $\overline{p}_i = 0.5$. Analysis is performed 
    under strongly coupled chemo-thermo-mechano degradation. 
    Note that the stress is increasing with time under degradation.
    \label{Fig:Spherical_Shell_T_theta_theta_given_p}}
\end{figure}

%-------------------------------------------------------------------------;
%  Figure-5: T_theta_theta across the cross section of the shell given t  ;
%-------------------------------------------------------------------------;
\begin{figure}
  \centering
 \includegraphics[scale=0.45,clip]{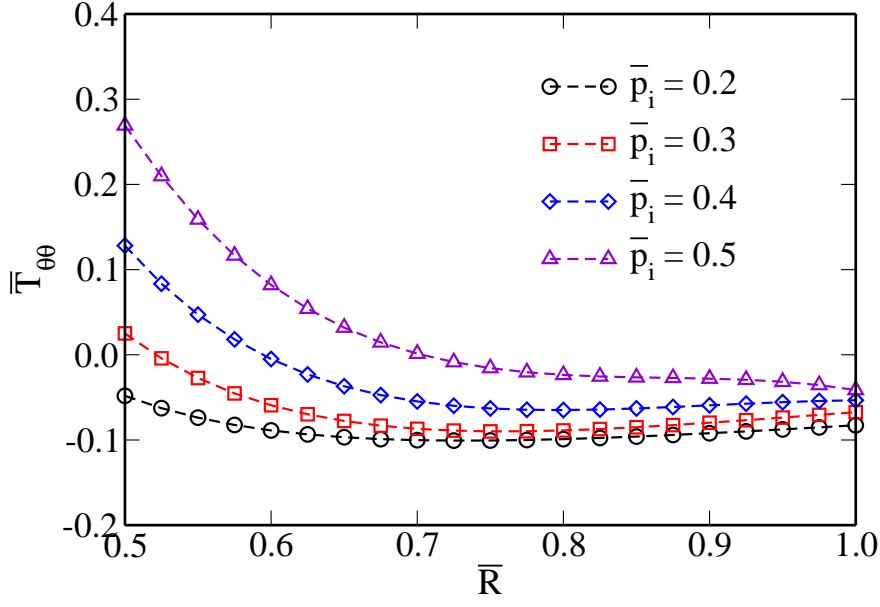}
  \caption{\textsf{Inflation of a degrading spherical shell:}~This 
    figure shows the hoop stress $\overline{T}_{\theta\theta}$ as 
    a function of $\overline{R}$ at $\overline{t} = 0.1$ for 
    various inner pressures $\overline{p}_i$. Analysis is performed 
    for strongly coupled chemo-thermo-mechano degradation. 
  $\overline{T}_{\theta \theta}$ increases in a non-linear 
  fashion as the pressure loading increases, which is different 
  from the case as time progresses. 
  \label{Fig:Spherical_Shell_T_theta_theta_given_t}}
\end{figure}

%----------------------------------------------------------------------;
%  Figure-6: Chemical potential across the cross section of the shell  ;
%----------------------------------------------------------------------;
\begin{figure}
  \centering
 \includegraphics[scale=0.33,clip]
 {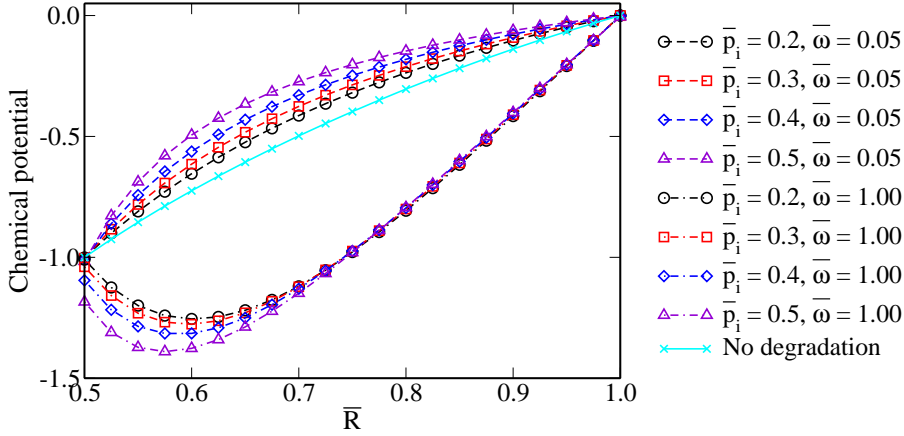}
  \caption{\textsf{Inflation of a degrading spherical shell:}~This 
    figure shows the chemical potential as a function of the 
    reference location $\overline{R}$ at $\overline{t} = 0.2$ 
    due to various inner pressures $\overline{p}_i$ under 
    different cases. One can see that for non-degrading 
    shell, the chemical potential is unchanged with respect 
    to pressure loading. However, for strong coupling, it 
    increases with $\overline{p}_i$ in a non-linear fashion 
    when $\overline{\omega}$ is small enough. This is 
    because for small $\overline{\omega}$, diffusion takes the 
    dominance in the coupling effect. When pressure loading 
    increases, the diffusivity is increasing due to the growing 
    $\mathrm{tr}[\mathbf{E}]$. For large $\overline{\omega}$, 
    the deformation is dominant in the coupling, which is $-
    \overline{I}_E$ term in chemical potential. Since the first 
    invariant $\overline{I}_E$ is always positive in this problem, 
    chemical potential is decreasing when the pressure loading 
    increases.
    \label{Fig:Spherical_Shell_chemical_potential_given_t}}
\end{figure}

%-------------------------------------------------------------------------------------;
%  Figure-7: r_i of the degrading shell under various p_i (weak and strong coupling)  ;
%-------------------------------------------------------------------------------------;
\begin{figure}
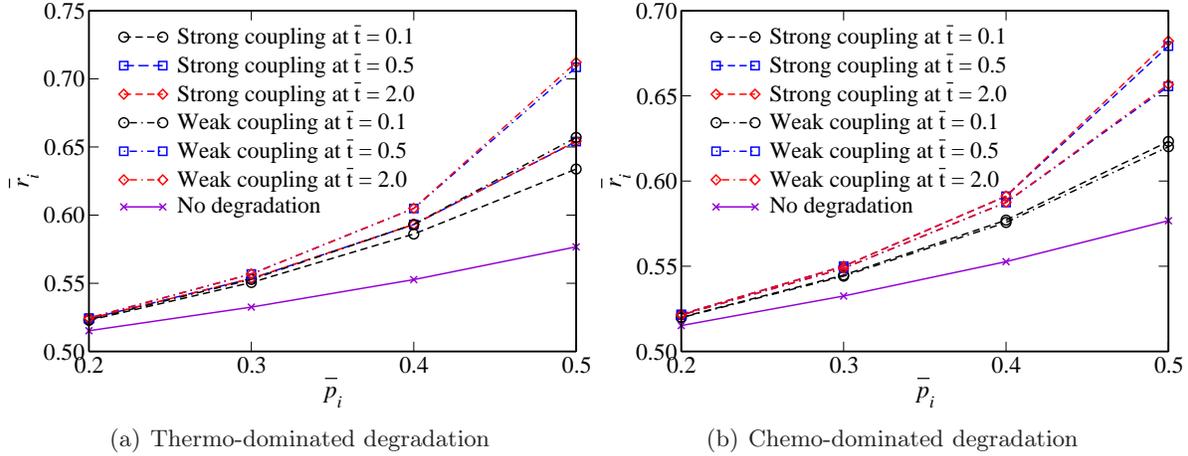

  \centering
    \subfigure[Thermo-dominated degradation]
    {\includegraphics[scale=0.3,clip]
    {Figures_Model/Spherical_Shell/r_vs_p.eps}}
  %%
  %%\vspace{0.5in}
  %%
  \subfigure[Chemo-dominated degradation]
    {\includegraphics[scale=0.3,clip]
    {Figures_Model/Spherical_Shell/r_vs_p_diff_dom.eps}}
  \vspace{-0.1in}
  \caption{\textsf{Inflation of a degrading spherical shell:}~This figure 
  shows the plot of $\overline{r}_i$ as a function of the inner pressure 
  $\overline{p}_i$ for strongly and weakly coupled chemo-thermo-mechano 
  degradation problem. Note that in weak coupling the heat conductivity 
  and diffusivity are both constants, while the Lam\'e parameters still
  depend on concentration and temperature. We take $\overline{\mu}_1 = 
  0.3$ and $\overline{\mu}_2 = 0.4$ for thermo-dominated degradation. For 
  chemo-dominated degradation, we have $\overline{\mu}_1 = 0.7$ and $\overline{
  \mu}_2 = 0.1$. For a given $\overline{p}_i$, one can see that $\overline{r}_i$ 
  for weak coupling is larger than strong coupling when thermal degradation 
  dominates. This is because $\overline{I}_{\mathbf{E}}$ is always 
  positive in this problem, the thermal conductivity decreases due to the 
  increase in $\overline{I}_{\mathbf{E}}$. However, when moisture-induced 
  degradation dominates, $\overline{r}_i$ for weak coupling is smaller than 
  strong coupling problem. From this figure, we can observe creep-like behavior 
  for all the case studies.
  \label{Fig:Spherical_Shell_WeakStrongCoupling_r_i}}
\end{figure}

%------------------------------------------------------------------------------;
%  Figure-8: Extent of damage across the cross section of the shell given p_i  ;
%------------------------------------------------------------------------------;
\begin{figure}
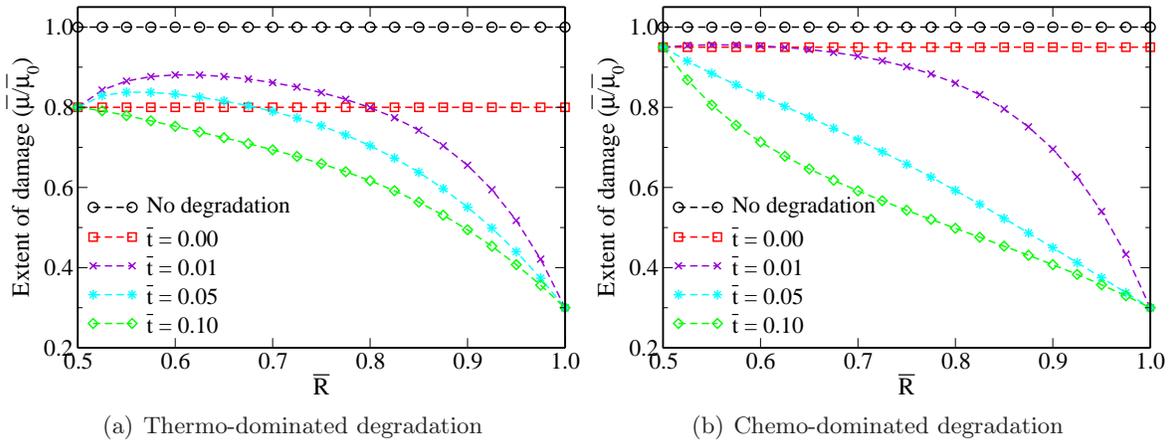

  \centering
    \subfigure[Thermo-dominated degradation]
    {\includegraphics[scale=0.3,clip]
    {Figures_Model/Spherical_Shell/mu_given_p.eps}}
  %%
  %\vspace{0.5in}
  %%
  \subfigure[Chemo-dominated degradation]
    {\includegraphics[scale=0.3,clip]
    {Figures_Model/Spherical_Shell/mu_given_p_diff_dom.eps}}
 \vspace{-0.1in}
  \caption{\textsf{Inflation of a degrading spherical shell:}~This 
    figure shows the extent of damage as a function of the reference 
    location at various instants of time due to inner pressure $\overline{p}_i 
    = 0.5$. Different values are chosen for $\overline{\mu}_1$ and 
    $\overline{\mu}_2$ for thermo-dominant and chemo-dominant 
    degradation.  Analysis is performed for strongly coupled case. 
    For thermo-dominated problem, healing-like behavior is observed 
    at early time steps. This is because at initial times, we have 
    variable heat sinks in the entire body. As $\overline{\vartheta} 
    \leq \overline{\vartheta}_0$, the material damage is less than 
    that of at time $\overline{t} = 0$ (but still remains below that 
    of the virgin material). However, this heal-like behavior becomes 
    less distinct (or even doesn't exist) when the chemo-degradation 
    achieves the dominance.    
    \label{Fig:Spherical_Shell_ExtOfDamage_given_p}}
\end{figure}

%--------------------------------------------------------------------;
%  Figure-9: Extent of damage across the cross section of the shell  ;
%            Given t: for weak and strong coupling                   ;
%--------------------------------------------------------------------;
\begin{figure}
  \centering
  \subfigure[Weak coupling degradation]
    {\includegraphics[scale=0.3,clip]
    {Figures_Model/Spherical_Shell/mu_given_t_weak.eps}}
  %%
 % \vspace{0.5in}
  %%
  \subfigure[Strong coupling degradation]
    {\includegraphics[scale=0.3,clip]
    {Figures_Model/Spherical_Shell/mu_given_t.eps}}
  \vspace{-0.1in}
  \caption{\textsf{Inflation of a degrading spherical shell:}~This 
    figure shows the extent of damage as a function of the reference 
    location at $\overline{t} = 1$ for various inner pressures `$\overline{p}_i$'. 
    Analysis is performed for thermo-dominated degradation. As the 
    pressure increases, for the weakly coupled problem, the extend 
    of damage decreases. This means that when the inflation pressure 
    $\overline{p}_i$ increases, the body degrades more significantly. 
    However, this is not the case for the strongly coupled problem. 
    In this particular case, thermo-mechano coupling dominates 
    and plays a vital role. As $\overline{I}_{\mathbf{E}} 
    \geq 0$, the strain-dependent thermal conductivity decreases as the 
    pressure loading increases. Hence, there is less damage in the material 
    due to the decrease in temperature values as compared to weakly coupled 
    chemo-thermo-mechano degradation problem.
    \label{Fig:Spherical_Shell_ExtOfDamage_given_t}}
\end{figure}

%--------------------------------------------------;
%  Figure-10: Pictorial Description: Beam bending  ;
%--------------------------------------------------;
\begin{figure}
  \centering
  \psfrag{R}{Reference configuration}
  \psfrag{C}{Current configuration}
  \psfrag{O1}{$\mathrm{O}_{\mathrm{ref}}$}
  \psfrag{O2}{$\mathrm{O}_{\mathrm{curr}}$}
  \psfrag{X}{$X$}
  \psfrag{Y}{$Y$}
  \psfrag{x}{$x$}
  \psfrag{y}{$y$}
  \psfrag{W}{$2L$}
  \psfrag{L}{\rotatebox{90}{$2W$}}
  \psfrag{p}{$\mathbf{x}$}
  \psfrag{t1}{$\theta$}
  \psfrag{rc}{$r_c$}
  \psfrag{r}{$r$}
  \includegraphics[scale=0.7]{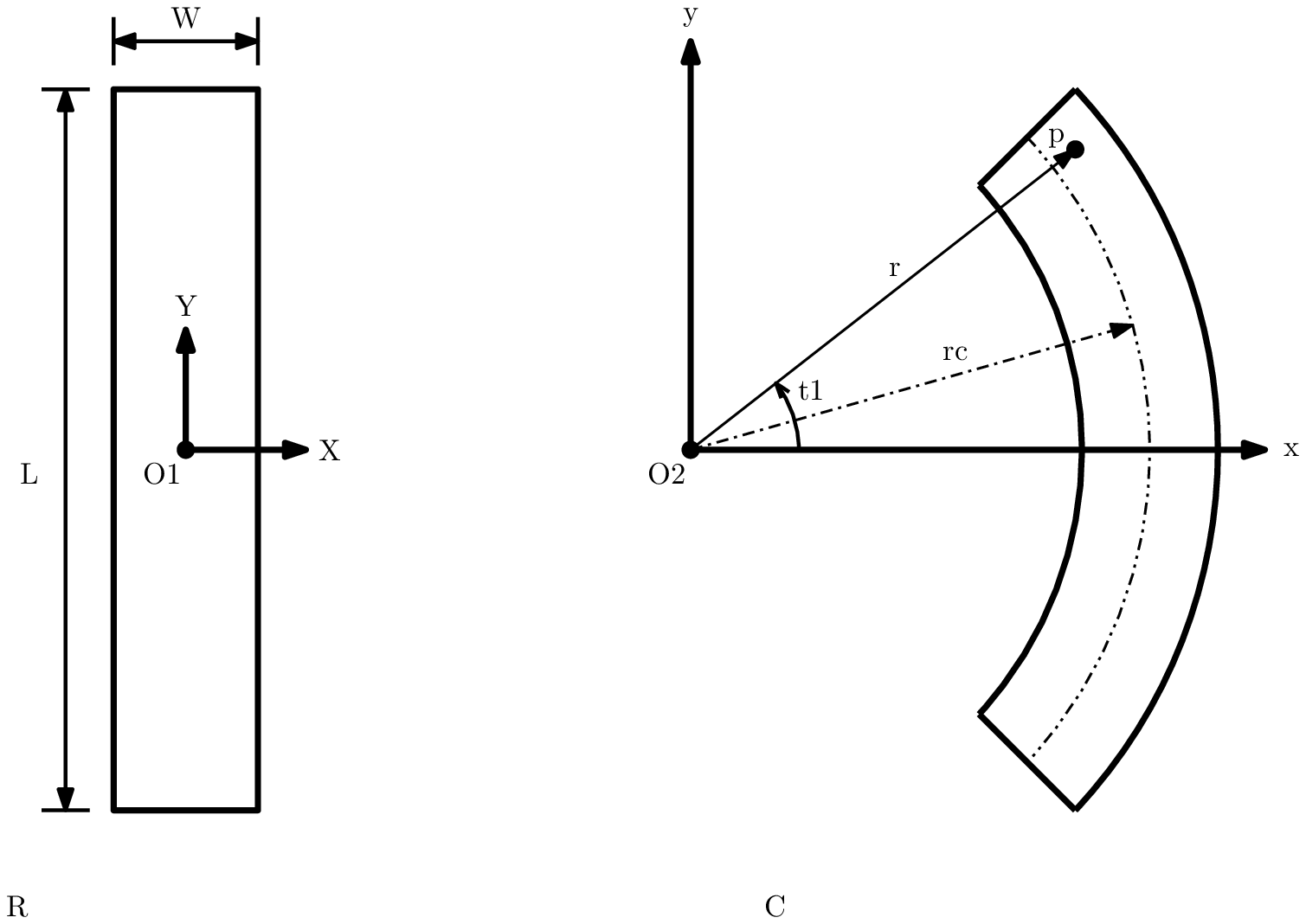}
  \caption{\textsf{Bending of a degrading beam:}~A 
    pictorial description of degrading beam in both 
    reference and current configurations. Bending moment 
    is applied at the two ends of the beam just after 
    time $\overline{t} = 0$. $\mathrm{O}_{\mathrm{ref}}$ 
    and $\mathrm{O}_{\mathrm{curr}}$ correspond to the 
    origin in the reference and current configurations.
    \label{Fig:Beam_Bending_PicDescription}}
\end{figure}

%----------------------------------------------------------------------------;
%  Figure-11: Neutral axis of the degrading beam (weak and strong coupling)  ;
%----------------------------------------------------------------------------;
\begin{figure}
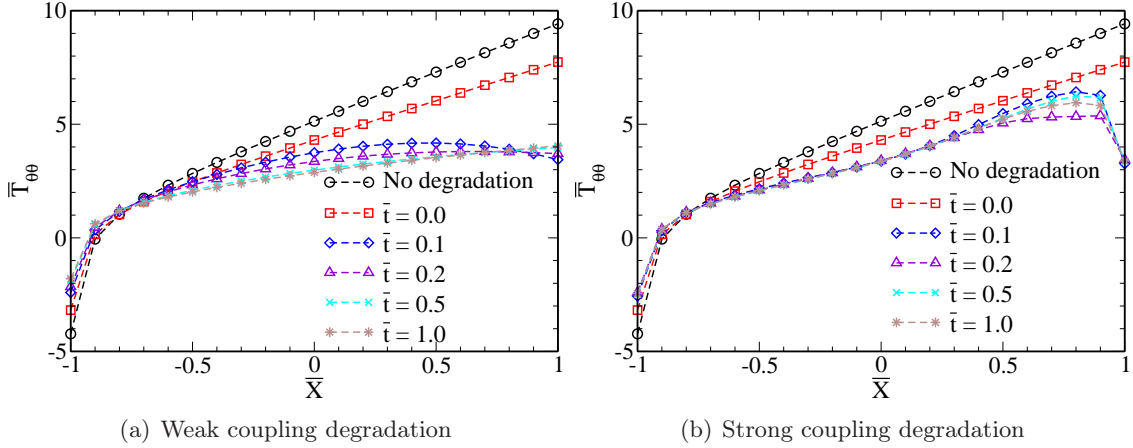

  \centering
  \subfigure[Weak coupling degradation]
    {\includegraphics[scale=0.3,clip]
    {Figures_Model/Beam_Bending/Weak_ThreeWay_Coupling.eps}}
  \vspace{0.5in}
  \subfigure[Strong coupling degradation]
    {\includegraphics[scale=0.3,clip]
    {Figures_Model/Beam_Bending/Strong_ThreeWay_Coupling.eps}}
  \vspace{-0.4in}
  \caption{\textsf{Bending of a degrading beam:}~This figure 
    shows the plot of $\overline{T}_{\theta \theta}$ as a function 
    of the reference location of the cross-section at various 
    instants of time. The stress distribution is not linear, 
    which is the case for finite deformation beam bending 
    problem. Herein, we observe that the neutral axis shifts 
    further to the left. Moreover, in case of weak coupling 
    for some instants of time the maximum stress does not 
    occur at either tensile or compressive sides of the beam 
    after the onset of degradation.
    \label{Fig:Beam_Bending_WeakStrongCoupling_NeutralAxis}}
\end{figure}
%
%--------------------------------------------------------;
%  Figure-12: Bending moment (weak and strong coupling)  ;
%--------------------------------------------------------;
\begin{figure}
  \centering
  \includegraphics[scale=0.45,clip]{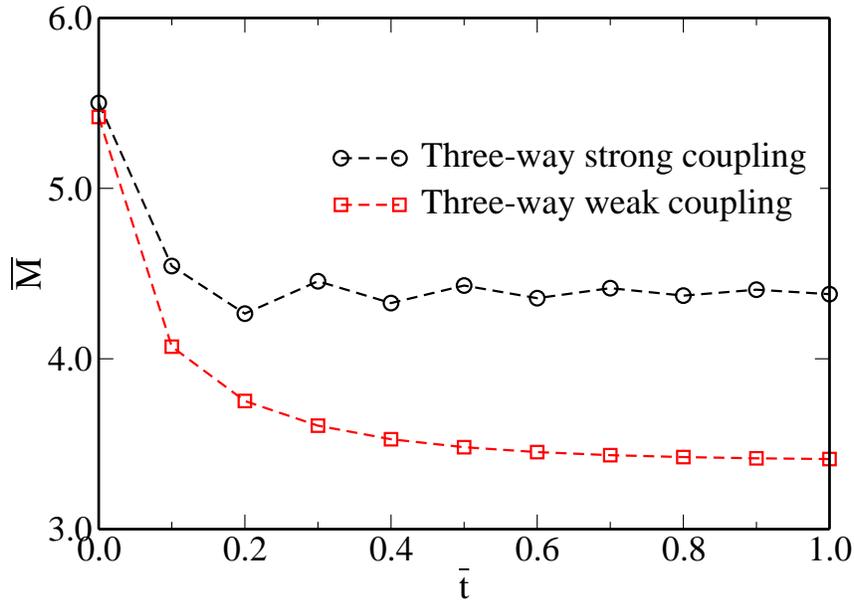}
  \caption{\textsf{Bending of a degrading beam:}~This 
    figure shows the plot of bending moment at various 
    instants of time for both strong and weak coupling 
    chemo-thermo-mechano degradation. Moment relaxation 
    is observed for both cases, however, in weak coupling 
    the moment declines at a much faster rate than that 
    of the strong coupling case. Note that the bending 
    moment is a constant without degradation.
    \label{Fig:Beam_Bending_StrongCoupling_Moment}}
\end{figure}
%
%---------------------------------------------------;
%  Figure-13: Chemical potential (strong coupling)  ;
%---------------------------------------------------;
\begin{figure}
  \centering
  \includegraphics[scale=0.38,clip]{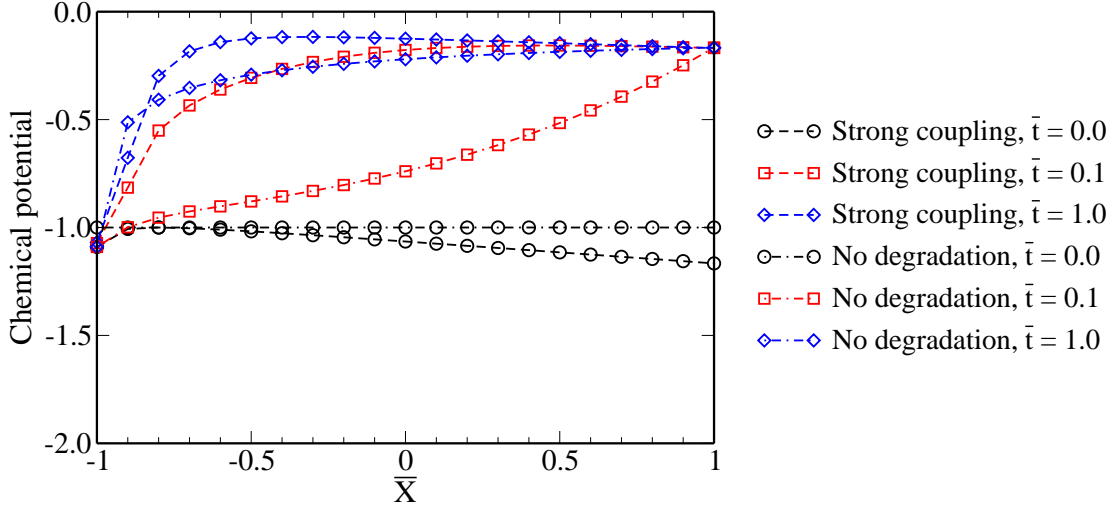}
  \caption{\textsf{Bending of a degrading beam:}~This 
    figure shows the plot of chemical potential as a 
    function of the reference location of the cross-section 
    at various instants of time when there is no degradation 
    and for strong coupling cases. In the strong coupling 
    scenario, although diffusion process is dominant, one 
    can still observe that the deformation has a significant 
    effect on chemical potential as compared with non-degradation 
    case.
    \label{Fig:Beam_Bending_StrongCoupling_chemical}}
\end{figure}
%
%-------------------------------------------------------------------;
%  Figure-14: Extent of damage across the cross section of the beam  ;
%-------------------------------------------------------------------;
\begin{figure}
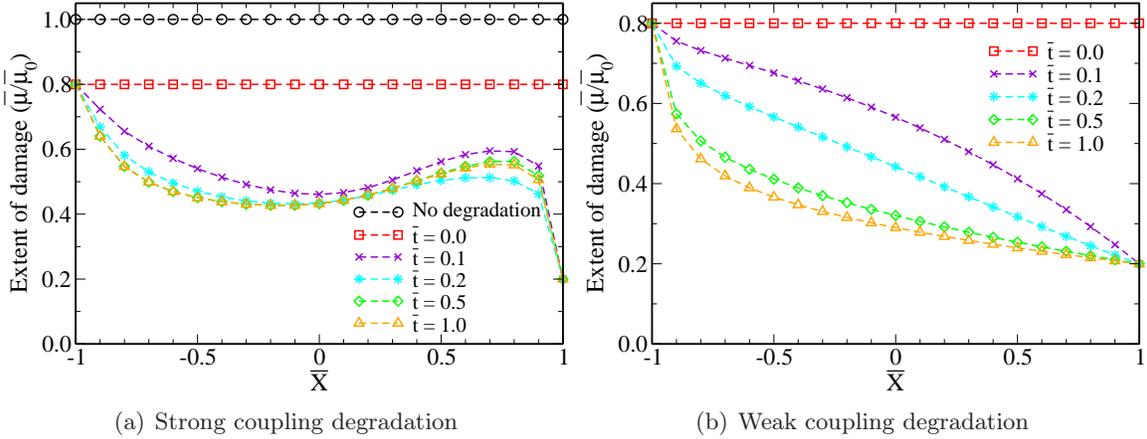

  \centering
  \subfigure[Strong coupling degradation]
    {\includegraphics[scale=0.3,clip]
    {Figures_Model/Beam_Bending/Damage_Along_CrossSection.eps}}
  %%
  %\vspace{0.5in}
  %%
  \subfigure[Weak coupling degradation]
    {\includegraphics[scale=0.3,clip]
    {Figures_Model/Beam_Bending/Damage_Along_CrossSection_Weak.eps}}
  \vspace{-0.1in}
  \caption{\textsf{Bending of a degrading beam:}~This figure shows 
    the extent of damage as a function of the reference location of 
    the cross-section at various instants of time (due to the application 
    of bending moment). Note that analysis is performed for both strongly 
    coupled and weakly coupled chemo-thermo-mechano degradation. One 
    can see that \emph{a virgin beam which is initially homogeneous 
    after degradation is not homogeneous anymore}. In addition, the extent 
    of damage is monotonic for weak coupling, which is not the case for 
    strong coupling. Such a phenomena has implications in damage control 
    and retrofitting of the degrading beams.
    \label{Fig:Beam_Bending_ExtOfDamage}}
\end{figure}
%----------------------------------------------------------;
%  Figure-15: Pictorial description of degrading cylinder  ;
%----------------------------------------------------------;
\begin{figure}
  \centering
  \includegraphics[scale=0.4]{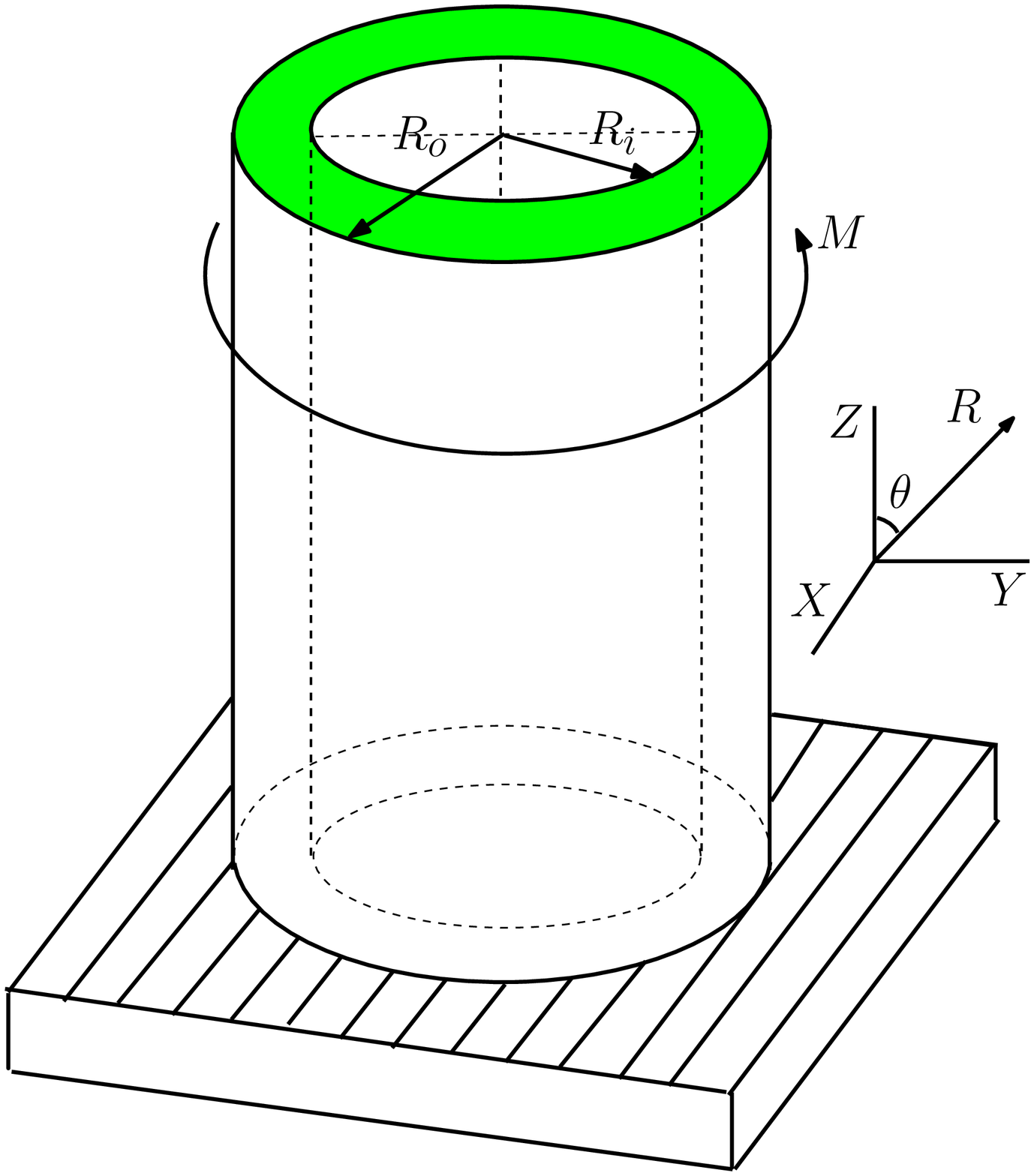}
  \caption{\textsf{Torsional shear of a degrading cylinder:}~A 
    pictorial description of the degrading cylinder under torsion 
    in the reference configuration. $R_i$ and $R_o$ are, respectively, 
    the inner and outer radii of the cylinder. $X$, $Y$, and $Z$ 
    are the Cartesian coordinates in the reference configuration. 
    The bottom of the cylinder is fixed and a twisting moment is 
    applied at the top of the cylinder for $t \geq 0$.
    \label{Fig:torsion_pictorial_description}}
\end{figure}

%-----------------------------------------------;
%  Figure-16: Moment vs t due to various \mu_1  ;
%-----------------------------------------------;
\begin{figure}
  \centering
   \subfigure[Moment under different $\overline{\mu}_1$]
    {\includegraphics[scale=0.4,clip]
    {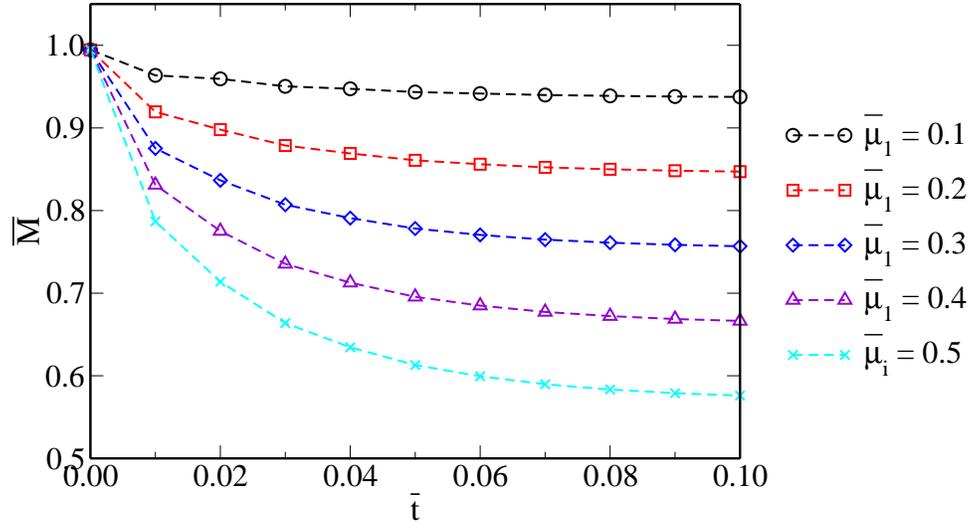}}
  \vspace{0.5in}
   \subfigure[Moment under different $\overline{R}_i$]
    {\includegraphics[scale=0.4,clip]
    {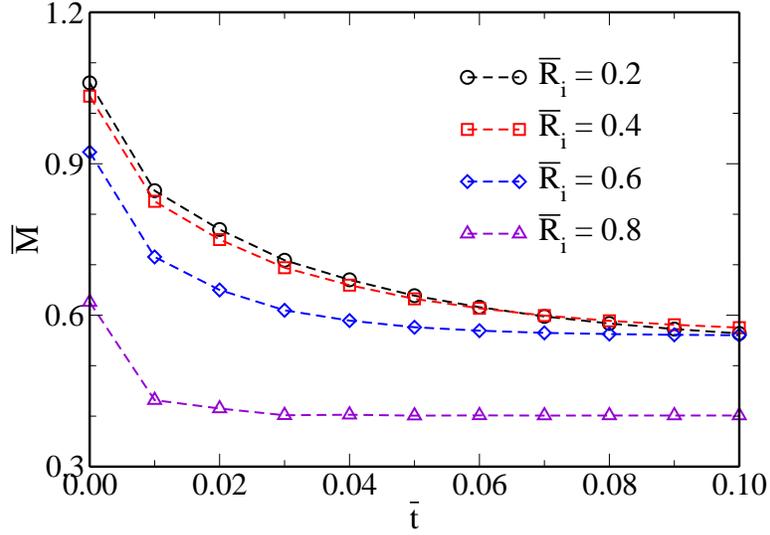}}
    \vspace{-0.4in}
  \caption{\textsf{Torsional shear of a degrading cylinder:}~This 
    figure shows the twisting moment at various instants of time 
    due to a given angle of twist per unit length of the cylinder,
    $\overline{\Psi}_1 = 0.75$. One can see that as $\overline{\mu}_1$ 
    increases the twisting moment required to keep $\overline{\Psi}_1$ 
    unchanged, decreases. Similar type of 
    behavior is observed when $\overline{\mu}_1$ is kept constant 
    and $\overline{\mu}_2$ is varied. \emph{Herein, the main observation is that 
    moment relaxation not only depends on material degradation but 
    also on the geometry of the degrading body}. 
    \label{Fig:Torsion_Moment_vs_t}}
\end{figure}

%----------------------------------------------------------------;
%  Figure-17: Normal force in torsional shear (Poynting effect)  ;
%----------------------------------------------------------------;
\begin{figure}
  \centering
  \includegraphics[scale=0.4]{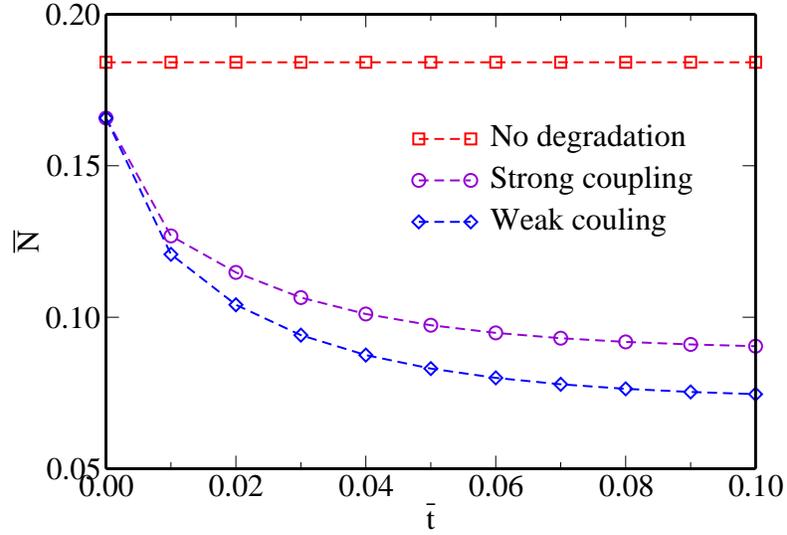}
  \caption{\textsf{Torsional shear of a degrading cylinder:}~This 
    figure shows the non-dimensionalized normal force $\overline{N}$ 
    due to Poynting effect at various instants of time. Analysis is 
    performed for a given angle of twist per unit length of the cylinder,
    $\overline{\Psi}_1 = 0.75$. When there is no degradation the normal 
    force is constant. However, due to degradation one can see that 
    the normal force relaxes over time. The decrease in this normal 
    force for weak-coupling is higher than that of the strong coupling.
    \label{Fig:Torsion_normal_force_vs_t}}
\end{figure}